\documentclass[10pt,journal,compsoc]{IEEEtran}
\usepackage{comment}
\usepackage{float}
\usepackage{tikz}
\usepackage{graphicx}
\usepackage[utf8]{inputenc}
\usepackage{mathtools}
\usepackage{amsmath}
\usepackage{pgfplots}
\usepackage{breqn}
\DeclarePairedDelimiter\ceil{\lceil}{\rceil}

\usepackage{subfiles}
\usepackage[justification=centering,font=scriptsize,skip=0pt, belowskip=-1pt,aboveskip=0pt]{caption}
\usepackage{amsmath}
\usepackage{epstopdf}
\usepackage{lipsum}
\usepackage[noadjust]{cite}
\usepackage[ruled, vlined, linesnumbered]{algorithm2e}
\SetAlFnt{\small\sffamily}

\SetCommentSty{mycommfont}
\usepackage{color}
\usepackage{pifont}
\newcommand{\subparagraph}{}
\usepackage{titlesec}

\usepackage{tabularx} % to break tables line to fit the page iwdth
\usepackage{multirow} % for multi-row cells
\usepackage{array}                                      %|
\newcolumntype{M}[1]{>{\centering\arraybackslash}m{#1}} %|
\usepackage{flushend}
\let\OLDthebibliography\thebibliography
\renewcommand\thebibliography[1]{
 \OLDthebibliography{#1}
 \setlength{\parskip}{0pt}
 \setlength{\itemsep}{0pt plus 0ex}
}
\title{DHT-based Communications Survey: Architectures and Use Cases \thanks{This work has been submitted to the IEEE for possible publication. Copyright may be transferred without notice, after which this version may no longer be accessible.}}

\author{Yahya Hassanzadeh-Nazarabadi\IEEEauthorrefmark{1}\IEEEauthorrefmark{4}, \textit{Member, IEEE}, Sanaz Taheri-Boshrooyeh\IEEEauthorrefmark{1}\IEEEauthorrefmark{5}, \textit{Member, IEEE}\\ Safa Otoum \IEEEauthorrefmark{3}, \textit{Member, IEEE}, Seyhan Ucar\IEEEauthorrefmark{2}, \textit{Member, IEEE}, Öznur Özkasap\IEEEauthorrefmark{1}, \textit{Senior Member, IEEE}\\
\IEEEauthorrefmark{1}Department of Computer Engineering, Koç University, İstanbul, Turkey \\ 
\IEEEauthorrefmark{4}DapperLabs, Vancouver, Canada \\
\IEEEauthorrefmark{5}Status Research and Development, Singapore  \\
\IEEEauthorrefmark{3}College of Technological Innovation (CTI), Zayed University, UAE \\
\IEEEauthorrefmark{2}InfoTech Labs, Toyota Motor North America R\&D, Mountain View, CA, USA \\
\{yhassanzadeh13, staheri14, oozkasap\}@ku.edu.tr, safa.otoum@zu.ac.ae, seyhan.ucar@toyota.com}

\pgfplotsset{compat=1.17}
\begin{document}

\IEEEtitleabstractindextext
{
\begin{abstract}

Several distributed system paradigms utilize Distributed Hash Tables (DHTs) to realize structured peer-to-peer (P2P) overlays. DHT structures arise as the most commonly used organizations for peers that can efficiently perform crucial services such as data storage, replication, query resolution, and load balancing. With the advances in various distributed system technologies, novel and efficient solutions based on DHTs emerge and play critical roles in system design. DHT-based methods and communications have been proposed to address challenges such as scalability, availability, reliability and performance, by considering unique characteristics of these technologies. In this article, we propose a classification of the state-of-the-art DHT-based methods focusing on their system architecture, communication, routing and technological aspects across various system domains. To the best of our knowledge, there is no comprehensive survey on DHT-based applications from system architecture and communication perspectives that spans various domains of recent distributed system technologies. We investigate the recently emerged DHT-based solutions in the seven key domains of edge and fog computing, cloud computing, blockchain, the Internet of Things (IoT), Online Social Networks (OSNs), Mobile Ad Hoc Networks (MANETs), and Vehicular Ad Hoc Networks (VANETs).
In contrast to the existing surveys, our study goes beyond the commonly known DHT methods such as storage, routing, and lookup, and identifies diverse DHT-based solutions including but not limited to aggregation, task scheduling, resource management and discovery, clustering and group management, federation, data dependency management, and data transmission.
Furthermore, we identify open problems and discuss future research guidelines for each domain.

\end{abstract}

}
\maketitle
\section{Introduction}
\label{dht_survey_section_intro}
Distributed Hash Tables (DHT)s \cite{stoica2001chord} are distributed key-value store architectures that realize the structured peer-to-peer (P2P) overlay paradigm. In such systems with $n$ nodes, each node (i.e., process) maintains communication channels to $O(\log{n})$ other nodes by following the DHT protocol, which shapes a connected distributed overlay graph of nodes. By utilizing this overlay graph, nodes can maintain addressable entities (e.g., data items or files) on DHT overlay graph, which resembles distributed \textit{put} query. The nodes can also query for each other and each others' entities, which resembles distributed \textit{get} query. Typically in DHTs, such distributed put and get queries are performed with a message complexity of $O(\log{n})$. Common examples of DHT overlays are Chord \cite{stoica2001chord}, Pastry \cite{rowstron2001pastry}, and Kademlia \cite{maymounkov2002kademlia}. Due to their scalability, fault tolerance, fast searching, correctness under concurrency, and load balancing, DHTs are widely used in various advanced distributed system technologies such as edge and fog computing \cite{simic2018edge, song2020smart, riabi2017proposal, sonbol2020edgekv, tanganelli2017fog, santos2018towards}, cloud computing \cite{xie2017design, gupta20172, nakagawa2015dht, govindarajan2017distributed, kavalionak2015integrating, benet2014ipfs, xiong2015full, hassanzadeh2015locality, hassanzadeh2016awake, hassanzadeh2016laras, hassanzadeh2018decentralized, hassanzadeh2019decentralized}, blockchains \cite{vora2018bheem, kalis2018validating, brunner2019sproof, cai2017hardening, iqbal2019trust, abe2019blockchain, abe2018mitigating, kaneko2018dht, huang2019zerocalo, yu2020virtual, ali2018blockchain, hassanzadeh2019lightchain, matsuoka2020blockchain, aniello2017prototype, cao2019building}, Internet of Things (IoT) \cite{tracey2019using,lua2005survey, zhang2015cloud, bolonio2013distributed, leiba2018incentivized, nguyen2017multiple, evdokimov2010comparison, fabian2009implementing, paganelli2012dht,fisher2009p2p, shrestha2010peer, manzanares2011efficient}, Online Social Networks (OSNs) \cite{gondor2015sonic, gondor2016distributed, xu2018oases, xu2018harnessing, nasir2015socially, badis2016routil, liu2013efficient, shen2014social, chakravorty2017ushare}, Mobile Ad Hoc Networks (MANETs) \cite{abid2015merging, shah2016cross, abid20153d, shah2014efficient, ramya2016deterring, manikandan2016stratified, tahir2017logical, ibrar2016stability, arunachalam2017broadcast, zahid2018distributed}, and Vehicular Ad Hoc Networks (VANETs) \cite{dressler2019virtual, dressler2014towards, hagenauer2016poster, hagenauer2018vehicular, ucar2019platoon, higuchi2019cooperative, malik2019trust, rowan2017securing, raj2016descriptive, jain2016rsu, lu2020federated}. 

% One paragraph per domain.
The \textit{cloud computing} paradigm offers on-demand availability of the computing and storage resources as services for the users \cite{tanenbaum2007distributed, androutsellis2004survey, fersi2013distributed, urdaneta2011survey}. A cloud computing system is typically comprised of a large-scale network of connected server nodes distributed across multiple data centers. The cloud computing systems are utilized by users through delegating their computing and storage tasks, without involving users for direct active management. DHTs are widely exploited in cloud computing systems to provide decentralized task scheduling \cite{xie2017design}, content aggregation \cite{nakagawa2015dht}, resource management \cite{govindarajan2017distributed, kavalionak2015integrating}, object storage \cite{benet2014ipfs, xiong2015full, hassanzadeh2016awake, hassanzadeh2018decentralized, hassanzadeh2016laras}, and load balancing \cite{gupta20172}.

In contrast to the cloud computing paradigm that spins around task delegation to remote data centers, the \textit{edge and fog computing} paradigm aims at moving the computation and storage resources from the cloud data centers closer to the end-users \cite{yousefpour2019all, varshney2017demystifying, karagiannis2019compute}. Processing data closer to the end-user is carried out with lower latency and higher efficiency by saving network bandwidth \cite{yousefpour2019all}. Hence, the edge and fog computing paradigm provides a more adaptable infrastructure for supporting the new emerging applications with latency requirements below what can be offered by the cloud computing systems, e.g., the $5^{th}$ Generation (5G)-based mobile network applications. The edge and fog computing solutions benefit from DHTs as a scalable distributed object storage platform \cite{simic2018edge, song2020smart, riabi2017proposal, sonbol2020edgekv} as well as a distributed resource discovery overlay \cite{tanganelli2017fog, santos2018towards}. 

The \textit{blockchains} are replicated state machines realizing an append-only database that establishes trust among a set of trustless participants, and enables a decentralized computation platform \cite{wood2014ethereum, hentschel2020flow}. DHTs are adopted in blockchain-supported systems to augment data integrity \cite{vora2018bheem,kalis2018validating, brunner2019sproof}, data privacy \cite{kosba2016hawk, wright2017sustainable},  trustworthy \cite{matsuoka2020blockchain, cai2017hardening, iqbal2019trust}, load balancing \cite{abe2019blockchain,abe2018mitigating, kaneko2018dht}, storage efficiency \cite{huang2019zerocalo, yu2020virtual, ali2018blockchain, hassanzadeh2019lightchain, hassanzadeh2020containerized, hassanzadeh2021lightchain, hassanzadeh2021smart}, and data management \cite{matsuoka2020blockchain, aniello2017prototype, cao2019building}. 

The \textit{Internet-of-Things (IoT)} is defined as the network of embedded connectors with other devices such as sensors and personal assistants that enables connecting and exchanging data with other devices and systems over the Internet \cite{cherbal2016survey, ghaleb2016mobility}. DHTs are implemented in the IoT infrastructures to support scalability \cite{tracey2019using,lua2005survey, zhang2015cloud, bolonio2013distributed, leiba2018incentivized, nguyen2017multiple}, and service discovery \cite{evdokimov2010comparison, fabian2009implementing, paganelli2012dht,fisher2009p2p, shrestha2010peer, manzanares2011efficient}. 

The \textit{Online Social Networks (OSNs)} are online platforms which connect people through social relationships with each other based on their personal or career interests, activities, backgrounds or real-life connections \cite{zuo2016survey,jameel2018wireless,siddula2018empirical,de2018survey}. In the context of OSNs, the DHTs are utilized to support distributed directory services \cite{gondor2015sonic, gondor2016distributed}, spam protection  \cite{xu2018oases, xu2018harnessing}, routing \cite{nasir2015socially, badis2016routil, liu2013efficient, shen2014social}, and data dependency management \cite{chakravorty2017ushare}.

The \textit{Mobile Ad Hoc Networks (MANETs)} are decentralized networks of mobile devices that do not rely on a pre-existing infrastructure (e.g., routers and access points). Rather, the network is established on-fly by nodes routing data for each other through forwarding to their directly connected nodes based on the routing algorithm in use \cite{abid2014survey}. In the context of MANETs, the DHTs are commonly utilized for efficient routing \cite{abid2015merging, shah2016cross, abid20153d, shah2014efficient}, data transmission \cite{ramya2016deterring, manikandan2016stratified}, handling the dynamic topology \cite{tahir2017logical, ibrar2016stability, arunachalam2017broadcast}, and mitigation of the traffic overhead \cite{zahid2018distributed}.

The \textit{Vehicular Ad Hoc Networks (VANETs)} realize the concept of MANETs in the domain of vehicles through vehicle-to-vehicle and vehicle-to-roadside connections, which enables relaying information among the vehicles, hence providing road safety, navigation, and other roadside services \cite{cooper2016comparative}. In the context of VANETs the DHTs are typically employed to provide service directory \cite{dressler2019virtual, dressler2014towards, hagenauer2016poster, hagenauer2018vehicular, ucar2019platoon}, scalable routing \cite{higuchi2019cooperative, malik2019trust, rowan2017securing, raj2016descriptive, jain2016rsu}, security and privacy \cite{lu2020federated}. For the different levels of the hierarchy, vehicles use various communication technologies to get benefit from DHTs. Vehicle-to-Vehicle (V2V) communication technologies (e.g., Dedicated Short Range Communication) are used, and the DHT overlay is cooperatively maintained among vehicles as the nodes. Vehicles use short-range communication technologies to communicate over the DHT. Vehicle-to-Infrastructure (V2I) and/or Vehicle-to-Cloud-to-Vehicle (V2C2V) are used to update the DHT at the edge level. The cloud/edge layer keeps the DHT, and it provides an access interface so vehicles can access and update the DHT accordingly.

% state the key contributions - new
\textbf{Original Contributions and Novelties:} To the best of our knowledge \textit{this is the first survey on state-of-the-art DHT-based solutions in domains of edge, fog, and cloud computing, blockchains, IoT, MANETs, VANETs, and OSNs from system architecture, communication, routing, and technological perspectives}. We also survey recently published papers, most of which have not been appeared yet in any relevant survey. As we discuss subsequently in this section, in contrast to the existing surveys in the various domains of distributed systems, our study sheds light on the more advanced, less recognized, and yet important capabilities empowered by DHTs. Our taxonomy in this paper goes beyond the commonly known DHT methods such as storage, routing, and lookup, and extracts more sophisticated and diverse DHT-based solutions including but not limited to aggregation, task scheduling, resource management and discovery, spam protection, clustering and group management, federation, data dependency management, data transmission, security, and privacy. Furthermore, we identify open problems and discuss future research guidelines for each distributed system domain. Moreover, we present the principles behind various types of DHTs in terms of their architecture, construction and routing, covering the most well-known DHTs (i.e., Chord \cite{stoica2001chord}, Kademlia \cite{maymounkov2002kademlia}, and Pastry \cite{rowstron2001pastry}), as well as \textit{the least surveyed ones} (i.e., Skip Graph \cite{aspnes2007skip} and Cycloid \cite{shen2004cycloid}). 

\textbf{Related Works:}
To the best of our knowledge, there is no comprehensive survey on DHT-based approaches from system architecture, communication, and technological perspectives that spans all the domains of edge, fog, and cloud computing, blockchain, IoT, OSNs, MANETs, and VANETs. Moreover, there is no existing survey study on the DHT-based applications in \textit{any} domains of edge \cite{toosi2014interconnected}, fog \cite{yousefpour2019all, varshney2017demystifying, karagiannis2019compute}, and cloud computing \cite{androutsellis2004survey, fersi2013distributed, urdaneta2011survey}. In the domain of blockchain \cite{sharma2020blockchain, elmamy2020survey}, the only DHT-based work is \cite{berdik2021survey}, which makes a cursory glance at the utilization of DHT-based data storage functionality while scoping out the aspects that we cover in this survey such as integrity, privacy, trustworthy, load balancing, and data management. In the context of IoT, the existing DHT-based surveys cover features such as the routing and lookup process \cite{cherbal2016survey}, and the mobility management \cite{ghaleb2016mobility}, while missing discussions on DHT-based solutions for supporting scalability and service discovery, which are covered in this survey. 
For the OSNs, none of the existing surveys \cite{zuo2016survey,jameel2018wireless,siddula2018empirical,de2018survey} adopt a DHT-oriented perspective in their classification, and solely provide a partial study on DHTs as a distributed routing and storage management candidate among the other OSN-based technologies, while leaving the aspects such as spam protection, service discovery, and data dependency management, that are covered in this survey. 
Considering the domain of MANETs, the only existing DHT-based survey \cite{abid2014survey} studies routing protocols, without going further in aspects that are covered by this survey such as data transmission, dynamic topology management, and traffic overhead migration. Similarly, in the field of VANETs, the only existing DHT-based survey \cite{cooper2016comparative} presents utilization of DHTs for distributed cluster management, yet leaving other aspects that are covered by our survey, i.e., service discovery, scalable routing, security, and privacy. Additionally, these existing surveys lack a thorough investigation of the DHT utilization challenges, open problems, and research guidelines in their relative domains. 

\textbf{Paper Organization:} The remainder of the paper is organized as follows. In Section \ref{section_routing_overlay}, we present the principles behind various types of DHTs in terms of their architecture, lookup table construction and routing. Based on our taxonomy we survey the system models, methods, applications and open problems of the DHT-based solutions in the respective distributed system domains of edge and fog computing (Section \ref{dht_survey:section_edge_computing}), cloud computing (Section \ref{dht_survey:section_cloud_computing}), blockchains (Section \ref{BC-DHT}), IoT (Section \ref{IoT-DHT}), OSNs (Section \ref{dht_survey:section_social_network}), MANETs (Section \ref{dht_survey:sec_manet}), and VANETs (Section \ref{section_vanet}). Section \ref{dht_survey:conclusion} presents the concluding remarks.
\section{DHT Overlays: A System Architecture Overview} \label{section_routing_overlay}

In this section, we present an architectural overview of the most typical DHT overlays (i.e., Chord \cite{stoica2001chord}, Kademlia \cite{maymounkov2002kademlia}, and Pastry \cite{rowstron2001pastry}), as well as some of the least surveyed ones (i.e., Skip Graph \cite{aspnes2007skip} and Cycloid \cite{shen2004cycloid}). We model a DHT as a distributed key-value store of entities. In the context of DHTs, we define entities as the processes running the DHT software (i.e., \textit{DHT nodes}) as well as the data objects they maintain. Each entity is represented by a unique key, which is named the identifier of that entity. The identifiers of the DHT nodes are determined by taking the collision-resistant hash value \cite{katz2014introduction} of their unique network (IP) address. While the identifiers of the data objects are typically determined using the collision-resistant hash value of their content. In practice, a collision-resistant hash function with a large identifier space is used to generate the identifiers, e.g., SHA-1 with $160$-bits identifier size. Having the hash function collision-resistant guarantees a one-to-one mapping from the DHT nodes as well as the data objects to their identifiers with a very high probability. In a DHT overlay with $n$ nodes, a DHT node commonly maintains connections to $O(\log{n})$ other nodes in a table, which is called its \textit{lookup table}. Remote Procedure Call (RPC) \cite{tanenbaum2007distributed} is a typical communication protocol between DHT nodes, however, TCP and UDP protocols \cite{kurose2010computer} are also utilized in some architectures, e.g., Kademlia \cite{maymounkov2002kademlia} nodes utilize UDP protocol to communicate with their lookup table neighbors. The lookup table neighbor relationship forms a connected graph of processes, which is called a \textit{DHT overlay}. The DHT overlay enables DHT nodes to store and maintain data objects in the DHT key-value store, as well as to efficiently search for each other and each others' data objects across the overlay. Both operations are done in a fully distributed manner within a message complexity of $O(\log{n})$. Storing a data object in a DHT is a distributed \textit{put} operation that lays the data object in the DHT node that has the closest identifier to it. The identifier closeness is domain-specific for each DHT and is defined in the identifier space of that DHT. We touch on this topic in more depth in the rest of this section. 
A lookup operation corresponds to a distributed \textit{get} operation, which is initiated by a \textit{lookup initiator} node for a \textit{target identifier}. The target identifier can be the identifier of a process or a data object. As the result of a lookup operation, the lookup message is routed collectively by the nodes on the lookup path from the lookup initiator to the node with the closest identifier to the lookup target, which terminates the lookup and announces itself as the lookup result back to the initiator. In the rest of this section, we introduce an overview of the architecture of different DHT types, their lookup table structure, and their lookup operation. We skip details of the DHT overlays construction as well as storing data objects on them for sake of space and refer the interested readers to their original papers. Also, in the rest of this section, we use the terms \textit{peer} and \textit{process} interchangeably both referring to a DHT node.

\subsection{Chord}

\begin{figure}
\centering
   \includegraphics[scale=0.35]{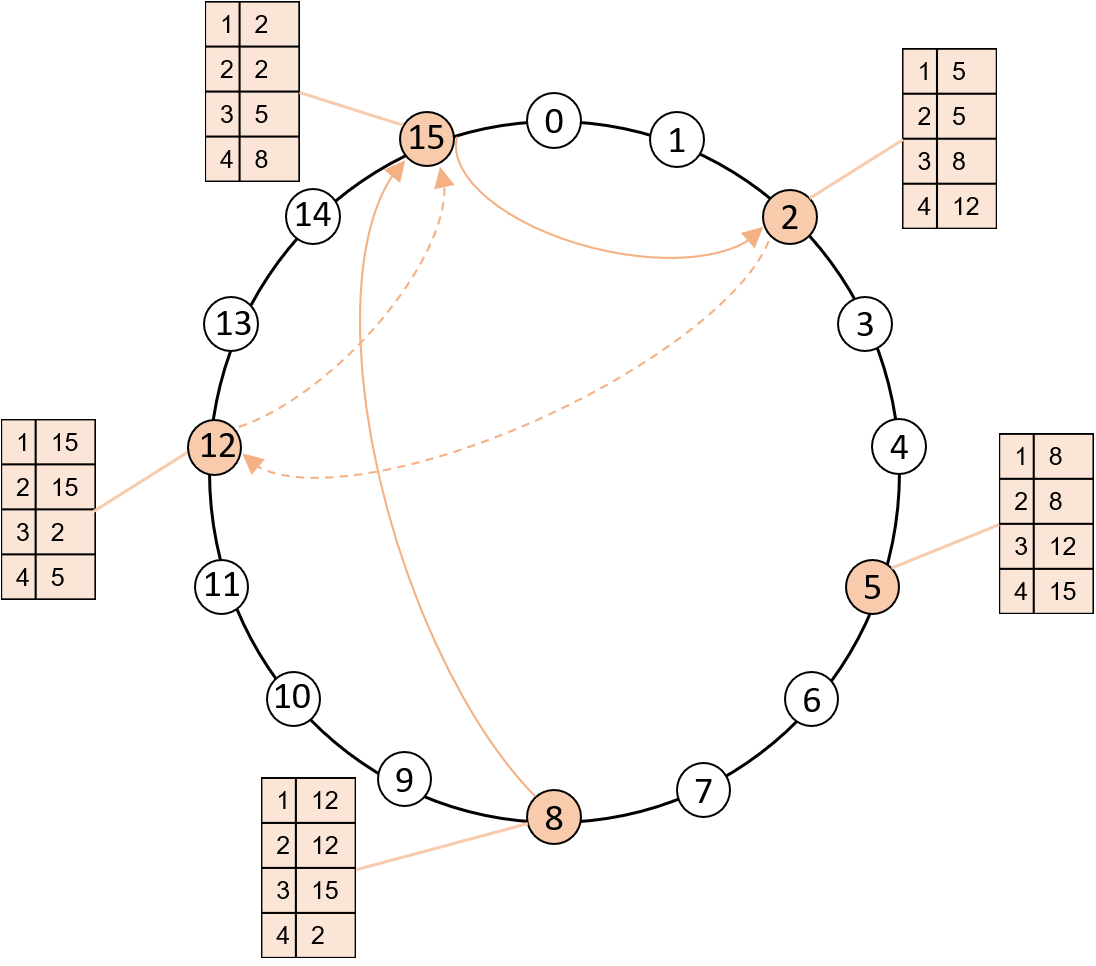}
\caption{An example Chord DHT with $m$=$4$ and $5$ peers. Peers are represented by colored circles with their finger tables attached, while the data items are represented by colored and white circles. Steps of resolving $lookup(1)$ initiated by peer $8$ and $lookup(13)$ initiated by peer $2$ are indicated by solid and dash arrows, respectively.}
\label{fig_chord_example}
\end{figure}

\textbf{Architecture:}
Utilizing consistent hashing with an $m$-bit identifier space, peers in Chord \cite{stoica2001chord} are assigned random identifiers ranging from $0$ to $2^m-1$ and organized in a logical ring overlay structure. A data item with key $d$ on the Chord ring is stored at the peer with the smallest identifier $p \ge d$, namely the $successor$ of $d$. Thus, a peer's responsibility of data item keys ranges from the identifier of its predecessor peer plus 1 to its own identifier. In general, with $k$ keys and $n$ peers in a Chord system, each peer stores $O(k/n)$ keys (that is $ < c \times k/n$  for some constant $c$) providing load balancing in the distribution of data items over the peers.
Figure \ref{fig_chord_example} shows an example Chord DHT overlay with a $4$-bit identifier space (i.e., $m$=$4$), and $5$ peers where the peer identifiers are $2$, $5$, $8$, $12$, $15$. For example, successor of peer $2$ is peer $5$, and successor of peer $8$ is peer $12$. Peer $5$ is responsible for data items with keys  $3$ and $4$, while peer $12$ is responsible for data items with keys  $9$, $10$, and $11$.

\textbf{Lookup Table:} 
The lookup table in Chord DHT is also called the \textit{finger table}. Each peer $p$ maintains a finger table $F$ with at most $m$ entries such that entry $i$ ($1\le i \le m$) of $F$ keeps the identifier of the first peer succeeding $p$ by the distance of at least  $2^{i-1}$ in the identifier space. That is, entry $i$ of peer $p$'s finger table $F$ would be $F[i] = successor (p+2^{i-1})$. Figure \ref{fig_chord_example} indicates example finger table of each peer in the system where the number of entries in each finger table is $m$=$4$.

\textbf{Routing:} 
If a peer with an identifier of $d$ exists in Chord, the lookup operation for $d$ (i.e., $lookup(d)$) returns its network address. Otherwise, the lookup operation returns the address of the successor of $d$ in the Chord ring. The latter case corresponds to looking up a data item, which returns the address of the responsible peer for it. A peer $p$ that initiates $lookup(d)$ forwards the message to the peer in index $j$ of its finger table with $F[j] \le d < F[j+1]$. There are two exceptions though: (1) when $p \le d < F[1]$, the forwarding is done to the peer in index $1$ of its finger table, and (2) when $d < F[i]\,, \forall{i}$, the forwarding is done to the peer in the largest index of the lookup table. The lookup query resolution continues in this manner by each peer on the lookup path forwarding the lookup message to the next peer based on the finger table, until the lookup message reaches the successor peer responsible for the lookup key $d$. Since a peer has a minimum distance of $2^{j-1}$ to the $j$ index of its finger table, at each step of the routing process, the distance between the lookup message and the lookup target drops by a factor of at least $2$, which resembles a binary search in a decentralized fashion. Therefore, in a Chord DHT with $n$ nodes, the routing of a lookup message takes a message and round complexity of $O(\log{n})$.
Figure \ref{fig_chord_example} illustrates the steps of routing $lookup(1)$ (solid arrows), and $lookup(13)$ (dash arrows) initiated by peer $8$ and $2$, respectively.

\subsection{Kademlia} \label{subsection_routing_overlay_Kademlia}
\begin{figure}
\begin{center}
    \includegraphics[scale=0.18]{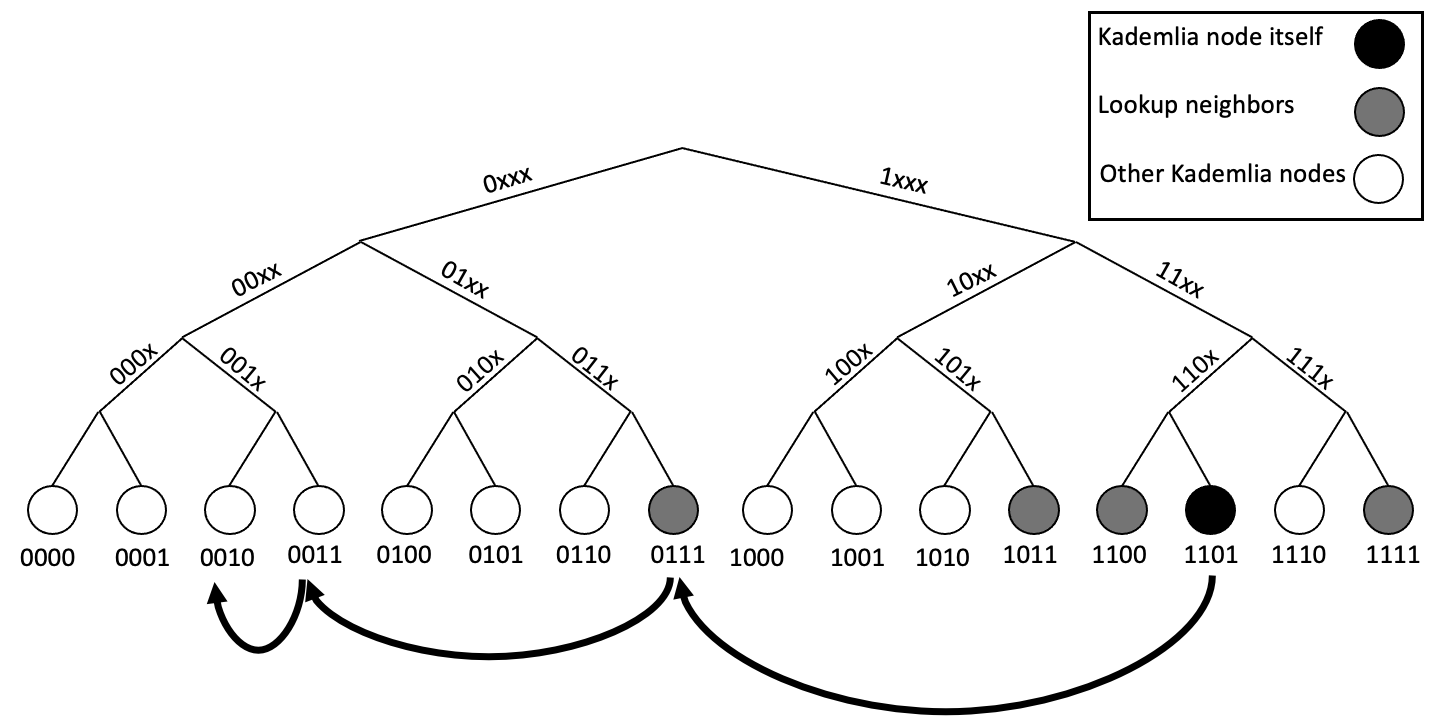}
\caption{The identifier space in a Kademlia overlay with $n = 16$ nodes. Arrows represent the lookup path for the target identifier of $0010$ that is instantiated by node $1101$.} \label{dht:fig_kademlia}
\end{center}
\end{figure}

\noindent\textbf{Architecture}: In a system with $n$ nodes, each Kademlia \cite{maymounkov2002kademlia} node is assigned an $\ceil{\log{n}}$-bit identifier and keeps a lookup table of $\ceil{\log{n}}$ neighbors. The identifier distance in Kademlia is measured as the length of the common prefix, where a longer common prefix corresponds to a shorter identifier distance in the identifier space. As shown by Figure \ref{dht:fig_kademlia}, the identifier space in Kademlia is typically illustrated as a binary tree, where the leaves are \textit{the actual Kademlia nodes}, and the intermediate nodes are the identifier sub-domains. It is worth noting that such binary tree representation of identifier space in Kademlia is merely an abstraction, and there is no real-world implication of such a centralized binary tree. Rather, as we explain later in this subsection, to establish a Kademlia overlay, it is sufficient for each Kademlia node to maintain connections to its lookup table neighbors.

\noindent\textbf{Lookup Table}: In a Kademlia overlay of $n$ nodes, each node is assigned an $\ceil{\log{n}}$-bit identifier. Accordingly, for every possible common prefix length $x \in [0, \ceil{\log{n}} - 1]$ bits, the node keeps one lookup table neighbor with exactly $x$ bits common prefix length. In the example illustrated by Figure \ref{dht:fig_kademlia}, the lookup table neighbors of node $1101$ are shaded in gray, which are $0111$ (zero-bit common prefix), $1011$ (one-bit common prefix), $1111$ (two-bits common prefix), and $1100$ (three-bits common prefix). 

\textbf{Routing:} Upon receiving a lookup message for a target identifier, the Kademlia node routes the message to its lookup table neighbor with the longest common prefix in its identifier to the lookup target. Note that based on the structure of the lookup table, each Kademlia node has more closer neighbors in the identifier space, than faraway ones. Thus, on each step of routing, the lookup message gets closer to the target identifier in the identifier space. Routing a lookup message is terminated at a node if either it has the lookup target identifier, or none of its lookup table neighbors have a closer identifier to the target than the node itself. The lookup path for the target identifier $0010$ initiated by node $1101$ is illustrated by arrows in Figure \ref{dht:fig_kademlia}. The lookup initiator $1101$ forwards the lookup message to its neighbor $0111$ that has the longest common prefix to the target (i.e., $1$ bit common prefix). On receiving the lookup message for $0010$, the node $0111$ forwards it to its neighbor $0011$ which has the longest common prefix to the target (i.e., $3$ bits common prefix), and results in the lookup message eventually reaches its target.

\subsection{Pastry} \label{subsection_routing_overlay_pastry}

\noindent\textbf{Architecture}: Pastry \cite{rowstron2001pastry} has a similar ring architecture to Chord \cite{stoica2001chord}. However, the identifiers of nodes are commonly represented as sequences of digits in the base $b$. A typical choice of $b$ in Pastry is $16$, which represents the identifiers in hexadecimal. The identifier distance in a Pastry overlay is gauged based on the common prefix length in the binary representation, i.e., the longer the common prefix of two identifiers are, the closer they are together in the identifier space. 

\begin{figure}
\begin{center}
    \includegraphics[scale = 0.20]{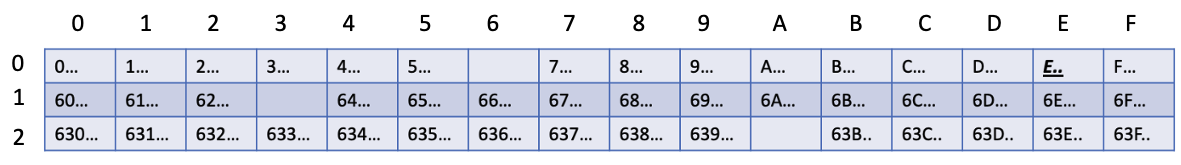}
\caption{The lookup table of a hypothetical Pastry node with the identifier of $63AB$ in a system where $n = 10,000$ and $b = 16$.} \label{fig_dht_pastry_lookup_table}
\end{center}
\end{figure}

\textbf{Lookup Table}: With the maximum number of $n$ nodes in the system, the lookup table of a Pastry node has $\ceil{\log_{b}{n}}-1$ rows and $b$ columns. Figure \ref{fig_dht_pastry_lookup_table} illustrates the lookup table of a hypothetical Pastry node with the identifier of $63AB$ in a system where $n = 10,000$ and $b = 16$. The lookup table consists of $\ceil{\log_{16}{10000}}-1 = 3$ rows and $16$ columns. The rows are numbered top-down starting from $0$, while the columns are numbered left to right starting from $0$ to $F$, each representing a hexadecimal digit. As a general principle, the lookup table neighbors at the row $i$ all have \textit{exactly} $i$ digits common prefix length in their identifiers with the current node, while the $i+1^{th}$ digit of their identifier is equal to the corresponding digit of their column (in base $b$). As shown by the example of Figure \ref{fig_dht_pastry_lookup_table}, the lookup table neighbors at the row $0$ (i.e., the top-most row) have exactly $0$-digit common prefix (that is no common prefix) with the node $63AB$ itself. Since all the identifiers starting with $6$ (i.e., $6...$) have at least one digit common prefix with $63AB$, the column representing the identifier prefix of $6$ in the row $0$ of the lookup table of node $63AB$ is left empty. 
Similarly, all the lookup table neighbors at the row $1$ have exactly 1-digit common prefix with the node $63AB$ itself, while their second digit varies from $0$ to $F$ depending on their corresponding column. Since all identifiers starting with the prefix of $63$ (i.e., $63..$) have more than one digit common prefix with $63AB$, the column representing the identifier prefix of $63$ in the row $1$ of the lookup table of the node $63AB$ is left empty. Finally, all lookup table neighbors in row $2$ (i.e., the bottom-most row) have exactly 2-digit common prefix of $63$ with node $63AB$, while their third digit varies from $0$ to $F$ depending on their corresponding column. Since, all identifiers starting with $63A$ (i.e., $63A.$) have more than two digits common prefix with $63AB$, the column representing the identifier prefix of $63A$ in the row $2$ of the lookup table of node $63AB$ is left empty.

\begin{figure}
\begin{center}
    \includegraphics[scale = 0.25]{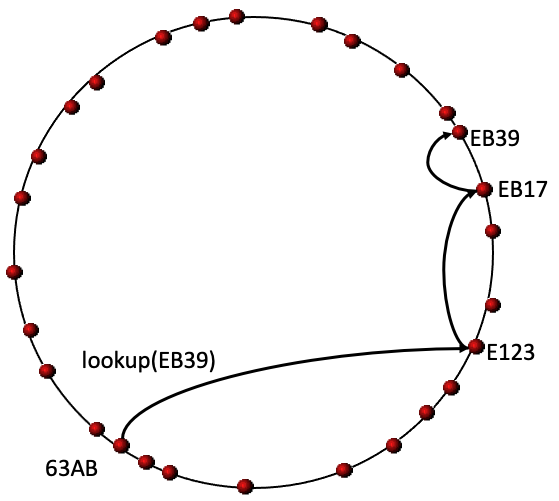}
\caption{A lookup example from node $63AB$ for the target identifier $EB3E$.} \label{fig_dht_pastry}
\end{center}
\end{figure}

\textbf{Routing}: Upon receiving a lookup message for a target identifier, a Pastry node routes the lookup message by forwarding it to the lookup table neighbor that has the longest common prefix with the lookup target. The lookup terminates when a node on the lookup path lacks any neighbor with a longer common prefix length to the lookup target than itself. Based on the structure of lookup tables in the Pastry, routing based on prefix guarantees that each node on the lookup path has at least one digit longer common prefix with the lookup target than the preceding node on the path. 
%Hence, the message complexity of a lookup operation in Pastry is $O(\log_{b}{n})$. 
An example lookup operation for the target identifier of $EB39$ from $63AB$ is represented by Figure \ref{fig_dht_pastry}, where at each step of the lookup, the identifier distance between the node on the path and the target decreases by at least one digit. 

\subsection{Skip Graphs}

\begin{figure}
\centering
   \includegraphics[scale=0.15]{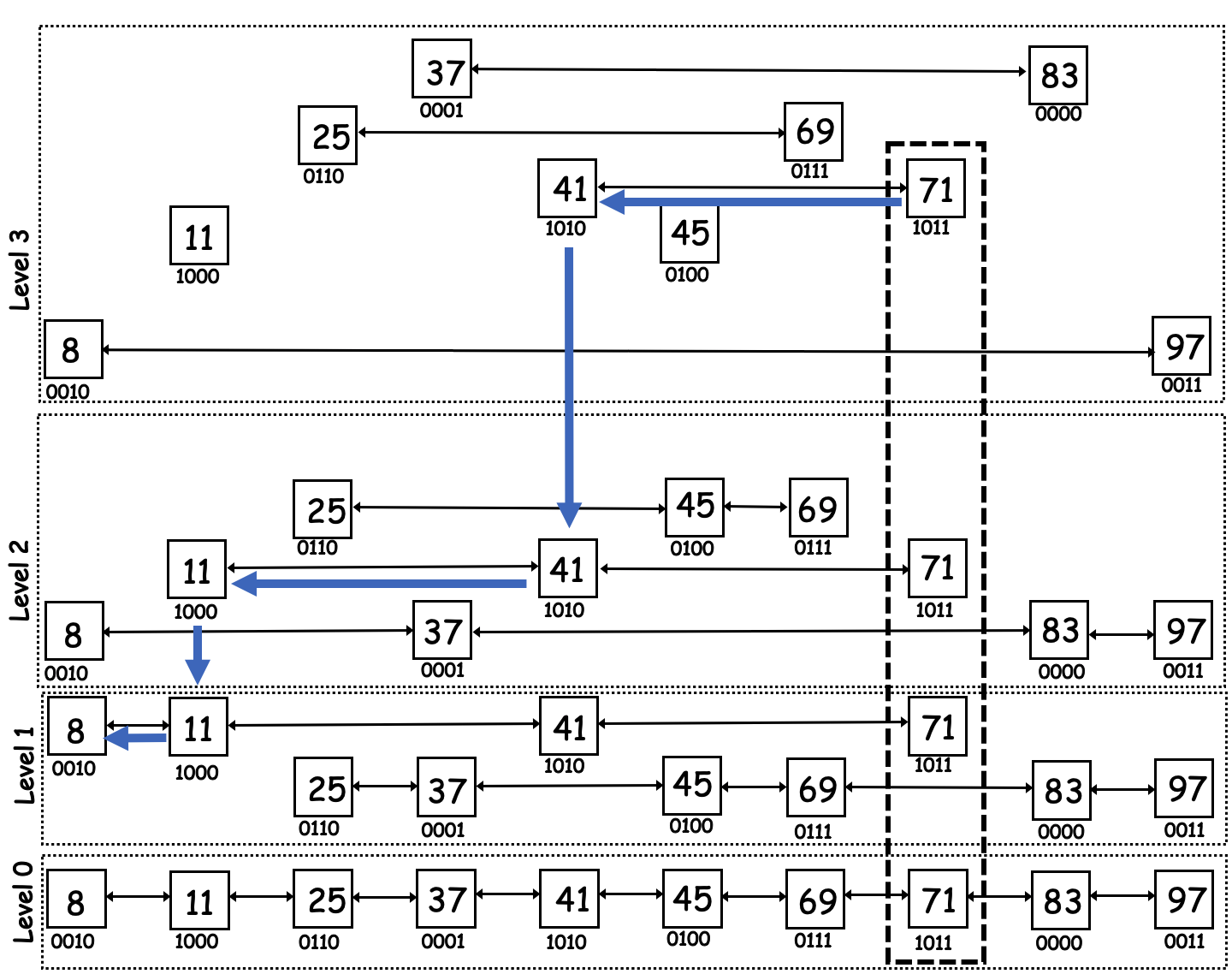}
\caption{An example Skip Graph routing overlay with $10$ nodes and $4$ levels. The Skip Graph nodes are illustrated by squares with the numerical IDs enclosed and name IDs beneath. The dashed rectangle groups all elements of the node with a numerical ID of $71$ together across all levels. The tick blue arrows show a lookup example for the numerical ID of $8$ that is initiated by node $71$.}
\label{fig_dht_skipgraph}
\end{figure}

\textbf{Architecture:}  
A Skip Graph with $n$ nodes has $O(\log{n})$ levels that are marked up starting from level $0$. Each node of Skip Graph is identified by two identifiers; a numerical ID and a name ID and has exactly one element at each level \cite{aspnes2007skip, hassanzadeh2020skip}. Numerical IDs are non-negative integers, and the name IDs are binary strings of length $O(\log{n})$ bits. Figure \ref{fig_dht_skipgraph} illustrates a Skip Graph DHT structure with $10$ nodes and $4$ levels. In this example, the elements of the Skip Graph node (with a numerical ID of) $71$ across all the levels of Skip Graph are inscribed in a dashed rectangle.
%In the level $0$ of Figure \ref{fig_dht_skipgraph}, there exists a single list (i.e., $2^{0} = 1$) that contains all nodes sorted based on their numerical IDs. There exists $2^1 = 2$ lists at the level $1$, one for the name ID prefix of $0$ and the other one for the prefix of $1$. Similarly, in the level $2$, there exist $2^{2} = 4$ lists that represent the name ID prefixes of $00$, $01$, $10$, and $11$. Note that since there is no node with the name ID prefix of $11$ in this Skip Graph, the linked-list representing this name ID prefix is left empty on the level $2$. Finally, there exists $2^{3} = 8$ lists at the level $3$ of the Skip Graph representing all $3$-bits name ID prefixes including two empty lists that represent the name ID prefixes of $110$ and $111$ as there is no node in the illustrated Skip Graph with these name ID prefixes. Likewise, as there is no other node of the Skip Graph with the name ID prefixes of $100$ and $010$, the nodes $11$ and $45$ solely represent their corresponding prefix linked-lists at level $3$, respectively.

\begin{figure}
\centering
   \includegraphics[scale=0.25]{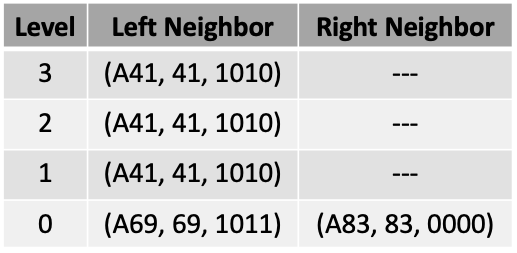}
\caption{Lookup table of node $71$ in the Skip Graph overlay of Figure \ref{fig_dht_skipgraph}. Each lookup table neighbor is represented by the tuple of $($address, numerical ID, name ID$)$. In this representation, $Axx$ is the (network) address of the node with the numerical ID of $xx$.}
\label{fig_dht_skipgraph_lookup_table}
\end{figure}

\textbf{Lookup Table:} Figure \ref{fig_dht_skipgraph_lookup_table} illustrates the lookup table structure of node $71$ from the Skip Graph overlay example of Figure \ref{fig_dht_skipgraph}. As shown by Figure \ref{fig_dht_skipgraph_lookup_table}, the lookup table of a Skip Graph node always has two columns (i.e., left and right) and as many rows as the number of the Skip Graph's levels. The lookup table neighbors of a node at the row $i$ are the left and right nodes it is connected to at the level $i$ of the Skip Graph. A lookup table neighbor is represented by the tuple of $($address, numerical ID, name ID$)$. For example, the left neighbor of node $71$ at the level $2$ of the Skip Graph of Figure \ref{fig_dht_skipgraph} is node $41$. This left neighbor is maintained at the row number $2$ of the lookup table of node $71$ as $(A41, \, 41,\, 1010)$ tuple, where $A41$ is its network address.
% Since a Skip Graph with $n$ nodes has $O(\log{n})$ levels, and the lookup table of the nodes have as many rows as the number of levels in the Skip Graph overlay, the number of lookup table neighbors of a node is also $O(\log{n})$. 

\textbf{Routing:}
In Skip Graphs, the lookup messages are routed either based on the name IDs of the nodes \cite{hassanzadeh2016laras} or their numerical IDs \cite{aspnes2007skip}. 
% In a Skip Graph with $n$ nodes, a lookup for a name ID or a numerical ID is done by traversing $O(\log{n})$ nodes in expectation.
%A lookup operation has an \textit{initiator} (i.e., the node that invokes an instance of the lookup protocol), and a \textit{target}, i.e., the name ID or numerical ID for which the initiator is intending to send the lookup message. The routing of the lookup message is done collectively by a subset of Skip Graph nodes. 
As the result of the lookup operation, the lookup message is routed from the initiator to the nodes that hold the most similar identifier to the lookup target. 
In the context of lookup for a target name ID, the most similar identifier is the one with the longest common prefix. 
For example in Figure \ref{fig_dht_skipgraph}, a lookup message for the target name ID of $0010$ reaches node $8$ that holds the target name ID. Similarly, a lookup for the target name ID of $0101$ stops at node $45$. Here, the node $45$ has the longest common prefix of $3$ bits with the target name ID (i.e., both the target name ID $\underline{010}1$ and $\underline{010}0$ that is the name ID of node $45$ start with $3$ bits common prefix of $010$). 
%Finally, a lookup for the target name ID of $1100$ terminates at the nodes $11$, $41$, and $71$, which have the longest common prefix of a single bit to the lookup target. 
In the context of a lookup for a target numerical ID, the most similar numerical ID is the greatest numerical ID that is less than or equal to the target numerical ID. For example in Figure \ref{fig_dht_skipgraph}, a lookup for the target numerical ID of $11$ reaches node $11$ that holds the target numerical ID. Similarly, a lookup for the target numerical ID of $70$ reaches node $69$ that holds the greatest numerical ID in the Skip Graph overlay that is less than the target numerical ID of $70$.
The blue arrows in Figure \ref{fig_dht_skipgraph} illustrate a lookup example based on the numerical ID that is initiated by the node $71$ for the target numerical ID of $8$. In this figure, the horizontal arrows reflect the exchanged messages between the nodes in the underlying network. The vertical arrows correspond to the internal computations of the nodes during the lookup. The lookup for a target numerical ID starts at the top-most level of the initiator. Since the target numerical ID is less than the numerical ID of the initiator (i.e., $8 < 71$), the lookup proceeds in the left direction.
% by routing the lookup message to node $41$. However, node $41$ does not have any neighbor at the level and direction of the ongoing lookup (i.e., a left neighbor at level $3$). Hence, it descends one level down on its lookup table and routes the message to node $11$, which is its left neighbor on level $2$. Similarly, node $11$ does not have any left neighbor on its lookup table at level $2$. Hence, it descends one level down to level $1$ on its lookup table and routes the lookup message to node $8$, which holds the lookup target. The lookup for the numerical ID of $8$ terminates by node $8$ receiving the lookup message. More examples for the lookup based on the numerical IDs are available in \cite{aspnes2007skip, hassanzadeh2020interlaced, hassanzadeh2015locality}. The lookup for name IDs is done similarly, and we skip providing an example for it for sake of space. We refer the interested readers to \cite{hassanzadeh2016laras, hassanzadeh2018decentralized} for examples on the lookup for name IDs. 

\subsection{Cycloid}
\begin{figure}
\centering
  \includegraphics[scale = 0.15]{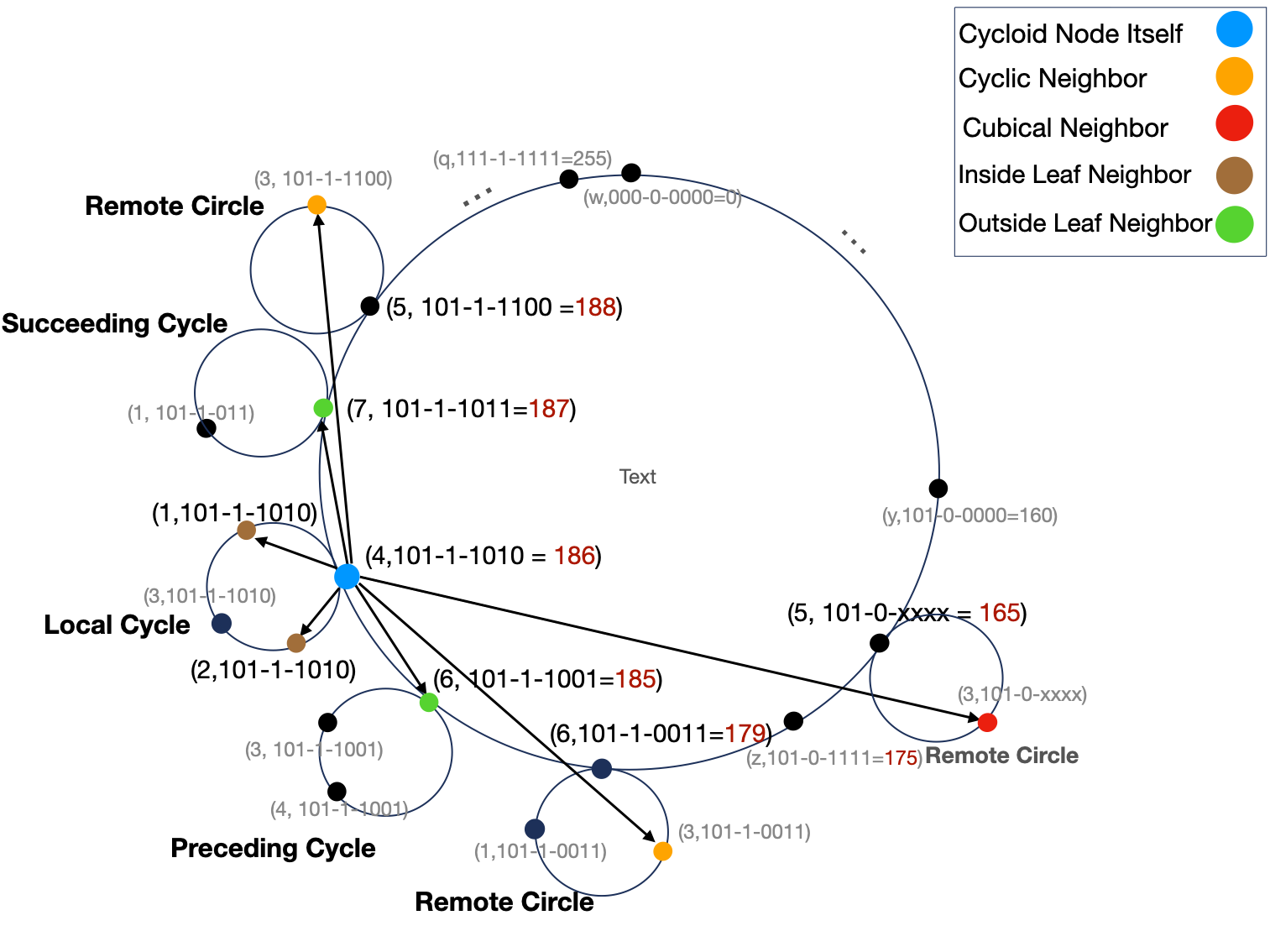}
\caption{An example of a Cycloid overlay with $8$ dimensions. The lookup table neighbors of the primary node $(4,101-1-1010)$ are determined by arrows. The decimal representation of cubical indices is shown next to them in red. $x$ means arbitrary bit value whereas other letters are arbitrary Cycloid indices.}
\label{fig:dht:cycloid-lookup}
\end{figure}

\noindent\textbf{Architecture}:
Cycloid \cite{shen2004cycloid} is a constant-degree DHT that resolves lookups with the message complexity of $O(d)$ where $n=d\cdot 2^{d}$, and $n$ is the maximum number of nodes in the system. This is in contrast to the typical DHTs that resolve a lookup operation with the message complexity of $O(\log{n})$ \cite{stoica2001chord}. The parameter $d$ is called the dimension of Cycloid overlay. Figure \ref{fig:dht:cycloid-lookup} shows an example of the Cycloid overlay with $d = 8$. Cycloid has a two-tier overlay: a larger ring with $d$ nodes at tier one, where each node represents a local cycle of size $2^{d}$ nodes at the second tier, resulting in $d\cdot 2^d$ nodes overall. A node identifier in the Cycloid is an ordered pair of two indices, i.e., $($\textit{cyclic index}, \textit{cubical index}$)$. The cubical index is an integer in the range of $[0, 2^{d}-1]$ that is commonly represented in $d$ bits as $a_{d-1} \cdots a_0$, and determines the second-tier local cycle that the node belongs to. As shown by Figure \ref{fig:dht:cycloid-lookup} all nodes with the same cubical index belong to the same local cycle on the second-tier. The cyclic index is commonly represented as an integer $k \in [0, d-1]$, which determines the position of the node within its local cycle. Within each local cycle, the node with the largest cyclic index is called the \textit{primary node}, and represents its local cycle on the first-tier ring. 
As a convention, the identifier of a peer in Cycloid is generated by taking the consistent hashing of its (IP) address. 
The identifier of a data object is computed by taking the hash value of its content. 
For a given hash value $h$, the Cycloid identifier is determined as $(h \% d, \frac{h}{d})$.
A data object is maintained by the peer with the closest identifier to it.
The identifier distance in Cycloid is measured as the common prefix length in the cubical indices, i.e., the longer the common prefix is the closer the identifiers are. 
In case cubical indices have the same common prefix length, the absolute difference between their cyclic indices determines the identifier distance. 
For instance in Figure \ref{fig:dht:cycloid-lookup}, $(4, 101-1-1010)$ is closer to $(3, 101-1-1011)$ (i.e., $7$-bits common prefix in cubical index) than $(4, 101-1-1100)$ (i.e., $5$ bits common prefix length).
In Figure \ref{fig:dht:cycloid-lookup}, the node $(4, 101-1-1100)$ is the primary node for the cycle that represents the cubical index of $101-1-1100$. 
As illustrated in this figure, for a certain node in a Cycloid overlay, the cycle it belongs to is called \textit{local cycle}, the immediate cycles to its local cycle on the larger ring are called the \textit{preceding} and \textit{succeeding} cycles, and any other cycle than these three is called a \textit{remote cycle}.

\textbf{Lookup table}: The lookup table of each node in Cycloid has a constant size of $7$ neighbors, which is independent of the system size $n$. The lookup table is composed of one cubical neighbor, two cyclic neighbors, and two outside and two inside leaf set neighbors. 
For a node with the identifier of $(k, a_{d-1} \cdots a_0)$, the cubical neighbor can be any node with a cyclic index of $(k-1)\%d$ that has an \textit{exact} $d-k-1$ bits common prefix length in its cubical index. 
The cyclic neighbors of the node are the first larger and smaller nodes (based on their cubical index) with the cyclic index of $(k-1)\% d$, which have \textit{at least} $d-k$ bits common prefix in their cubical index. 
The inside leaf set neighbors are the preceding and succeeding nodes on its local cycle based on their cyclic index.
The outside leaf set neighbors are the primary nodes of the preceding and succeeding remote cycles. The lookup table neighbors of node $(4,101-1-1010)$ in Figure \ref{fig:dht:cycloid-lookup} are colored based on their type.

\begin{figure*}
\centering
  \includegraphics[scale = 0.28]{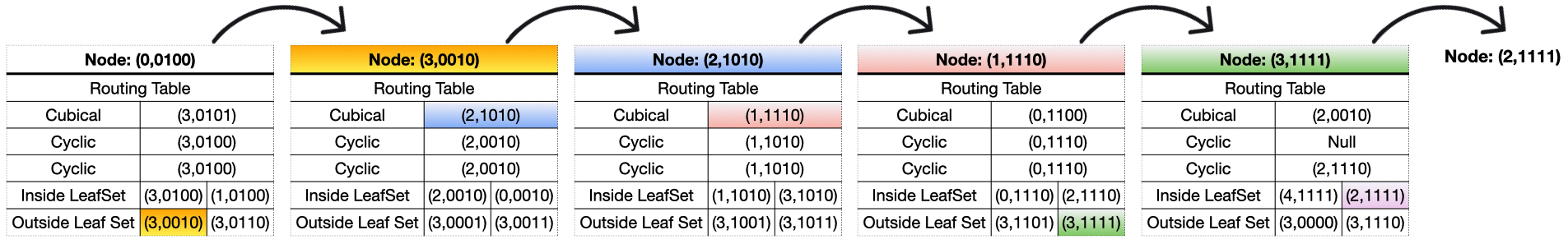}
\caption{An example of routing lookup message $(2,1111)$ initiated by $(0, 0100)$ in a hypothetical Cycloid DHT.}
\label{fig:dht:cycloid-routing}
\end{figure*}

\textbf{Routing}: 
The principal rule in routing a lookup message in Cycloid is to forward it to the lookup table neighbor that has the closest identifier to the lookup target.
If the Cycloid node itself has the closest identifier to the lookup target than all of its lookup table neighbors, it terminates the lookup and announces itself as the lookup result to the lookup initiator. 
If both the Cycloid node and all its neighbors have the same identifier distance to the lookup target, the node forwards the lookup message to one of its outside leaf set neighbors aiming at the lookup message is circulated around till it reaches a cycle that is closer to it on the identifier space. 
An example of routing in a hypothetical $4$-dimensional Cycloid DHT is illustrated in Figure \ref{fig:dht:cycloid-routing}, where the node $(0,0100)$ initiates a lookup for the target identifier of $(2,1111)$. 
Both the lookup initiator and all its lookup table neighbors have the same identifier distance to the lookup target (i.e., $0$-bit common prefix in cubical index). 
Hence, the lookup message is forwarded to one of the outside leaf set neighbors for circulation, i.e., to $(3, 0010)$. 
Upon receiving the lookup message, node $(3, 0010)$ forwards it to its lookup table neighbor $(2, 1010)$ which has the closest identifier to the lookup target among its other neighbors, i.e., $1$-bit common prefix length in the cubical index. 
The lookup message is then forwarded by node $(2, 1010)$ to its closest neighbor to the lookup target, i.e., node $(1, 1110)$ that has $3$-bit common prefix in its cubical index to the lookup target. 
On receiving the lookup message for $(2, 1111)$, node $(1, 1110)$ forwards it to its lookup table neighbor $(3, 1111)$ that has the maximum common prefix length of $4$-bits to the lookup target. 
The lookup message is finally routed to the target by node $(3, 1111)$ that resides on the same cycle as the target.

\section{Edge and Fog Computing}
\label{dht_survey:section_edge_computing}
%%% Author: Yahya
\subsection{System Model}
\label{dht_survey:section_edge_computing_definition}
In the edge and fog computing paradigm, processing and storage power is distributed among the resources (e.g., routers) located closer to the end-users in the interest of saving bandwidth and boosting response time. This is in contrast with the traditional cloud computing paradigm, which concerns with consolidating the computation and storage power into a powerful and centralized data center that is normally distant from the users. Hence, the computation of raw data in the edge and fog computing is done closer to its origin, which is in contrast to the remote data centers in cloud computing. This migration of resources towards the end-users in the edge and fog computing paradigm results in a faster response time and higher throughput \cite{satyanarayanan2017emergence}. 

\begin{figure}
\begin{center}
    \includegraphics[scale=0.18]{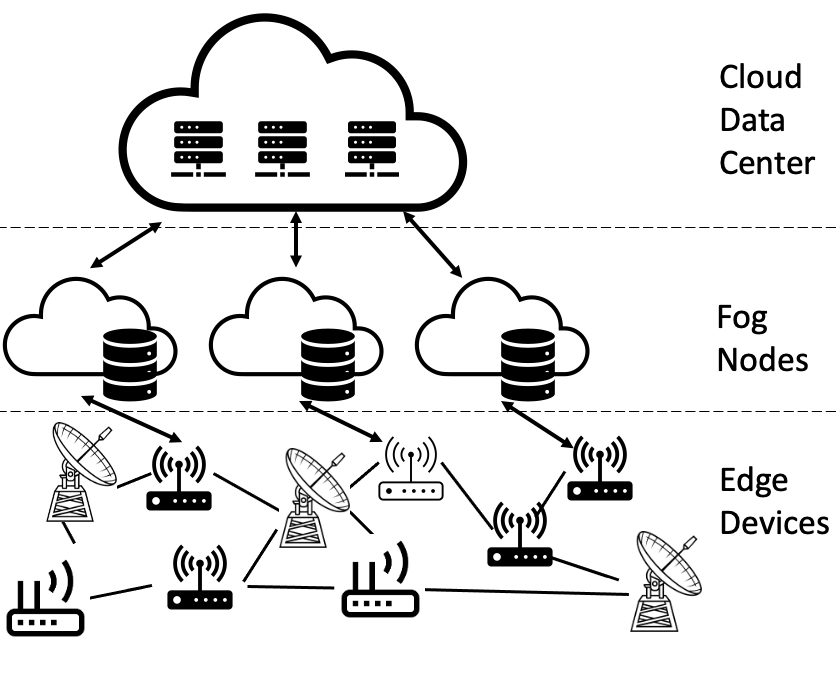}
\caption{An overview of system model in edge and fog computing paradigm.} \label{fig_edge_model}
\end{center}
\end{figure}

The primary difference between the edge computing system and the fog computing systems is the placement of resources in the network. The edge computing systems aim at performing the computations on the edge devices close to the source of data \cite{dolui2017comparison}, e.g., routers and base stations. On the other hand, the fog computing systems act as the midpoint processing of data by processing the aggregated data from the edge devices at the fog nodes \cite{bonomi2012fog}. Compared to the edge nodes that are generally directly connected to the source devices (i.e., the users), the fog nodes are more powerful nodes that are located distant from the source devices. However, by lying between the edge devices and the cloud data center, the fog computing nodes provide more decentralization and move the processing and storage power at replicated nodes closer to the edge devices. Figure \ref{fig_edge_model} illustrates the hierarchical organization in the edge and fog computing paradigm. 

Table \ref{dht_survery:table_edge_comparison} summarizes the DHT-based edge and fog computing solutions that we survey in the rest of this section. In this table, the \textit{Type} column presents whether the solution contributes to the infrastructure of the edge and fog computing systems, or enables an edge- or fog-based application. The \textit{Nodes} column refers to the primary representation of the nodes once a DHT overlay of physical devices (e.g., servers) is shaped. %For example, in the DHT-based task management application \cite{simic2018edge} once the DHT overlay is shaped among the physical devices, it is primarily utilized for storing computation results as the DHT nodes. 
Also, the \textit{Identifiers} column represents the (unique) attribute of the DHT nodes that is used as the input to a collision-resistant hash function \cite{katz2014introduction} for generating their unique identifiers. The \textit{DHT Utilization} column represents the primary purpose of DHT utilization in the solution. The \textit{Domain} column represents how the DHT nodes are controlled by the administrative domain(s). In a centralized solution, all nodes are managed by a single monolith administrative domain. In a distributed solution, nodes are managed by several agents all belonging to the same administrative domain. In a decentralized solution, the nodes are partitioned among several independent administrative domains, where each domain controls its subset of nodes.

\begin{table*}
\centering
\setlength{\tabcolsep}{3pt}
{
    \begin{tabular}{ |l|l|l|l|l|l|l|l|  }
    \hline
    Solution &
    Type &
    DHT &
    Nodes & 
    Identifiers & 
    DHT Utilization &
    Domain \\
    \hline
    SA-Chord \cite{d2018sa} & 
    Inf &
    Chord & 
    Resources & 
    - & 
    Routing Overlay & 
    Decentralized \\
    GRED \cite{xie2019efficient} &
    Inf & 
    General &
    SDN Switches &
    Delaunay Triangular &
    Object Storage &
    Centralized \\
    EdgeKV \cite{sonbol2020edgekv} &
    Inf & 
    Chord &
    Super Peers &
    Content &
    Object Storage &
    Decentralized \\
    Access Control Management \cite{riabi2017proposal} &
    App &
    Chord &
    Access Control Lists &
    - &
    Object Storage &
    Decentralized \\
    %
    %
    % Video Streaming  \cite{nakagawa2015dht} &
    % App &
    % Chord &
    % Edge servers &
    % IP address &
    % Routing Overlay &
    % Centralized \\
    %
    %
    Service Discovery \cite{santos2018towards} &
    App &
    General &
    Services &
    Content &
    Routing Overlay &
    Decentralized \\
    Resource Discovery \cite{tanganelli2017fog} &
    App &
    General &
    Resources &
    Attribute Values &
    Routing Overlay &
    Decentralized \\
    Resource Management  \cite{song2020smart} &
    App &
    General &
    Data Objects &
    Content &
    Object Storage &
    Decentralized \\
    Task Management  \cite{simic2018edge} &
    App &
    General &
    Computation Results &
    Content &
    Object Storage &
    Distributed \\
    \hline
    \end{tabular}
    
}
\caption{A summary of DHT-based solutions in Edge/Fog computing. In this table \textit{Inf} and \textit{App} are abbreviations for \textit{Infrastructure}, and \textit{Application}, respectively.}
\label{dht_survery:table_edge_comparison}
\end{table*}
%%% Author: Yahya
\subsection{DHT-Based Infrastructure Solutions}
\label{dht_survey:section_edge_computing_infrastrucutre}

\subsubsection{Routing Overlays}
\begin{figure}
\begin{center}
    \includegraphics[scale=0.20]{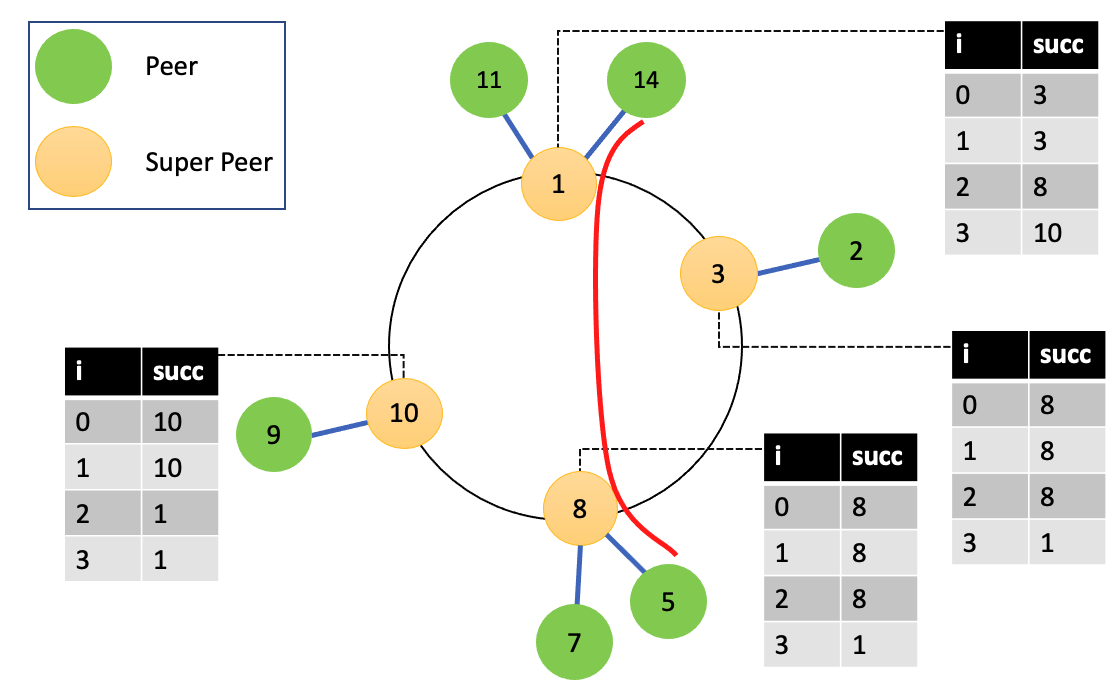}
\caption{An example of DHT overlay structure in SA-Chord \cite{d2018sa}. The lookup tables of super-peers on are represented by grey tables connected to them. An example of a lookup path from peer $14$ to peer $5$ is illustrated by the red line.} \label{fig_edge_sachord}
\end{center}
\end{figure}

To minimize the energy consumption and response time in the resource heterogeneous edge computing infrastructures, SA-Chord \cite{d2018sa} provides a two-tier routing overlay of peers and super-peers. The peers are assumed as the edge computing resource-constraint devices with a low energy level and limited bandwidth. These regular peers only send and receive messages without participating in the routing, which preserves their energy level to last longer in the system. On the other hand, the super-peers are strong peers with higher energy levels and bandwidth. The super-peers can last longer in the system, and hence help the overlay with routing the queries at a higher speed. The core routing functionality in SA-Chord is established over the super-peers, which route queries between each other as well as the regular peers.
% The regular peers, on the other hand, are the consumers of this infrastructure and exchange their queries through super-peers. 
An example of an SA-Chord system with $4$ super-peers and $6$ peers is shown in Figure \ref{fig_edge_sachord}. Each peer is assigned to the super-peer that immediately succeeds it on the identifier space, i.e., the super-peer with the smallest identifier that is greater than the peer's identifier. In the example of Figure \ref{fig_edge_sachord}, the peers $5$ and $7$ are assigned to the super-peer $8$ that immediately succeeds them on the identifier space, i.e., $8$ is the super-peer with the smallest identifier greater than both $5$ and $7$. The regular peers should only know their corresponding super-peer. The super-peers establish a DHT-based routing overlay among themselves, where each super-peer maintains a routing table of a subset of other super-peers based on the Chord protocol \cite{stoica2001chord}. The inter-peer communications are routed through the super-peers. For example, the traffic of communications between peer $5$ and $7$ in Figure \ref{fig_edge_sachord} is handled via their shared super-peer $8$. Similarly, as shown by the red line, the lookup query of peer $14$ for peer $5$ is submitted to its connected super-peer $1$, and is routed based on Chord protocol through super-peer $8$ to the peer $5$. 
%The decision on being a super-peer or a regular peer is done via a weighted average over an objective function. Each peer evaluates the objective function on its resources and finds its objective function value. The super-peer with the minimum identifier on the Chord (i.e., super-peer $1$ in the example of Figure \ref{fig_edge_sachord}) is responsible for aggregating the objective function values of other peers. This is done by periodically circulating an aggregation message over the Chord ring, collecting other peers' values, performing the aggregation, and sharing the result with other peers in the next circulation. Based on the aggregated value, each node can decide whether it is a regular peer or a super-peer. 
%SA-Chord is experimentally shown to be faster than Chord in average latency of queries. Also, the network of SA-Chord nodes is shown to last longer than the Chord network due to minimizing the energy consumption of nodes. 

\subsubsection{Object Storage: }
\textbf{Greedy Routing for Edge Data (GRED) \cite{xie2019efficient}} is a low-latency Software Defined Network (SDN)-based \cite{assefa2019survey} DHT overlay for object storage. A GRED system is composed of an SDN controller, SDN switches, edge servers, and data objects.  
The SDN controller constructs and maintains a DHT overlay of SDN switches in a centralized way, i.e., it builds the topology graph of the DHT among the switches and programs them with their corresponding lookup tables. Each edge server is assigned to connect to one of the SDN switches in the DHT overlay based on its identifier. 
The identifier of edge servers is assigned by the SDN controller in a way that the identifier distance between edge servers reflects their latency. The identifier of data objects is determined by taking the hash value of their content. 
Each data object is then stored on the closest edge server based on its identifier, i.e., the edge server with the minimum identifier distance to it. 
The identifier assignment in this way enables greedy routing, i.e., routing for an arbitrary identifier always stops at the closest existing identifier to it on the identifier space. Since the identifier of edge servers is assigned based on their relative latency, GRED provides a low-latency routing infrastructure for the edge computing data objects.

\textbf{EdgeKV \cite{sonbol2020edgekv}} is a fault-tolerant and consistent object storage for edge computing systems. Similar to SA-Chord \cite{d2018sa}, EdgeKV is also organized into a two-tier routing overlay of super-peers and ordinary peers. The super-peers form a DHT overlay and are responsible for handling a group of distinct edge servers (i.e., the ordinary peers) that are assigned to them. The edge servers of each group form an independent component of the system, which are organized into a complete overlay graph among themselves, and are responsible for storing the same data objects. A data object is represented by a key and value pair. Similar to SA-Chord \cite{d2018sa}, the messages to and from an edge group are routed through their assigned super-peer. A read or write query for a data object is submitted by the user to the DHT overlay of super-peers. The query is then routed by the super-peers to the corresponding group of edge servers responsible for that key. 

The edge servers of a group form a replicated state machine \cite{bessani2014state} of key-value pairs assigned to that group, by utilizing the Raft consensus protocol \cite{ongaro2014search} for writing on the keys. Hence, a write request routed to the group is processed through a consensus and is written on all edge servers of that group, which results in a consistent storage state among all edge servers of the group. Due to this consistent state, a read request for a key can be handled by any of the edge servers in that group. Since all edge servers keep the same state of data objects, failing some does not make the data objects unavailable, which provides a fault-tolerant storage.

%%% Author: Yahya
\subsection{DHT Applications in Edge/Fog Computing}
\label{dht_survey:section_edge_computing_applications}

\subsubsection{Access Control Management}
Intelligent Transport System (ITS) is defined as the integration of communication, control, and processing of information to the transportation systems \cite{riabi2017proposal}. An ITS provides a set of resources (e.g., computation and storage power) for the users, which facilitates their transportation in terms of time and cost. To provide a fine-grained, scalable, and low-latency access control management for the ITS devices (e.g., sensors, computing resources, and storage systems), a DHT-based access control mechanism over the fog servers is proposed in \cite{riabi2017proposal}. Each fog server is in charge of a subset of ITS devices and acts as a gateway between the users and its set of ITS devices. Fog servers represent and maintain the Access Control List (ACL) of their ITS devices over the DHT overlay. An ACL for a fog server represents its set of ITS devices accompanied by the list of legitimate users that can access those devices. The key of each ACL on DHT is the identity of its corresponding fog server, and the value is the ACL itself. To provide low-latency query resolution for the mobile users, a user always gets connected to the closest fog server while it is mobilizing in the city. Access queries of the user are then submitted to the DHT of fog servers through that closest fog server, and the response is returned to the user, i.e., whether the access is granted or denied. This approach provides scalable storage of ACLs over fog servers under the assumption that the ITS devices are uniformly distributed among the fog servers. This results in the storage maintenance load of ACLs for millions of ITS devices to get distributed evenly across the DHT overlay of fog servers.

\subsubsection{Video Streaming}
An edge computing-based video streaming service is proposed in \cite{nakayama2017peer} to address the withdrawal problem in P2P streaming. The withdrawal problem is defined as the distribution of content is getting stopped once no user is watching it anymore, which results in content getting unavailable once the users shift interest away from it. The solution is comprised of three entities: cloud, edge servers, and users. Each user is both a content receiver as well as a content distributor and is connected to an edge server. The edge servers act as the brokers and provide available bandwidth to the system for streaming the content of the users to each other. The cloud is only responsible for managing the edge servers. A DHT is established among the edge servers in a centralized way by the cloud, i.e., the cloud creates the lookup tables of the edge servers on their behalf.  The edge servers are responsible for routing the users' content among each other. The assignment of users to the edge servers is done by the cloud entity. The content distribution is done by each edge server broadcasting the content of its connected users to its neighbors on the DHT overlay, which is also broadcast by them to their neighbors. Hence, in a system with $n$ edge servers eventually, a user's content is available at every edge server within the round complexity of $O(\log{n})$. Content distribution in this way stands against the traditional on-demand ones, where an edge server would stop receiving a stream once its users stop viewing it. 

%This is to address the withdrawal problem in P2P streaming where the distribution of content is getting stopped once no user is watching it anymore.%  Withdrawal problem results in content getting unavailable once the users shift interest away from it. In this solution, the load balancing among the edge servers is provided by the cloud, which adaptively adds extra edge servers if there is no one available with a load below a threshold. The load of an edge server is defined as the number of users connected to it. 

\subsubsection{Service and Resource Discovery}
\textbf{Service Discovery \cite{santos2018towards}:} 
%In environments such as smart cities and ITS, where the nodes are subject to a high degree of mobility, fog-based resource discovery shows a better efficiency than the centralized cloud-based solutions. Fog computing systems are distributed in their nature and inherently dispersed geographically. This allows each mobile node to choose the closest fog node based on their relative latency to it, and query the system through that node. 
This solution proposes a service-based smart city, where the fog nodes offer their computational resources as services to the mobile users. To provide a decentralized and scalable discovery mechanism for the services offered by the fog nodes, a DHT overlay is established among them. Each service offered by a fog node is registered as a DHT node. The key of the node is the service identifier, which is the hash value of the service instance. The value of the node is the service template that contains all information to mount, register, and instantiate the service. Each mobile user in this solution chooses the closest fog node based on the relative latency, and efficiently retrieves the information needed to mount and utilize its desire service by querying the DHT overlay of fog nodes. 
% The service templates are represented by Network Function Virtualization (NFV) \cite{mijumbi2015network}, which enables a service template to be executed remotely on the virtual machine of its user fog node on DHT. 

\textbf{Resource Discovery \cite{tanganelli2017fog}:}
This solution supports resource discovery in the mobile environments based on the specific range of attributes of provided resources, e.g., all charging stations for electronic vehicles between a certain latitude and longitude. To support the range queries, a two-tier DHT structure is utilized. Each tier-one DHT node is a fog node, which represents a resource type (e.g., the charging station resource type), and is attached to a tier-two DHT of resource nodes (e.g., all charging stations). The identifier of a fog node in the tier-one DHT is the resource name it represents. The identifier of a resource node in the tier-two DHT is the value of its attribute (e.g., geographical location). Looking up resources with a specific range of attribute is done in two steps. In the first step, the query is dispatched in the tier-one DHT to lookup the fog server responsible for that resource type. Then the outer DHT attached to the responsible fog server is queried for the desired range of the attribute it represents.

\subsubsection{Resource and Task Management}
\textbf{Smart Contract-based Resource Management \cite{song2020smart}:} This solution aims at fully decentralized storage and computing resource provision for the end-users in the edge computing systems. The proposed architecture is composed of three roles: consumer, storage provider, and computation provider. The consumers are the end-users aiming at exploiting the edge computing platform for executing their tasks. The storage providers form a DHT overlay and represent the data objects stored on them by DHT nodes. 
The primary purpose of DHT utilization is to establish a decentralized key-value storage for the data objects. 
Each computation task of the consumers is assigned to a subset of storage and computation providers through the invocation of a resource trading smart contract \cite{wang2019blockchain} over the Ethereum blockchain \cite{wood2014ethereum}. Upon the assignment is done, the consumer breaks its task into chunks, and stores the chunks over the DHT of the storage providers. This makes the chunks accessible by the computation providers, to fetch and apply computations on them.
%To manage the resources in full decentralization, this solution utilizes a smart contract to perform the resource assignment for the computation tasks of the consumers. A smart contract is a self-executing code on blockchains. It is registered by a transaction to a blockchain and is assigned a unique address, which resembles the function names in programming languages \cite{wang2019blockchain}. The registered smart contracts are invoked by any transaction that requests their execution on specific inputs, and extend the state of the blockchain accordingly.
%In this solution \cite{song2020smart}, the consumer submits a computation task as a transaction to the Ethereum blockchain \cite{wood2014ethereum}, which invokes a resource trading smart contract on it. The execution of the resource trading smart contract assigns the storage and computation providers for the consumer's task. The utilization of a resource trading smart contract over the blockchain provides a decentralized trust between the consumers and providers. The smart contract receives a deposit from the consumer and pays the providers based on the unit of time that their resources are occupied on the delegated computation task of the consumer. 

\textbf{Task Management \cite{simic2018edge}:} To provide a task management service for the edge computing devices, this solution utilizes a system of clusters. Each cluster is comprised of two components: a cluster master, and edge servers. The cluster master is responsible for the task management and scheduling on the edge servers of that cluster. The edge servers are responsible for storing data and running the tasks. The master of a cluster receives the tasks from the users, breaks them into smaller sub-tasks, and assigns them to the edge servers of its cluster, for computation. For a certain task, the sub-tasks should be executed in a specific order of precedence, i.e., the execution of each sub-task is done on top of the result of the predecessor sub-tasks. However, the sub-tasks of a task are distributed among the edge servers, and there is no centralization of the execution. Hence, the edge servers of each cluster in this solution form a DHT overlay to share their computation results. The result of each sub-task is stored on this DHT overlay, which makes it available for the edge servers that are in charge of executing the subsequent sub-tasks. 

\subsection{Open Problems}
\label{dht_survey:section_edge_computing_open_problems}

\textbf{Query Load Balancing:} The solutions utilizing DHTs as an object storage platform benefit from its inherent storage load balancing by distributing the objects over the DHT nodes based on the hash value of their content \cite{xie2019efficient, sonbol2020edgekv, riabi2017proposal, song2020smart, simic2018edge}. However, such solutions do not necessarily scale as the query load of their users gets heterogeneous across the DHT overlay, e.g., non-uniform locality distribution of users in ITS \cite{riabi2017proposal} or smart city \cite{santos2018towards} scenarios, which applies a heterogeneous query load on their corresponding closest DHT nodes. This results in a non-uniform available bandwidth and quality-of-service distribution on DHT nodes, which degrades performance of the system at the overloaded nodes and increases their failure likelihood. Balancing the load of DHT nodes in heterogeneous scenarios \cite{hassanzadeh2019decentralized} may be considered as a potential research direction. % to address the query load balancing in these scenarios. 

% should be moved to cloud
\textbf{Decentralized Maintenance:} Maintenance of the DHT overlay and its connectivity in the edge computing solutions like \cite{nakagawa2015dht}, is done through a centralized registry. Each node registers itself to the registry server upon joining the system. The server performs the insertion protocol of the node in the DHT overlay completely on behalf of the node, and shares the final routing table with the node. The DHT construction protocols, however, are meant for decentralization \cite{stoica2001chord, maymounkov2002kademlia}, i.e., a DHT overlay is capable of being constructed in a fully decentralized way by its participating nodes invoking the insertion protocol independently of each other. However, the fundamental challenge is to stabilize the constructed DHT overlay addressing the heterogeneous dynamics of the nodes in their online and offline states, which is basically handled by the centralized registry in the aforementioned solutions (i.e., \cite{nakagawa2015dht}). Addressing the centralization problem and providing decentralized overlay construction and maintenance through the decentralized DHT stabilization algorithms \cite{kaur2020churn, hassanzadeh2020interlaced} may be considered as a potential research direction.

\textbf{Range Queries:} 
Querying a range of keys on the DHT overlay in solutions like \cite{tanganelli2017fog} is done by searching for each key of the query range individually. In a DHT overlay with $n$ nodes, performing a range query for $k$ keys in this way imposes a message complexity of $O(k \log{n})$. When the size of queried range outnumbers the DHT nodes (i.e., $k > n$), such scenarios apply a linear message complexity to the system, which degrades the efficiency of DHT utilization. Efficient DHT-based range query solutions \cite{zheng2006distributed, gao2004adaptive} may be considered as a potential research direction to address this issue.

\section{Cloud Computing}
\label{dht_survey:section_cloud_computing}
%%% Author: Yahya
\subsection{System Model}
\label{dht_survey:section_cloud_system_model}

\begin{figure*}
\begin{center}
    \includegraphics[scale=0.29]{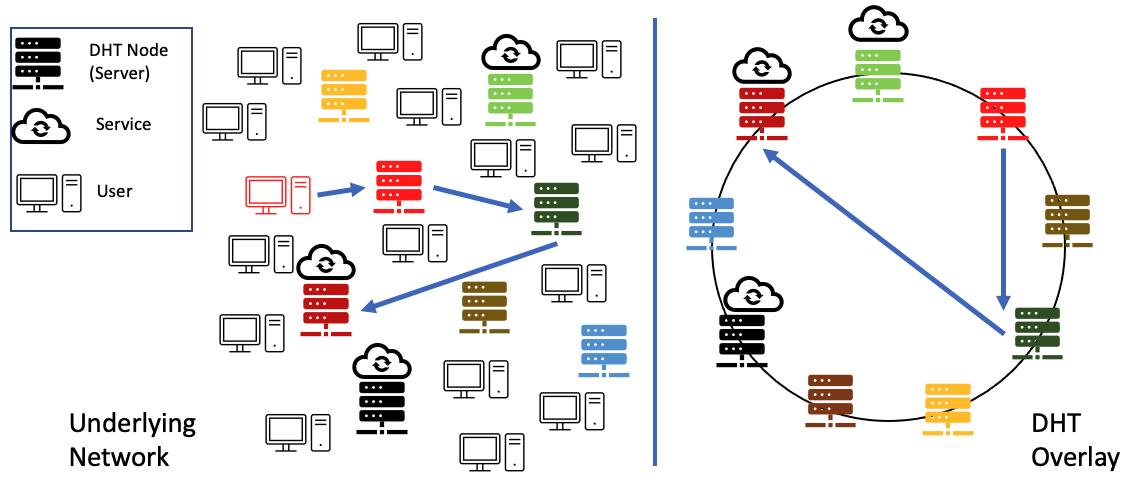}
\caption{DHT-based cloud computing system model. A sample service is replicated three times. Routing a request from a user to a replica of the service is illustrated in both the underlying network, as well as its corresponding DHT overlay.} \label{fig_cloud_model}
\end{center}
\end{figure*}

As shown by Figure \ref{fig_cloud_model}, in this paper, we model the DHT-based cloud computing systems with two entities: DHT nodes and services. An entity is defined as a resource that can be uniquely identified using an identifier. For example, in the storage applications, the services are modeled as storage and retrieval components that are identified by a unique service identifier (e.g., the hash of the service name). A service component is hosted and maintained by a DHT node. The DHT nodes are processes that run an instance of the DHT software and collaboratively build a DHT overlay. The users are external processes that query the DHT nodes via some client access interface. The DHT-based cloud computing systems enable the nodes to exploit the DHT overlay for looking up other nodes as well as the services they host. Users are also able to look up the services by submitting their queries to a DHT node to route on their behalf. The lookup query is initiated by a DHT node, routed by intermediary nodes, and reaches a target DHT node. This end-to-end communication from an initiator to the target on the overlay is made possible via the hop-to-hop communications between nodes in an underlying network \cite{kurose2017computer}. Figure \ref{fig_cloud_model} depicts the example of routing on a DHT overlay and its corresponding hop-to-hop representation in the underlying network. As shown in the figure, there exists a one-to-one correspondence between the DHT nodes in the underlying network (left side) and the DHT nodes on the overlay (right side). 

As the group of users interested in a specific service scales up, the query rate on the DHT node hosting the service increases. This degrades the performance of the host, increases its response time, and elevates its failure likelihood. As the DHT node hosting service fails, its service becomes unavailable to the users. To provide query load balancing on the DHT nodes, service availability, and avoid a single point of failure for the services, a DHT node hosting service makes exact copies of its service component on other DHT nodes, which are called the corresponding \textit{replicas} for that service \cite{hassanzadeh2016awake}. The procedure of determining the replica nodes is called \textit{replication}. The replication paradigm is illustrated in the example of Figure \ref{fig_cloud_model}. 

Table \ref{dht_survery:table_cloud_comparison} summarizes the DHT-based cloud computing solutions we survey in this section. The description of columns is similar to the comparison table of edge and fog computing solutions in Section \ref{dht_survey:section_edge_computing_definition}. 

\begin{table*}
\centering
\setlength{\tabcolsep}{3pt}
%\resizebox{\columnwidth}{!}
{
    \begin{tabular}{ |l|l|l|l|l|l|l|l|  }
    \hline
    Solution &
    Type &
    DHT &
    Nodes & 
    Identifiers & 
    DHT Utilization &
    Domain \\
    \hline
    Storage Cost Optimization \cite{zhou2016cost} &
    Inf &
    Voldemort &
    Servers &
    Address & 
    Storage &
    Centralized \\
    Compute Cost Optimization \cite{zhou2014novel} &
    Inf &
    Voldemort & 
    Virtual Machines &
    Address &
    Storage & 
    Centralized \\
    Locality-Aware Replication \cite{hassanzadeh2018decentralized} & 
    Inf &
    Skip Graph & 
    Data Objects & 
    Address &
    Storage & 
    Decentralized \\
    Multi-objective Replication \cite{hassanzadeh2019decentralized} & 
    Inf &
    Skip Graph & 
    Data Objects & 
    Address & 
    Storage & 
    Decentralized \\
    Cluster task scheduling \cite{xie2017design} &
    App &
    Chord &
    Virtual Machines &
    Content & 
    Task Management&
    Decentralized \\
    m-Cloud \cite{nakagawa2015dht} &
    App &
    General &
    Cloud domains &
    Content & 
    Aggregation &
    Decentralized \\
    Game Object Storage \cite{kavalionak2015integrating} &
    App &
    Chord & 
    Game objects &
    Content & 
    Storage& 
    Distributed \\
    IPFS \cite{benet2014ipfs} & 
    App &
    Kademlia & 
    Data objects &
    Content & 
    Storage &
    Decentralized \\
    Time-sensitive object storage \cite{xiong2015full} &
    App &
    General &
    Symmetric keys &
    Name & 
    Access Control &
    Decentralized \\ 
    P2P Streaming \cite{gupta20172} &
    App &
    General &
    Video chunks &
    Content & 
    Load Balancing &
    Distributed \\
    \hline
    \end{tabular}
    
}
\caption{A summary of DHT-based solutions in cloud computing. In this table \textit{Inf} and \textit{App} are abbreviations for \textit{Infrastructure}, and \textit{Application}, respectively.}
\label{dht_survery:table_cloud_comparison}
\end{table*}

%, where a service is replicated three times in the system. The red user is interested in utilizing the service queries its closest DHT node based on its corresponding latency in the underlying network. The query is routed within the DHT overlay and is answered by one of the service replicas. The routing path of the query in the underlying network and its corresponding overlay network path are shown in Figure \ref{fig_cloud_model}. The performance of a DHT-based cloud computing system is measured in terms of its query processing and response time \cite{hassanzadeh2018decentralized, hassanzadeh2019decentralized}. 
%%% Author: Yahya
\subsection{DHT-based Infrastructure Solutions}
\label{dht_survey:section_cloud_infrastructure}
\subsubsection{Resource Provisioning }
\textbf{Storage Cost Optimization \cite{zhou2016cost}:} This solution minimizes the cost of establishing DHT-based cloud storage given a specific set of objectives, e.g., desired average response time. Here minimizing the cost stands for minimizing the number of required DHT nodes and their storage capacity. The inputs to the algorithm are the distribution of total data in the system over the time, the access frequency of users, the required processing power of DHT nodes, and the desired average response time. Given this set of inputs, the algorithm optimizes the number of DHT nodes and their storage capacity. The solution is specifically proposed and implemented for Voldemort \cite{auradkar2012data}, which is an open-source DHT-based key-value store designed by LinkedIn. 

\textbf{Compute Cost Optimization \cite{zhou2014novel}:} This solution optimizes the computation cost of the DHT-based cloud computing systems that are based on the Infrastructure-as-a-Service (IaaS) platforms, e.g., Amazon Web Services (AWS). In such platforms, each DHT node is represented as a process on an IaaS virtual machine. The DHT-based cloud provider is getting charged per amount of resources its DHT nodes use over time, e.g., processing, disk, and memory. The allocated resource by each DHT node is modeled as a fixed-capacity queue, which maintains the queries of users (i.e., computation requests) at that node pending to be processed. A longer queue implies a higher amount of allocated resources, which results in a higher cost for the cloud provider. Once the queue hits its capacity, the node denies any further requests, which degrades the service availability of the system. To provide a trade-off between the resource cost, service availability, and response time, the solution models the resource provisioning problem as a non-linear programming model \cite{cormen2009introduction}. Inputs to the model are the range of available computation capacities of the virtual machines (i.e., range of their available input queue capacity), the cost associated with each capacity, and the service availability and response time thresholds. The programming model then determines the number of IaaS virtual machines (i.e., DHT nodes) and their input queue capacity such that the overall cost of provisioning the virtual machines is minimized while the desired service availability and response time thresholds are met. Similar to \cite{zhou2016cost}, this solution is also specifically proposed and implemented for Voldemort DHTs \cite{auradkar2012data}.

\subsubsection{Replication}
\textbf{Locality-Aware Replication \cite{hassanzadeh2018decentralized}:} To optimize the response time in DHT-based cloud computing systems, the notion of locality-aware replication is defined, which places the replicas of a service owner in a way that the average access delay between each user and its corresponding replica is minimized \cite{hassanzadeh2016laras}. The corresponding replica for a user is the one with the minimum average latency to it. The average access delay of replication is defined as the average pairwise latency between each user and its corresponding replica. GLARAS \cite{hassanzadeh2018decentralized} is the first fully decentralized locality-aware replication for DHTs where the primary node hosting a service (i.e., the service owner) can place its replicas in a locality-aware way.
% A service owner independently runs GLARAS without the need of any special (super) node. 
To run GLARAS, the service owner only needs some public parameters of the system, which are time-independent, and determined at the bootstrapping of the system, e.g., the number of desired replicas. GLARAS models the locality-aware replication by an Integer Linear Programming (ILP) model \cite{cormen2009introduction}, solves it, and returns the identifier of replica nodes to the service owner. GLARAS introduces a novel system modeling approach that logarithmically shrinks the size of the ILP model and makes it be solved efficiently.

\textbf{Multi-objective Replication \cite{hassanzadeh2019decentralized}:} The performance of the cloud computing systems with heterogeneous resources of nodes (e.g., bandwidth, storage, computation, etc.) is a function of both the propagation and transmission delays \cite{kurose2010computer} as well as the computation and storage capacities. In such scenarios, both the average access delay, as well as the average response time play key performance roles. These performance metrics are tightly coupled with the availability behavior of the nodes, i.e., a (temporarily) offline or failed replica negatively affects both the average access delay as well as the response time of the system. The former happens as once a replica goes offline, its corresponding users should find another closest replica, which inherently has a higher propagation delay. Likewise, the query load of an unavailable replica is distributed among other replicas, which degrades their response time. Pyramid \cite{hassanzadeh2019decentralized} is the first fully decentralized multi-objective replication framework for heterogeneous DHT-based cloud computing systems. It optimizes both the average access delay as well as the response time of the replicas under the time-variant availability behavior of the nodes. In the system model of Pyramid, nodes periodically report their availability and resources over an efficient and light blockchain system \cite{hassanzadeh2019lightchain}. Hence, each node can obtain the aggregated average availability of the resources of other nodes across the system over the time. A service owner then invokes Pyramid on this aggregated information. Pyramid models the replication as a multi-objective Integer Linear Programming problem (ILP) \cite{cormen2009introduction}, solves it, and returns the identifiers of replicas for that service owner in the system. Similar to GLARAS \cite{hassanzadeh2018decentralized}, Pyramid also takes a novel problem size reduction approach that makes the ILP model solvable efficiently.

%%% Author: Yahya
\subsection{DHT-based Applications}
\label{dht_survey:section_cloud_applications}

\subsubsection{Decentralized Task Management}
A container \cite{turnbull2014docker} is an entire runtime environment, which contains an application accompanied by all its dependencies and libraries needed to run it. With its encapsulation provided, containerization aims at software portability across different computing environments. A container cloud \cite{pahl2017cloud} is a platform that maintains and orchestrates the computation tasks in the containers. Kubernetes \cite{kubernetes} is one of the widely-used examples of container clouds. The container cloud orchestrates containers into clusters where each cluster is managed by a single master node that performs the task management on the nodes of its cluster. Having a single cluster master leads to a single point of failure as well as a performance bottleneck as the cluster size scaling up. To address these reliability and scalability problems, a DHT-based decentralized task management system is proposed in \cite{xie2017design}. The entire cluster is completely replaced by a decentralized DHT overlay of nodes. The nodes utilize the overlay to find each other as well as each other's available resources. A node advertises itself to the others by representing its state as the key and its address as the value on DHT. 
The management protocol running at each node then pulls the state of other nodes of the cluster and decides on scheduling the tasks either on itself or on the other nodes accordingly. 

\subsubsection{Privacy Preserving Aggregation}
\begin{figure}
\begin{center}
    \includegraphics[scale=0.14]{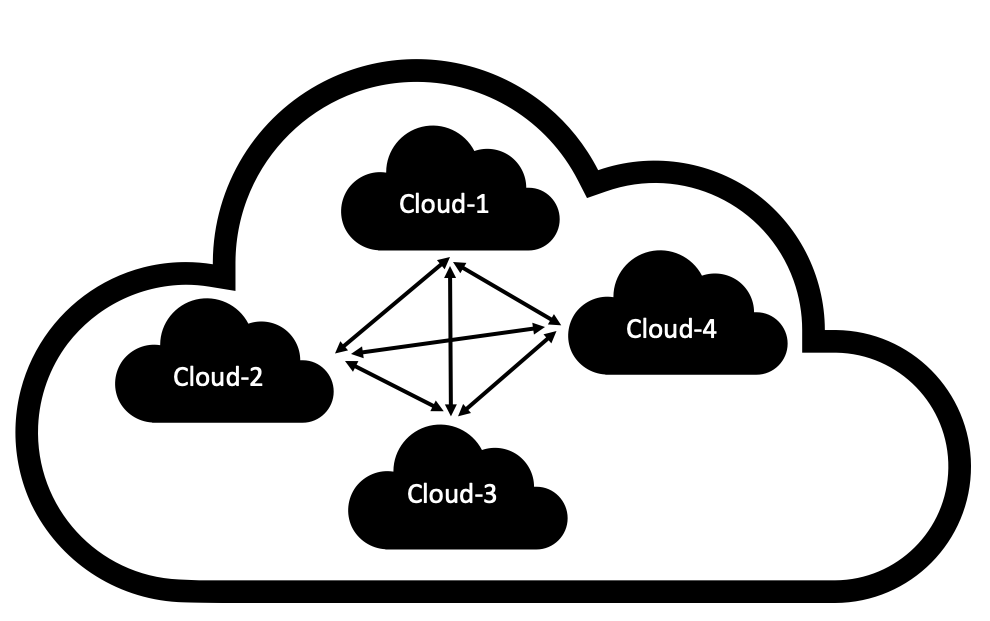}
\caption{An example of cloud federation of four independent cloud domains. The outer-most cloud constitutes of the federation of the inner cloud domains.} \label{fig_cloud_federation}
\end{center}
\end{figure}

m-Cloud \cite{nakagawa2015dht} is a DHT-based privacy-preserving aggregation for sensor data, which utilizes a federation of several cloud domains. Cloud federation is defined as the system model where different administrative domains share data on their private clouds with each other \cite{tanenbaum2007distributed}. Figure \ref{fig_cloud_federation} shows an example of the cloud federation system model, where the outer federated cloud domain constitutes of four inner private cloud domains. Each cloud domain is responsible for collecting a share of each sensor data. The data in each cloud domain is aggregated before sharing with others to preserve the privacy of individual sensor data. Having $m$ cloud domains, each sensor splits its data into $m$ chunks and shares the $i^{th}$ chunk with the $i^{th}$ cloud domain. Each cloud domain applies the computation function on the aggregated pieces of data and returns the result to a single operator. The operator aggregates the individual cloud results and computes the final result. Each cloud domain runs a DHT node, participates in a DHT overlay with other domains, and is responsible for a range of keys. The assignment of sensors' data chunks to the cloud nodes is done by computing the chunk's key using the hash of the chunk and finding the DHT cloud node that is responsible for that key. m-Cloud however is only capable of aggregated computation on specific polynomials. The paper lists up the set of all polynomial functions that can be computed in this way. An example of such polynomials is $x + x^2 + x^3$, where m-Cloud operates by chunking and sharing $x$ term with one DHT node, $x^2$ with another, and so on.

\subsubsection{Object Storage}
\textbf{Online Gaming \cite{kavalionak2015integrating}: } This solution aims at establishing a scalable infrastructure for the massive multi-user P2P online gaming platforms. In the centralized counterpart, a game server hosts all the game objects (e.g., characters), and users query the server for the game objects they interact with. However, this approach is not scalable, as the load on the game server increases linearly with the number of interacting users. In the proposed distributed solution, the query load of users is distributed across a DHT overlay of the game servers, where the game objects are also represented by the DHT nodes. Using the hash value of game objects as their identifier uniformly distributes them across the game servers.

\textbf{IPFS \cite{benet2014ipfs}:} The InterPlanetary File System (IPFS) is a P2P cloud storage system. IPFS uses content-addressing to uniquely identify a data object within the entire system. It utilizes the Kademlia DHT  \cite{maymounkov2002kademlia} for representing both peers and data objects. The small objects with sizes around $1KB$ are directly maintained on the DHT, while the large objects are maintained on the peers and the reference to those peers are stored over the DHT. In the former case, the object is stored by the key as its hash value and the value as the content of the object itself. In the latter case, however, the key remains the same, while the value is replaced by the address of the owner peer. Objects are immutable in IPFS, i.e., new versions of an object have different content than the older version and hence a different hash value. However, to support versioning, the new versions point back to their old version, which builds up an object graph in a decentralized way. The very first version of an object instantiates a graph for that object with only a single vertex. Children of a vertex are new versions of that vertex that are generated by applying some changes directly to that vertex. Having each object hence, one can traverse the object graph and visit and retrieve all the older versions. Since each object is represented by a DHT node, the corresponding graph for that object is maintained in a fully decentralized way over different peers holding objects of the graph vertices. With this architecture, IPFS is aimed to be utilized in scenarios like a global mount file system, personal sync folder, encrypted data sharing system, and the root file system of distributed Virtual Machines. 

\textbf{Time-sensitive object storage \cite{xiong2015full}:} This solution provides time-sensitive access control for the data objects over cloud storage. The time-sensitive access control means that sensitive data objects can be retrieved only after a specified release time, and are no longer available to new users after an expiration time. The user encrypts the data objects it owns using a symmetric key scheme \cite{katz2014introduction}, and stores them over a cloud storage platform. 
To provide time-based access control, the user also encrypts the symmetric key into chunks and distributes it across another DHT-based cloud storage of non-colluding nodes. This DHT-based cloud storage is solely meant to store the symmetric keys' chunks and is separate from the cloud storage platform responsible for maintaining the users' encrypted data objects. Each node on this DHT-based cloud storage keeps an encrypted chunk of the symmetric key of the user. The encrypted chunks of a key have an expiration time, and the honest DHT nodes discard the expired chunks. Thus, even though the data is stored over the cloud, it becomes unreachable to new users after the expiration time, as the key will no longer exist over the DHT nodes. A polynomial-based secret sharing algorithm \cite{ito1989secret} is utilized to distribute the symmetric keys of data objects in chunks over the DHT nodes. The distribution is done in a way, which reconstructing the symmetric key of a data object from its chunks relies upon both recovering its chunks from the DHT nodes and a secret timestamp that is only made available by a trusted time server at the released time of its corresponding data object.

\subsubsection{Video Streaming}
To provide load balancing in the Peer-to-Peer (P2P) video streaming systems, a 2-tier DHT-based architecture is proposed in \cite{gupta20172}. The first tier nodes in this solution are the supernodes, which belong to the service provider, and are all controlled by the same administrative domain. The second tier nodes are the users' devices. Each service provider supernode at the first tier maintains a set of video files and streams them to the second tier user nodes upon their request. The load of a supernode is defined as the number of videos it streams concurrently. To balance the load among the supernodes, all users watching the same stream construct a distinct DHT overlay. The DHT of users watching a stream is constructed and maintained by the service provider supernode that originally streams the content. Each stream is divided into chunks of identical sizes and stored over the DHT overlay of user devices interested in watching it. Each user then looks up and resolves its local DHT overlay for the chunks of the video, instead of querying the content provider supernode directly. By the local DHT, we mean the corresponding DHT of the video stream that the user is watching. In this way, the load of supernodes is balanced among its DHT-overlay of user devices.

\subsection{Open Problems}

\textbf{Dynamicity and Decentralization:} The infrastructure-oriented resource optimization solutions like \cite{zhou2016cost, zhou2014novel} tend to centralization by requiring to input almost the entire behavior of the system apriori, e.g., the time-based distribution of data objects over the entire system \cite{zhou2016cost}. Such solutions are also static as they do not aim on reacting to the dynamics of the system, e.g., sudden changes in the data object distribution. Some efforts have been made to converge towards dynamic and fully decentralized solutions \cite{hassanzadeh2016laras, hassanzadeh2018decentralized}, which operate based on the local view of the nodes, and react to the dynamics by predicting the future behavior of the system based on its present and past states \cite{hassanzadeh2020interlaced, hassanzadeh2016awake}. However, relying on the local view of the nodes for decision-making in a distributed system is prone to some margin of error. Investigating the error margin in decentralized and dynamic approaches for finding the operational trade-off points may be a potential research direction. 
% Additionally, solutions like \cite{kavalionak2015integrating} rely on a single master node, which is prone to a single point of failure \cite{tanenbaum2007distributed}. Applying a decentralized DHT-based task management solution like \cite{turnbull2014docker} to the centralized task management system models \cite{kavalionak2015integrating}, may also be considered as a potential research direction.

\textbf{Practical Abstractions:} In the infrastructure-oriented resource optimization algorithms like \cite{zhou2014novel}, an IaaS virtual machine is abstracted by a single computation parameter (i.e., the queue capacity for incoming requests). This abstraction is far from the real-world IaaS platforms where the virtual machines have the computation, communication, and storage capacities. Optimizing the DHT-based infrastructure solutions under a more realistic IaaS abstraction may be another potential research direction. 
% As a potential research direction, representing virtual machines with an attribute vector instead of a single capacity pictures a more realistic model to optimize the resource provisioning problem for the computation-, communication-, or storage-intensive tasks. 

\textbf{Trusted Third Party Mitigation:} The entire security of solutions like \cite{xiong2015full} rely on the existence of a trusted third party that shares a common reference of time. Hence, although such solutions are assumed decentralized in their operation, they are not decentralized in providing their security objectives. Securely applying decentralized time references through utilities like vector clocks \cite{lamport2019time} and blockchains \cite{fan2019blockchain} can be of potential research interest.

\textbf{Consistency Model:} The replicas in \cite{kavalionak2015integrating} are assumed on an eventual consistency model with an optimistic approach, i.e. if the writes on a game object stop, its primary and backup replicas eventually converge. Additionally, the synchronization interval in \cite{kavalionak2015integrating} between the backup and the primary is $30$ seconds, which incurs a high inconsistency interval considering continuous real-time changes in the state of the online game objects. Applying a stronger consistency model \cite{tanenbaum2007distributed} may be a potential research direction in such scenarios.

\textbf{Uniform Load Distribution:} In P2P streaming solutions like \cite{gupta20172}, the demand distribution of video chunks is time-sensitive, i.e., all the users involved in watching a stream, follow the corresponding chunks in the same order. Storing video chunks over a DHT in such a solution does not provide a time-invariant uniform load distribution across the nodes. Rather, the load is moving over the DHT nodes, i.e., each node experiences a spike in its load of queries once all users converge towards the chunk of videos it holds. The load then subsides once users move from its chunk to the next chunk of the stream. Investigating uniform load distribution in DHT-based P2P streaming services may be a potential research direction.

\textbf{Free-riders:} The liveness of DHT-based P2P solutions like \cite{gupta20172} is directly tied to the active participation of peers in the protocol. However, without an incentive mechanism in place, the peers may not maintain their availability in the system. Especially that the peers do not belong to the administrative domain of the system, hence, they do not necessarily follow the same interest as the administrative domain intends. In \cite{gupta20172} where stream chunks are stored over a DHT overlay of peers involved in watching it, a peer may switch offline once it gets all the chunks of its interest, a behavior that is known as free-riding the system \cite{tanenbaum2007distributed}. Applying an incentive mechanism that deters nodes from free-riding in P2P streaming scenarios may be a potential research direction.

\section{Blockchain} \label{BC-DHT}

\subsection{System Model}

Blockchain has become an extremely powerful and innovative technology with virtually unlimited applications across a wide range of fields, including finance, energy resources, IoT, healthcare, supply chain and many others \cite{chen2020incentive}.

The structures of blockchain methodologies are similar, despite the use of different tools and infrastructures across many different domains. It is simply a distributed ledger, however, the transactions are designed to be almost tamper-proof by using hashing and distributed mechanisms. Entities can typically access blockchain historical transactions, however, changing historical transactions in the ledger is nearly impossible. This is owing to the fact that it is distributed and other considerations \cite{fanning2016blockchain}.

The key blockchain technology features can be summarized as follows.

\begin{itemize}
    \item \textit{Transparency and tractability}: All ledger records are accessible and traceable by a pre-defined group of members who have the right to access the records;
    \item \textit{Consensus decentralization}: The nature of blockchain that requires unauthorized members to reach a consensus;
    \item \textit{Automation and smart contracts}: Blockchain does not need human interaction and verification. The software is implemented so redundant transactions and conflicts are not used;
    \item \textit{Security perspective}: Blockchain is replicated, shared and tamper-proof technology, which makes it vulnerable to security issues;
    \item \textit{Immutability}: The records in blockchain are irreversible and cannot be repudiated. 
\end{itemize}

\begin{figure}[H]
\centering
   \includegraphics[scale = 0.48]{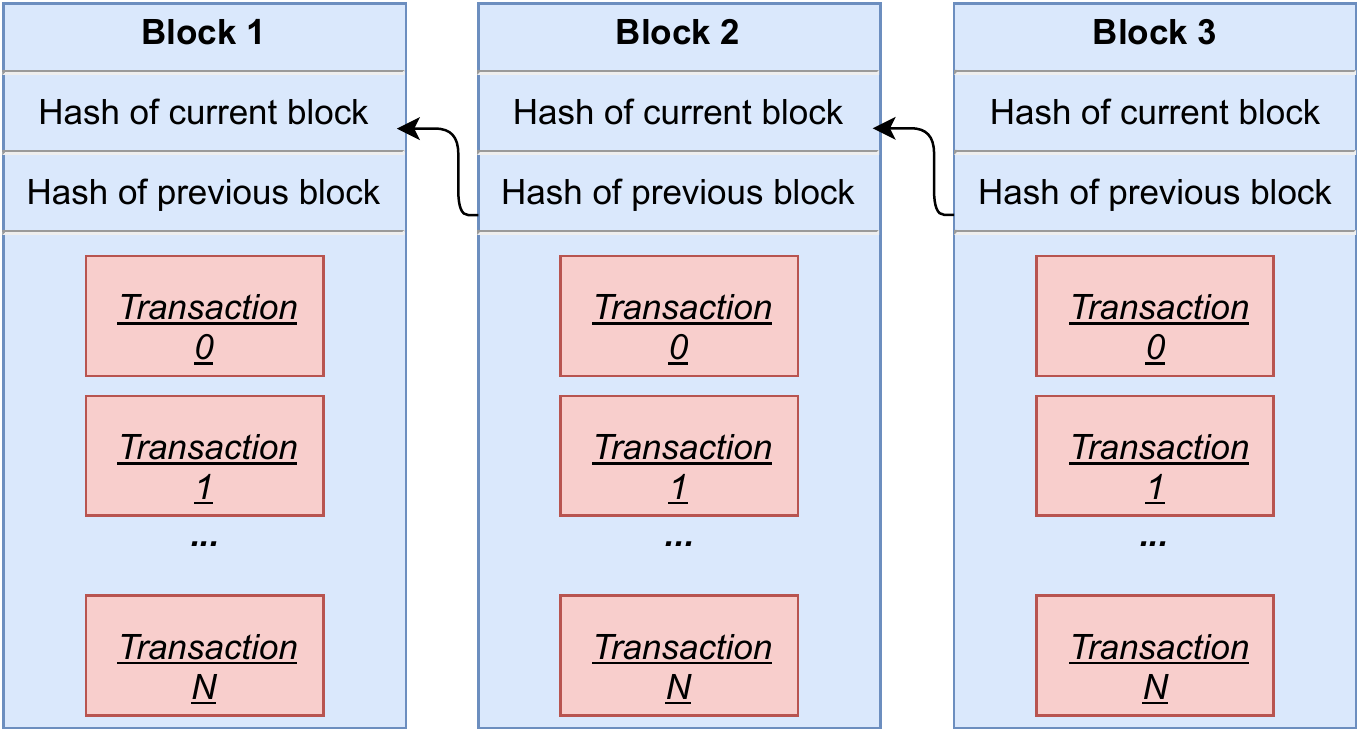}
\caption{Architecture of Blockchain}
\label{BCM}
\end{figure}

Blockchain scheme is presented in Fig. \ref{BCM}. A blockchain ledger is built in a P2P manner, and whenever an alteration to historical transactions occurs, the modifications would be available to the ledger holders. Any updates over blockchain should meet a proof-of-work theory that restricts devices with high processing power overriding historical instances. Indeed, blockchain has established reliability and security capabilities in many information systems.

Many researchers have adopted blockchain technology to improve interoperability and reduce third party interventions. For example, the decentralized infrastructures rely on hashing mechanisms for all transactions. Within the various devices and their resources, ease-of-use, reliability, privacy and accessibility are effortlessly remarked. 
Table \ref{blockchainTable} presents a summary of the blockchain-supported DHT solutions that are examined in the following subsections.

\begin{table*}[]
\resizebox{\textwidth}{!}{%
\begin{tabular}{|l|l|l|l|l|l|}
\hline
Solution                                                                & Type           & DHT             & Nodes             & DHT Utilization                  & Domain                                                                                 \\ \hline
Decoupling Validation \cite{bernardini2019blockchains} & Infrastructure & General         & Regular           &  Broadcast validation                                & Distributed                                                                            \\
Access control \cite{shafagh2017towards} & Model & General & Blockchain & Data management &Distributed\\
Lookup system \cite{matsuoka2020blockchain}                                                 & Infrastructure & Kademlia        & Regular           & Efficient data propagation and storage                & Distributed                                                                            \\
Tamper-Resistant \cite{aniello2017prototype}              & Prototype      & General         & Regular           & Blockchain layer fragmentation                                 & Decentralized                                                                          \\
ZeroCalo \cite{huang2019zerocalo}                & Infrastructure & General         & Miner nodes       & Nodes construction and message broadcasting                                 & Distributed                                                                            \\
Virtual Block Group \cite{yu2020virtual}      & Model          & General         & Data block        & Keeps the blockchain storage index                               & Distributed                                                                             \\
LightChain \cite{hassanzadeh2021lightchain}            & Infrastructure &  Skip Graph       & Regular           & Improves the communication and storage scalability & decentralization                                                                            \\

PingER \cite{ali2018blockchain}                            & Infrastructure & General         & Regular           & Files saving                                  & \begin{tabular}[c]{@{}l@{}}Decentralized \end{tabular} \\
Hash validation technique \cite{kalis2018validating}               & Method         & General         & Blockchain        & Blockchain assistant hash validation                                 & Distributed                                                                            \\
BHEEM \cite{vora2018bheem}                                  & Framework      & General hashing & Blockchain        &  blockchain-based secure health records                                & Decentralized                                                                          \\

SPROOF \cite{brunner2019sproof}                              & Platform       & General         & Blockchain        & Blockchain data storage                                 & Decentralized                                                                          \\Supply Chain Physical Distribution \cite{wu2017distributed}                           & Framework      & General hashing & Blockchain        & Public to point at private ledger                                  & Distributed                                                                            \\

Hawk compiler \cite{kosba2016hawk} & System &General & Regular & Data privacy by enabling transaction anonymity & Decentralized \\
Enigma \cite{zyskind2015enigma} & Model & off-chain DHT & Regular & Limited accessibility on data & Distributed \\
Security enhancement \cite{wright2017sustainable} & Method & General & Regular & Secure and optimal repositories & Decentralized\\

Trust Management in Social IoV \cite{iqbal2019trust}               & Framework      & General         & Vehicles nodes    & Trust information storage for trustworthy                                & Distributed                                                                            \\

Encrypted Keyword Search \cite{cai2017hardening}            & Infrastructure & Standard        & Regular           & Monitoring and attesting of new nodes                                  & Distributed                                                                            \\
KARAKASA \cite{abe2019blockchain}                                & Scheme         & Chord           & Regular           & Verification of new transactions or blocks                                 & Distributed                                                                            \\
Bitcoin Node Solution \cite{abe2018mitigating}           & Scheme         & Standard        & Clustered nodes   & Transactions verification           & Distributed   \\        

%Decentralizing Privacy solution \cite{zyskind2015decentralizing}         & System         & General         & Blockchain        & Automated access-control manager & Decentralized                                                                          \\

DHT Clustering \cite{kaneko2018dht}                          & Method         & Kademlia        & Mining/Blockchain & Load balancing                                  & Distributed                                                                            \\
  \hline                            
\end{tabular}%
 }
 \caption{A summary of DHT-supported solutions in Blockchain \label{blockchainTable}}
\end{table*}

\subsection{DHT-Based Infrastructure Solutions}

\subsubsection{Block Validation and State Storage}
In \cite{bernardini2019blockchains} the authors introduced two roles for each node: a storage role in which each node acts as a DHT that keeps values for a subset of the state elements; and a validation role where each node is validated and broadcast throughout the network.
In their model for the storage role, each defined node acts as a DHT node that retains the blockchain state and replies to each query for a state element and its value.
As for the validation role, each node keeps only the latest received blocks, in which a truncated blockchain of the nodes list could differ with time due to new block propagation. When a node receives a block, it will check the validity of the new block by considering the root-hash idea.

\subsubsection{{Data Management: Scalability and Efficiency}}

Lack of scalability is considered one of the major hurdles in blockchain. The issue regarding lack of scalability has grown with the increased adoption of cryptocurrencies, and transactions considered conventional increase exponentially. Various solutions have been proposed to overcome this scalability drawback.

The authors implemented a protocol that turns a blockchain into an autonomous access-control director without the need for third-party trust. Following Kademlia’s \cite{maymounkov2002kademlia} persistence applying LevelDB2, an interface to the blockchain was adopted. The DHT has been aided by a network of nodes that satisfy approved read/write transactions, and randomizing the data across the nodes to ensure the high availability.
The work in \cite{shafagh2017towards} introduced a blockchain solution for IoT networks that provide data management and distributed access control, it used blockchain that serve as a verifiable and access control overlay to the storage layer. The proposed system does not rely on physical storage nodes, but supports utilization of cloud storage resources.  
DHT has been adopted in a framework that targets book search—namely the Peer Book Search and Library (PBSL)—by establishing virtual libraries based on blockchain. The search and transmission of different resources is achieved by using DHT and peer-to-peer technologies. Blockchain technology has been used for the copyright protection and trading purposes, and by introducing the aforementioned technologies PBSL has overcome the drawbacks of existing library systems, and facilitated knowledge production, distribution, acquisition and consumption \cite{cao2019building}.

In \cite{matsuoka2020blockchain}, blockchain and DHT were adopted to assist a lookup system in which blockchain enables trusted communications between nodes without the need for third parties. DHT helps the nodes store the data distribution and propagate the data efficiently. Adopting a technique that sets a data expiration period does not raise the maximum capacity of the blockchain in this system.

Some researchers applied consensus mechanisms for scalability and data availability enhancement, such as the work presented in \cite{aniello2017prototype}. The authors implemented an architecture that employs a Byzantine Fault Tolerant (BFT) consensus algorithm on the first layer of blockchain, as well as a DHT solution to partition the first layer blockchain ledger amongst the nodes. The authors evaluated a two layered blockchain-supported database by focusing on scalability and availability.

\subsubsection{{Storage Efficiency}}

The proposed approach in \cite{bernardini2019blockchains} helped minimize the time required by a node of a blockchain-supported network to download the data needed for particular goals. The proposed approach assigned two roles for each node: storage and validation. For storage, each node acts as a node of a DHT to store the values of the state elements. The main aspect of validation is speed; as it allows execution of the initial synchronization of nodes in seconds.

The proposed \textbf{ZeroCalo} model in \cite{huang2019zerocalo} improves its ability to add thousands of nodes to the DHT structure. The model fulfills a consensus-based proof-of-work algorithm for the distributed ledger to provide efficient message broadcast criteria, and introduces efficient energy consumption which make it possible to be used on IoT networks.

The model addresses the node scalability issues, and a virtual block group (VBG) was proposed in \cite{yu2020virtual}. Each node in the VBG system stores only a portion of the block data and maintains the storage index to DHT by treating the block data as resources, which improves query efficiency. This methodology assured the security and dependability of block data while also saving hard drive space by reducing data acquisition time.

The LightChain solution, which adopts a DHT overlay, has been proposed in \cite{hassanzadeh2021lightchain} to improve storage scalability while also permitting decentralization in blockchain architectures. It provides addressable transactions, peers, and blocks, all of which are fully accessible on demand. It also allows each peer to replicate a subset of the transactions in order to respond to inquiries from other peers.

In \cite{ali2018blockchain}, using DHT and the legitimate blockchain, the researchers designed a blockchain access control and data storage architecture for PingER, an end-to-end Internet service management project, in order to eliminate total reliance on a single repository.
In their framework, the blockchain is used to retain the metadata of the files, while DHT saves the actual files off-chain at numerous locations by forming a network of the proposed Agents. The suggested PingER framework benefits from DHT and blockchain because they enable distributed storing, parallel computing, and effective lookup abilities.

\subsubsection{{Third Parties Reduction}}

Many proposals have been made to use blockchain to promote interoperability and eliminate external parties. For example, the decentralized apps concentrate on hash processes for all interactions by combining the concepts of different network nodes and data storage mechanisms, as well as managing the accessibility, dependability,and the privacy.

%\begin{enumerate}

\textbf{Data integrity}: refers to the accuracy and validity of collected data, and maintaining data integrity is a key focus of system security solutions. Data validation and error checking mechanisms are examples of methods to address data alteration. Many blockchain-based solutions supported by DHT have been proposed, including the solutions presented in \cite{kalis2018validating}, \cite{brunner2019sproof}, and others.

The work in  \cite{vora2018bheem} used the blockchain for implementing a secured medical information project. The authors created three sorts of nodes: the archive nodes, the light nodes and the full nodes. These different nodes manage storage overload issues, and the database and cipher managers help handle the stored hashes and ensure the level of integrity.

Kalis and Belloum \cite{kalis2018validating} provided another technique to guaranteeing data integrity using blockchain technology. They proposed a blockchain-supported hash validation mechanism. Their model separates the aggregated data from the blockchain, allowing a data identity and a hash of the collected data to be transmitted to the blockchain.

They also go over some of the possible applications for such a system. The researchers believe that the blockchain hash verification assists the detection of the unintentional and malicious acts that done to the acquired data, based on the information gained from evaluating various use scenarios \cite{kalis2018validating}.
\cite{brunner2019sproof} is another approach that leverages blockchain technology to assure data integrity.
They presented a blockchain-supported hash validation technique in which the model is maintained independently and the collected data from the blockchain is used to direct identifiers and hashes from the data to the blockchain.
The authors found that the blockchain-supported hash validation approach facilitates the detection of malicious activities based on all of the use cases mentioned in the study \cite{brunner2019sproof}.

\textbf{Data privacy}: Some proposals require privacy for transactions. There is, however, a mechanism for implementing public and private ledgers, as well as their connection to individual transactions. Hashes in the public ledger can lead to an entry in the private ledger that contains an individual event linked with that hash. Which is in some how a distributed implementation of event storage. Even with the distinction of private and public ledgers, the structure is still constrained by the general benefits of blockchain \cite{wu2017distributed}.

The Hawk compiler, a decentralized smart contract system, was presented in \cite{kosba2016hawk} as a means to achieve data privacy by enabling transaction anonymity. It functions by translating the generic codes written by programmers, into cryptographic primitives. 
Another project, known as Enigma, was presented in \cite{zyskind2015enigma}. This idea relies on splitting the data into unrecognizable chunks and distributing them over the network, thereby ensuring that no node has access to the data. Enigma applies off-chain DHT to store the data references. DHT has been proposed to build secure and optimal repositories by \cite{wright2017sustainable}. The authors introduced a method to automate contract management with hierarchical conditionality structures within a hierarchy of intelligent agents, using hierarchical cryptographic key-pairs. In addition, they proposed a method to produce a hierarchy of common secrets in order to facilitate hierarchical communication channels of increased security.

\textbf{Trustworthiness}: The work in \cite{iqbal2019trust} considered trustworthy perspectives at the network-level of a node, By collecting the trust information of every node in a single DHT that is managed by Pre-Trusted Objects (PTOs). Every single node can request the used DHT for knowledge on some other nodes' trustworthiness, with the PTOs ensuring that no malicious nodes are participating in the information exchange. 
In \cite{matsuoka2020blockchain}, using blockchain and DHT, the authors suggest a novel lookup method. DHT allows nodes in the system to store dispersed data and propagate it effectively, whereas Blockchain facilitates communication between nodes without engaging a trusted third party. Adopting a mechanism to establish a data expiration period does not raise the maximum capacity of a system's blockchain.
The researchers in \cite{cai2017hardening} proposed a protocol that helps monitor and attests  to the nodes in distributed networks. With this, each node in a network will be continuously monitored and attested to by the search queries in DHTs. The output of each attestation process has been recorded to the blockchain which identifies compromised nodes that can be removed. This solution achieves high levels of security, efficiency and robustness. 
%\end{enumerate}
\subsection{DHT Applications in Blockchain}

\subsubsection{{Load Balancing}}
Load balancing refers to the process of distributing incoming traffic from groups of servers or devices.
Many studies have considered load balancing.

\textbf{KARAKASA\cite{abe2019blockchain}} is a load balancing function proposed by \cite{abe2019blockchain} that enables storage load balancing by using DHT as a distributed storage to the storage of the bitcoin adopted nodes. It is based on the idea of maintaining node independencies of the nodes to keep the whole blockchain among DHT networked nodes. The proposed model proves that the DHT cluster's nodes can act as complete nodes without accessing the entire blockchain, and that the user can manage a Bitcoin node without the need to trust other users.

With KARAKASA, the users in a Bitcoin network who run the storage resource-constrained machines, cooperate with each other to form clusters. Every Bitcoin node carries a part of the blockchain that is allocated according to the adopted DHT process. Once the new transactions and blocks are verified, each node will query  the needed block to the DHT cluster. It is important to note that with DHT clusters, each node can handle new transactions or block verification processes in the same way full nodes do.
The proposed Bitcoin nodes include two storage functions: the BlockStorage which keeps the original data that will be stored among the DHT clustered nodes, and the ChainState that has the UTXOs. 
Each block ID in the Chord will be established by the cryptographic hashing process in order to map blocks to the accepted Chord ring.
The same authors have proposed a storage load balancing technique based on DHT \cite{abe2018mitigating}. They coined the phrase "full node", which refers to a Bitcoin node that can store all historical transaction data in order to check whether new transactions are valid or not.

In the proposed model, the DHT cluster's nodes behave as full nodes with no need to involve the entire blockchain \cite{abe2018mitigating}.
Another load balancing solution that has been considered is proposed in \cite{kaneko2018dht}. This involves a state where all participant nodes are divided into mining nodes in a P2P fashioned network. On the other hand, blockchain data also have nodes in a (DHT)-based network.

\subsubsection{Data verification}
\textbf{SPROOF \cite{brunner2019sproof}}, a platform for verifying documents in blockchain has been proposed in \cite{brunner2019sproof}. With SPROOF, all data is stored in decentralized and transparent functions. In addition to data integrity, the platform helps manage scalability and privacy issues.Within a transaction, SPROOF adds hashes of data to the blockchain, and the raw data is then kept in a DHT. If a cryptographically secure hash function is employed to create the hash, the stored data will gain the integrity and ordering qualities from the blockchain.

\subsection{Open Problems}

\textbf{Security Measures}: 
% The blockchain and cryptocurrency spheres have given rise to a number of innovative security measures to address cryptocurrency theft, as well as promising applications for other areas of computer security. 
% The inclusion of DHT in Blockchain has spawned several innovative security procedures and presented promising applications to many other fields.
Constructing and maintaining a DHT overlay in blockchains can lead to numerous security concerns such as routing attacks, Sybil attacks, Eclipse attacks, and storage/retrieval attacks. With DHT, request routing is the main component and each node must keep its own routing table correctly and update it accordingly. However, many \textit{routing attacks} could target this major component in different forms. For example, a compromised node may target the lookup routing function by forwarding lookups to incorrect nodes, or corrupt the nodes’ routing tables by forwarding faulty updates. Blockchains are also vulnerable to the \textit{Sybil attacks} where a malicious node might self-control a portion of node identifiers. Under the \textit{Eclipse attack} some nodes could mislead other nodes by providing them with false neighbors hence directing newly joining nodes to a different network under malicious control, and blocking or manipulating the honest traffic for the victim nodes. A compromised node may also \textit{attack DHT storage layer} and refuse to provide data to clients when they request it. % Kademlia proved its ability to counter this attack since it performs parallel searches; that is, all searches come together on the same path. 

\textbf{Permission-less Environments}: Though blockchain is a permission-less ledger, some users could use centralization that requires the proof-of-work algorithm to be replaced by a few trusted users. In permission-less conditions node behavior is unpredictable, and a malicious node could deny processing or directing messages to its own partitioned network. A permission-less system also depends on normal user behavior and can guarantee how well a system is working by rewarding normal behavior and disciplining negative activity. 
Without a procedure for recognizing the malicious behaviors, the DHT cannot be considered a secure system for permission-less environments.

\section{Internet of Things (IoT)} \label{IoT-DHT}
\subsection{System Model}

\begin{figure}[H]
\centering
   \includegraphics[scale = 0.34]{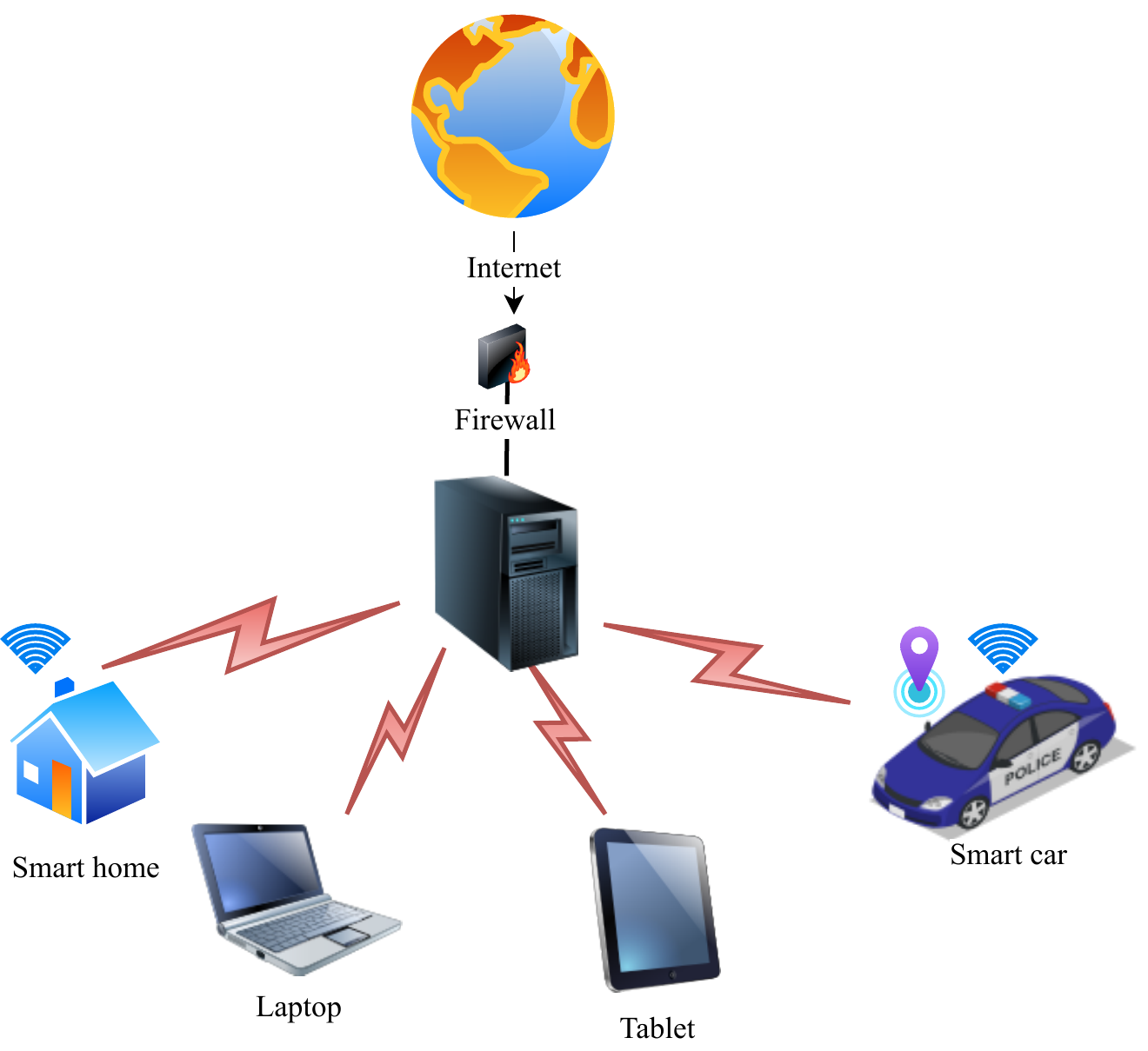}
\caption{Architecture of IoT Network}
\label{IoT_arch_M}
\end{figure}

The Internet of Things (IoT) is a fundamental component of industrial evolution, where all devices (things) connect to each other in order to collect and share data. 
Currently, innovations in IoT networks have revealed new potential for developers to create IoT-based services using various facilities, including processing, storage and communications. Consequently, IoT solutions have been proposed in many different fields, such as traffic applications (e.g. smart traffic control), healthcare (e.g. smart health monitoring) and smart management infrastructures. Blockchain has been introduced as a promising solution to managing IoT applications \cite{tseng2020blockchain}.
Fig. \ref{IoT_arch_M} represents the architecture of the IoT network, and the characteristics are summarized as follows.

%\begin{itemize}
\textbf{Heterogeneity}: The IoT network includes various services, devices, sensors and operating systems that interrelate through different protocols. Blockchain technology has recently been adopted to manage heterogeneous IoTs \cite{tseng2020blockchain}.
    \textbf{Energy limitation}: IoT devices are typically small and lightweight with limitations on resource usage. Hence, the devices are designed to function with minimal energy.
    \textbf{Vast amount of collected data:} Since the number of IoT devices is huge (i.e. into billions), the generated amount of data from these devices is also enormous.
    \textbf{Unique identity}: IoT devices have unique IDs and identifiers provided by manufactures that can be used to upgrade the devices to different platforms.
    \textbf{Intelligence}: IoT devices are considered smart devices that can interact intelligently according to combinations of hardware and software. 
    \textbf{Large scale}: As IoT devices are produced in billions, management of their generated data is critical.
    \textbf{Dynamic environment}: The IoT considers that, as a dynamic network, devices will continually join the network and others leave.
    \textbf{Complex system}: IoT has a prodigious number of devices, and coordination at this massive scale is highly complicated due to resource limitations.
%\end{itemize}

The (IoT) faces the same issues as any other communication system in terms of providing an architecture that can handle various stakeholders, a large number of devices distributed globally, and their limited connectivity.
An architecture like this must be scalable and allow for smooth operation across networks and devices with minimal human interaction. DHT has been adapted by many studies to handle such items.

Table \ref{tab:IoTSummary} summarizes the DHT-supported solutions in IoT networks.

% \usepackage{graphicx}
\begin{table*}[]
\resizebox{\textwidth}{!}{%
\begin{tabular}{|l|l|l|l|l|l|}
\hline
Solution                                                                        & Type         & DHT         & Nodes         & DHT Utilization                            & Domain               \\ \hline
HPP protocol \cite{tracey2019using}                                    & Architecture & Kademlia    & IoT nodes     & The basis for the P2P overlay in HPP       & Decentralized        \\
Data-centric system for IoT applications \cite{zhang2015cloud} & Framework    & General& Regular nodes & DHT has addressed the scalability challenge & Distributed\\
Software updates to IoT devices \cite{leiba2018incentivized}                  & Architecture & General & IoT nodes     & DSN peer-discovery scheme                  & Decentralized        \\
Decentralized control-plane \cite{bolonio2013distributed}                     & Architecture & General & IoT nodes     & Leverages the scalability                  & Decentralized        \\
Overlay management architecture \cite{nguyen2017multiple}           & Architecture & Chord       & IoT nodes           & Heterogeneous devices management           & Decentralized        \\
Distributed discovery service \cite{paganelli2012dht}                   &  Framework            & General           &        IoT nodes       & Multi-attribute and range queries                                        &       Distributed               \\
 MAAN \cite{cai2004maan}   & Framework            & Chord            &  Regular nodes             & Load balancing                                  &    Decentralized                \\
Squid \cite{schmidt2008squid} & Architecture  & General& Regular nodes & Information discovery model & Decentralized
\\Information exchange \cite{fisher2009p2p} & Architecture &Pastry&Regular nodes & Service discovery and scalability & Distributed
\\ 
P2P name service discovery \cite{fabian2009implementing} &Architecture & Bamboo& Regular nodes & Service discovery model & Distributed
\\ Devify framework \cite{b7}&Architecture &General&IoT nodes  & IoT interoperability  &Distributed
\\
Dynamic and self-configurable solution \cite{santos2017secure} & Architecture & Kademlia & IoT nodes & Infrastructure setup and messages routing & Distributed \\
Unique Identity (UID) \cite{shen2010distributed} & Scheme &General&IoT nodes & ID management & Distributed\\

IDS alerts management \cite{nasir2020prioritization} & Framework & Kademlia  &IoT nodes & Minimizing the sensors' alerts number
&Decentralized
\\
IoT flexibility \cite{fersi2015distributed} & Architecture & Chord &IoT nodes & Mobility and flexibility solution & Distributed \\
\hline
\end{tabular}%
}\caption{A summary of DHT-supported solutions in IoT \label{tab:IoTSummary}}\end{table*}

\subsection{DHT-Based Infrastructure Solutions}

\subsubsection{{Data Management (scalability and flexibility)}}

A common issue in most IoT-solutions is scalability, and the need for highly dynamic adjustment capabilities on such a large scale is complex. Each device (thing) in IoT is known by a unique distinctive ID. 
Many solutions apply DHT to IoT-based infrastructures to manage scalability issues.

The work in \cite{tracey2019using} presented an architecture that applied DHT to describe the flow of data in IoTs, from tiny sensors to entire applications. The main focus of this research was scalability perspectives, and the authors proposed a Holistic Peer-to-Peer Protocol (HPP) for low capability devices as a reliable means of independently exchanging core network. They also introduced the HPP channel as a link among peers that can conceal the network details, as well as the HPP endpoint that refers to the communication endpoint involving the HPP channels. HPP adopts DHT with Kademlia k-buckets as the basis for P2P overlay. Kademlia was chosen because it has proven scalability and robustness for different applications, and it reduces the configuration number. The authors used Kademlia-DHT buckets for node identification and XOR-based routing, and started with one bucket in order to reduce the required storage \cite{tracey2019using}.

P2P networks help achieve scalability related to data sharing and data distribution.
A study comparing P2P networks is presented in \cite{lua2005survey}. DHT-supported P2P networks, with DHT as a substrate, can locate the data values using the peers' identifiers that match the object’s key. DHT-supported networks feature a property that assigns random IDs to peers in a variety of identifiers on a regular basis.

A DHT-supported overlay network with packets routed within an overlay network that uses DHT is presented in \cite{zhang2015cloud}. Using DHT in such networks overcomes scalability challenges due to the increasing number of intermediate nodes, and also achieves placement flexibility and controllable replication.
The authors focused on data placement, information durability and access latency, and presented the Global Data Plane (GDP), which adopts the location-independent routing in large spaces in order to support the heterogeneous platforms and help with a variety of storage policies.
DHT has addressed the scalability challenge that arise with the increased number of overlay hops \cite{zhang2015cloud}.

An IoT software-update delivery network, where individuals (nodes) are compensated by vendors with digital currency for update-delivery purposes is proposed in \cite{leiba2018incentivized}. In their model, the decentralized storage network (DSN) is accessible by all individuals in the network. The DHT is adopted with a tracker-less peer discovery scheme to achieve optimal accessibility.

IoT needs a decentralized control plane that can manage multiple stakeholders, large numbers of devices, restricted connectivity and power limitations among devices. \cite{bolonio2013distributed} proposed a distributed control plane based on DHT that leverages the scalability and flexibility of P2P-DHT. The DHT is between the IoT devices and their masters , and the proposed solution enables remote sensor and device control without limitations involving security and bootstrapping. The authors used a RELOAD/Chord in their implementation, and enhanced the RELOAD specifications so it can run securely and efficiently.

A Chord algorithm-based on overlay management architecture for heterogeneous IoT device management is presented in \cite{nguyen2017multiple}. The proposed model adopts the three components of rings, nodes and sensors, as well as multiple peer Chord rings, and considers each ring in a smart context which uses chord to discover and organize things. The nodes are divided into two types: single nodes that participated in one ring and shared nodes that simultaneously participated in at least two rings. The sensors refer to actual devices with computation, communication and storage limitations that prevent them from joining the rings as nodes; thus, they are connected to specific gateways.
The authors identified each gateway, sensor and ring using unique hashed IDs, and some peer rings could intersect with others via shared nodes. 
The proposed model proved its ability to perform identification, key insertion and removal, key lookup and node joining and leaving. The model also met scalability and robustness requirements.
In IoT, everything (e.g. device) is assigned a unique identity, and management of these identities plays an important role in evaluating IoT system efficiency. The authors in \cite{shen2010distributed} proposed a correlated lookup scheme to enhance identity management performance. DHT has been implemented as the fundamental structure to build the basic methodology.

The work in \cite{nasir2020prioritization} proposes a framework that controls false alarms generated by Intrusion Detection Systems (IDSs), by considering prioritization that allocates a priority indicator to each alert. Kademlia-based DHT has been adopted for efficient alert transportation, event correlation has been utilized to discover similarities among events collected by various sensors in order to reduce false alarm rates.
In their proposed architecture, the generated messages are directed to the collectors via the DHT for processing, and Kademlia is used to store and convey alarm information. The collector nodes use clustering and correlation mechanisms to investigate anomalous behaviours and notify neighboring nodes of specified attacks. In addition, Kademlia the nodes form an overlay network and keep a routing table to help find nodes, and inquiry nodes for communication.
The results showed that DHT-supported solutions offer more accommodations than others solutions, and the Kademlia is adopted for information routing due to the high number of generated messages.
By increasing false positive accuracy to more than $\approx80\%$, the suggested framework reduced message generation by $\approx62\%$  \cite{nasir2020prioritization}.
Another management function that targets IoT flexibility, was proposed in \cite{fersi2015distributed}. The authors presented a DHT-based design and overlay architecture that meets IoT requirements (e.g. mobility and flexibility).  

\subsubsection{{Service Discovery}}

Service discovery refers to the automatic process of finding services on a network. DHT has been investigated by numerous researchers as a potential service discovery approach, since it offers essential identifier lookup functionality, and can also perform essential roles  for highly scalable and robust global discovery services, as presented in \cite{fisher2009p2p}, \cite{fabian2007distributed}, \cite{evdokimov2010comparison}, \cite{shafagh2017towards} and \cite{schmidt2008squid}. In \cite{evdokimov2010comparison}, a DHT-supported framework for service discovery in IoT has been proposed. The authors analyzed the various requirements for service discovery, and presented a summary of five approaches summarized from industrial and literature works. They presented structured P2P systems using DHT as an approach that can provide high fault tolerance, load balancing, and the elimination of single points of failure.

\textbf{OIDA} , a DHT-based discovery service, was introduced in \cite{fabian2009implementing}. OIDA is a Bamboo-based DHT, it has been tested on 350 highly distributed nodes. The experiments showed that OIDA is capable of meeting all of the basic functional and non-functional requirements for IoT name services (e.g. security). The DHT inclusion also promoted the scalability and load balancing advantages.

The authors in \cite{shafagh2017towards} proposed a Blockchain assisted IoT system that employs data management and distributed control. They presented a centralized trustworthy capability  and gave users control over their data. By employing blockchain as an auditable and decentralized control layer for the storage layer, their proposed system enabled secure sharing of data and robust and reliable access control management. This proposed work introduced blockchain-based locality-aware decentralized storage that facilitates the data storage at the networks' edge.

Another discovery service scheme that employs DHT is presented in \cite{paganelli2012dht}. This method employs a P2P network scheme to ensure system expansion, resilience, and easy maintenance. The authors created a layered method by separating three key features: range query support, multi attribute indexing,  and p2p routing \cite{paganelli2012dht}. To ensure ease of design and execution, they adopted an over-DHT indexing method.
The authors adopted different techniques for managing different layer functions, such as the SFC linearization technique to map the multiple dimension field as a one dimension field, they also used a search structure namely Prefix Hash Tree (PHT) that leverages a generic-based DHT to get an interface for the second layer and the Kademlia-based DHT implementation. 

%\textcolor{green}{Chord algorithm is considered one of the well-known DHT mechanisms that minimizes the number of hops used for P2P' network updates.
%Chord algorithm has been applied in large IoT networks to manage what the authors proposed \cite{b2}, an approach that considers the environment as a variety of peer Chord rings intersected by share nodes. In their work, nodes are treated as IoT gateways and are considered transmitters/connectors by sensors. DHT has been adopted to identify gateways, sensors and rings by specific hashed IDs; enabling them to recognize each other. The same work proposed an object discovery operation that can help solve connection issues between things,  and enable the IoT systems to efficiently manage and control things.

Other studies have adopted P2P DHT-based systems to implement their discovery services, such as the works of \cite{fisher2009p2p}, \cite{shrestha2010peer} and \cite{manzanares2011efficient}. 
The authors in \cite{fisher2009p2p} proposed a P2P-based model to support information exchanges between members of a supply chain. The authors adopted the Pastry DHT, and their work highlighted its feasibility for large P2P networks with twenty thousand nodes, and proved its scalability for large networks.
The work in \cite{shrestha2010peer} proposed a P2P network in which members of the supply chain operate the network nodes and, thereby, form a structured P2P entire network with a distorted view of the of all the other nodes. In \cite{manzanares2011efficient}, the authors focused on the EPCglobal network, and proposed a discovery service that provides item track and trace capabilities along the entire supply chain.
The work in \cite{santos2017secure} proposed a dynamic and self-configurable infrastructure on top of a structured P2P network. The authors also considered the heterogeneity issue and data access by introducing a set of communication protocols. Some mechanisms that achieved information flow security and privacy were introduced as well. 
The proposed infrastructure relied on a DHT to set up the infrastructure, along with route messages among peers.

\subsection{DHT-based Applications in IoT}
\subsubsection{Information Discovery}
\textbf{Multi-Attribute Addressable Network (MAAN)} has been proposed in \cite{cai2004maan}. This extends the Chord to enable multi-attribute and range queries by mapping the attribute values to the Chord identifier using the uniform locality preserving hashing. It operates by resolving multi-attribute questions using a single attribute query routing algorithm.
In MAAN, the Chord utilizes a consistent hash to map keys to nodes and allocates an m-bit identification for each node using a base hashing function (ex. SHA1). By using SHA1, which generates randomly dispersed identifiers, this mapping aids in load balancing.
The authors suggested employing a consistent locality-preserving based hashing function to provide uniform hashing value distributions.
They analyzed the complexity, and found that each node in MAAN has only $O(logN)$ neighbors for $N$ nodes. However, the number of routing  is $O(logN +N\times S_{min})$, where $S_{min}$ is the minimum selectivity range for all aspects \cite{cai2004maan}.

\textbf{Squid \cite{schmidt2008squid}} is a P2P system that enables efficient information discovery, and has been used for flexible keyword searches by implementing a DHT-supported structured keyword search. 
The authors introduced the definition of multi-dimensional information spaces with locality maintenance in the spaces. This dimensionality reduction scheme maps the multi-dimensional information space to the physical peers. 
Each data component in Squid has a set of keywords associated with it in a multi-dimensional keyword space. The data components are points in space, and the keywords are the coordinates \cite{schmidt2008squid}.
The Devify framework, a new IoT software architecture, has been proposed in \cite{b7}. It is intended to address P2P IoT networks and inter-operable IoT development. The authors adopted the flow-based programming (FBP) paradigm in the IoT in order to consider the application as a data exchange network.

\subsection{Open Problems}

With the growth of IoT devices, the networks’ management still has open issues. Many challenges arise with managing IoT networks from the devices’ management to connectivity. The following items represent the main challenges of managing IoT networks.

\textbf{Lack of trust in peers behaviours:} IoT newly connected devices might be a possible entry point for intruders to target the IoT network. It is an important role to prevent unlicensed devices from joining the network in order to retain network safe.  Reflecting the newly connected devices to the DHT record with securing the routing is still an open issue.

\textbf{Self-management}: With the expanding number of IoT devices, power management, device management and data management grow with it. such as some IoT devices operate on AC power, however, other devices operate on batteries. As well as, each device in IoT networks needs to be installed, configured, maintained and updated to tackle issues. Additionally, IoT devices produce a huge amount of data and such an amount creates a challenge, and appropriate data management grows to be essential. DHT solutions helped somehow in data management hence the power management and device management still have open issues.   

%\textbf{Connectivity:} All IoT devices have connectivity with the internet which differentiates them from the other technologies. 

\section{Social Networks}
\label{dht_survey:section_social_network}
%%% Author: Sanaz
% \clearpage
\subsection{System Model}

%what is social network
%Online Social Networks (OSN) are prevalent and vary from Facebook, as a friendship-based network, to Linkedin which is focused on the professional network. Among the services provided by these networks are data sharing, storage, and contact/data discovery. Existing OSNs deploy a centralized architecture with a central provider managing the social networking services. Subsequently, the provider gets access to a large amount of data which can be monetized for business-related purposes \cite{taheri2015security,boshrooyeh2018ppad,boshrooyeh2020privado,nasir2015socially}. This leads to various privacy issues which made researchers design alternative decentralized architectures such as P2P OSN to put users in full control of their personal data \textcolor{black}{\cite{nasir2015socially}}. 

% why P2P social network
In a Peer-to-Peer (P2P) Online Social Network (OSN), the social networking services are implemented relying on the users' storage and computation capacity without the deployment of a central service provider. For example, users are able to share information with their friends as well as to discover new friends while having no complete knowledge of the network. DHT-based social networks are proposed to realize the P2P OSNs by providing efficient decentralized storage management and routing services.

\begin{figure}
\centering
   \includegraphics[scale = 0.38]{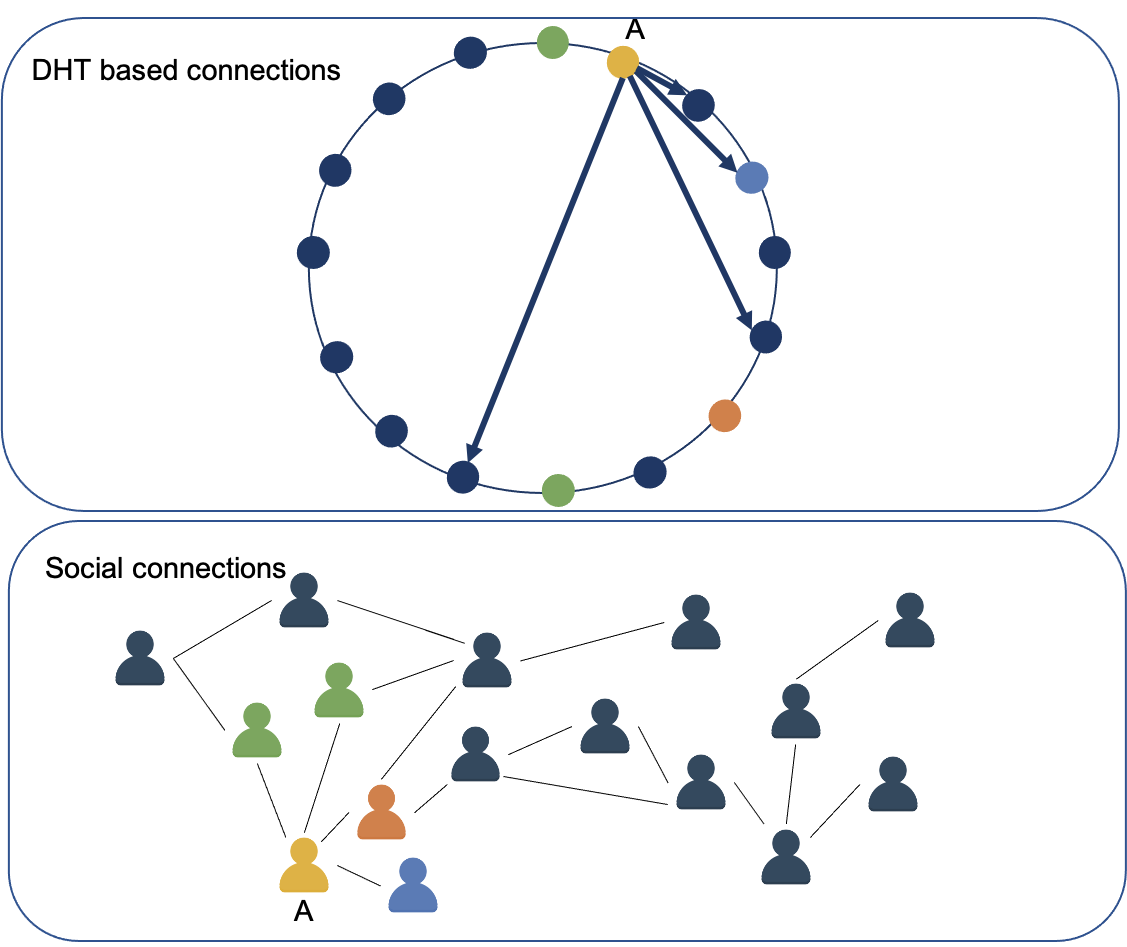}
\caption{Overview of DHT-based P2P OSN. Peers connections in the DHT overlay rely on different metrics and do not necessarily mirror the users' social links.}
\label{OSN:fig:model}
\end{figure}

A sample system model of P2P OSNs is depicted in Figure \ref{OSN:fig:model}. In a DHT-empowered OSN, nodes are usually OSN users. The users' connections in the DHT are managed through identifiers, which are either randomly selected by peers or upon various performance-concerned metrics with no central mediator. On the DHT overlay, the identifier of a node does not necessarily reflect a social factor, hence, the DHT connections may not necessarily reflect the social connections. Rather, the services provided by DHTs are empowering users to discover their social connections and their relative activities on the OSN, e.g., each others' posts. As explained in Section \ref{section_routing_overlay}, the DHT is utilized to make both the nodes as well as their resources addressable and efficiently accessible within the network. The resources in a DHT-based OSN are the social network data objects such as posts and news feeds, which makes users enabled to efficiently locating each others' as well as each others' social interactions. 

% In DHT-based OSNs, peers and information may be associated with identifiers. Using identifiers, peers can locate each others and data items e.g., walls and posts via DHT search query.% for a target identifier. DHT facilitates this by having peers and data values associated with (mostly) unique identifiers. % are unique in most DHT-based designs. Storage and retrieval of social network information e.g., walls and posts also usually take place through DHT queries. %To this end, a small unit of data e.g., object is defined and the entire wall of a user is managed as a collection of objects \cite{jahid2012decent,zahak2015collaborative}. Objects are treated as key-value pairs  where the key component constitutes the object's DHT identifier (usually from the same domain as user's identifiers) by which it gets inserted to the DHT overlay and become accessible through search operations. A sample of such structure is illustrated in Figure \ref{OSN:fig:profile}. Objects may contain links to other objects (i.e., keys of other objects) to form a linked-list type of structure which will convey a canonical order among them.  Typically, a \textit{Root Object} is designated as an entry point to a user's profile inside which links to other contents of the profile can be found (like an indexing file). Given the root object, one can retrieve a profile in an iterative manner, by following the links provided in the root object and the subsequent fetched data.

Apart from efficient and distributed routing, storage management, and query processing, DHT-based P2P OSN solutions provide more advanced tools and services concerning security, privacy, access control, and service efficiency. A comprehensive list of such services with their impact on P2P OSNs is presented in the remaining of this section and is summarized in Table \ref{dht_survery:table_osn_comparison}. 
In this table, the \textit{Nodes} column refers to the primary representation of the nodes once a DHT overlay of physical users' devices is shaped.
The \textit{Identifiers} column signifies the nature of the DHT identifiers in the corresponding service. The \textit{Architecture} indicates the structure of DHT connections, which can be P2P, super-peer, or decentralized. In a P2P architecture, all the DHT nodes exist at an equal hierarchical level and collaboratively supply the DHT-relevant services. In a super-peer architecture, a resourceful subset of participants gets involve in the DHT construction and services. In a decentralized architecture, the DHT operations are moderated by a known set of administrative domains where each domain has custody of its nodes.

\begin{table*}
\centering
\setlength{\tabcolsep}{3pt}
{
\begin{tabularx}{\textwidth}{|M{4cm}|M{4cm}|l|X|l|M{2cm}|M{2.5cm}|}
    \hline
    Solution & 
    DHT Service &
    Nodes & 
    Identifiers &
    DHT Type &
    Architecture \\
    \hline %============================================
    OSN Federation & %Application
    %& %DHT primary feature
    Verifiable Distributed Directory Service \cite{gondor2015sonic},  \cite{gondor2016distributed} & %DHT advance service
    Providers & %Nodes 
    Hash of user verification key & %Identifiers
    Kademlia &%TomP2P ()& %DHT
    Super-peer %Architecture 
    \\ \hline %============================================
    Spam protection & %Application
    %& %DHT primary feature
    Group Management \cite{xu2018oases,xu2018harnessing} & %DHT advance service Using Scribe 
    Users & %Nodes 
    Hash of group title \& creator&% name concatenated with its creator’s name & %Identifiers
    Pastry  & %DHT
    Decentralized %Architecture 
    \\ \hline %============================================
    \multirow{3}{*}{\shortstack[c]{Efficient \& Trustworthy Routing \\ \& File Discovery} } & %Application
    %& %DHT primary feature
    Socially-aware DHT  \cite{nasir2015socially} & %DHT advance service
    Users & %Nodes 
    Random & %Identifiers
    % \begin{tabular}{@{}l@{}l@{}}
    %     Flexibility\\
    %     Fault-tolerance \\
    %     Stable lookup w. $log(N)$ complexity
    % \end{tabular} &  %DHT-enabled properties
    Symphony & %chord & %DHT
    Super-peer %Architecture 
    \\ %\cline{2-6} %\hline %============================================
     & %Application
    %& %DHT primary feature
    Auxiliary lookup tables \cite{badis2016routil}  & %DHT advance service
    Users & %Nodes 
    Random & %Identifiers
    % %\begin{tabular}{@{}l@{}l@{}}\end{tabular}
    % &  %DHT-enabled properties
    Chord & %DHT
    P2P %Architecture 
    \\ %\cline{2-6} %\hline %============================================
    & %Application
    %& %DHT primary feature
    One-way Clustering \cite{liu2013efficient} & %DHT advance service
    Users & %Nodes 
    Hilbert value of interests &% vector of a user& %Identifiers
    % \begin{tabular}{@{}l@{}l@{}}
        
    %     \end{tabular} &  %DHT-enabled properties
    - & %DHT
    Super-peer %Architecture 
    \\ %\cline{2-6}%\hline %============================================
    & %Application
    Two-way Clustering \cite{shen2014social}    & %DHT advance service
    Users & %Nodes 
    Physical location \& interests &
    Cycloid  & %DHT
    Super-peer %Architecture 
    \\ %\hline %============================================
    \vspace{3pt} Data Dependency Management & %Application
    Auxiliary Identifier  \cite{chakravorty2017ushare} & %DHT advance service
    Users & %Nodes 
    Hash of original \& re-encrypted data& 
    - & %DHT
    P2P %Architecture 
    \\ \hline %============================================
    \end{tabularx}
    
}
\caption{A summary of DHT-based solutions in P2P OSNs }
\label{dht_survery:table_osn_comparison}
\end{table*}
\subsection{DHT applications in P2P Social Networks} \label{OSN:Sec:DHT_application}
\subsubsection{Federated Social Networks}
OSN providers mostly follow a closed design and do not provide a common platform for the exchange of information across different providers  \cite{gondor2016distributed}. Users willing to participate in two different OSN platforms have to create separate accounts on each platform. The OSN federation allows users to communicate their information across different OSN platforms in a transparent manner. The federation would lead to an open and decentralized ecosystem of OSN platforms in which users get to freely decide on their preferred platform and yet not losing their established connections in other networks.

\textbf{Verifiable Distributed Directory Service \cite{gondor2016distributed,gondor2015sonic}:}
The key to achieve OSN federation is to provide a common protocol connecting separate OSN platforms and enable them exchange and share information about their users. The communication protocol should handle the heterogeneity of the OSNs implementations and the fact that users' profiles are hosted by different servers on each platform.  At the heart of it is the ability to uniquely, globally and verifiably locate a user account and retrieve its profile content. 
SONIC \cite{gondor2015sonic} features a cross-platform communication protocol that allows different OSNs to federate. SONIC promotes a DHT-based directory service for user identification \cite{gondor2016distributed}. Each peer generates a unique pair of signature and verification keys and shares the verification key with her connections. A user's information is stored on the DHT overlay of federated providers as a key-value pair, where the key is derived from user's verification key and the value is the certified (i.e., cryptographically-signed) network address of the users' profile. This structure allows unique user identification by only knowing the verification keys. Moreover, the key-value bundle has verifiable authenticity due to the associated author-generated signature \cite{franchi2019blogracy,gondor2016distributed}. As such, no malicious storing service provider would be able to block accessibility to a profile by falsifying its network address without having access to the corresponding signature key. 
% SONIC \cite{gondor2015sonic} utilizes TomP2P which is a Kademlia-based DHT implementation written in Java. The rationale behind this choice is that Kademlia is based on on a reactive key-based routing protocol where users update and stabilize their routing tables based on the other node’s search queries. This mitigates the necessity of a separate stabilization mechanism and makes Kademlia-based DHTs very robust and performant. 

\subsubsection{Spam Protection}
Users' activities in a social network e.g., data exchange are vulnerable to spammers. Social spammers are entities that generate invalid data and advertise commercial spam messages and disseminate malware.
Observations prove that spammers mostly spread their malicious posts or victim links in a short period. Therefore, the prime challenge to defeat spammers and spam content is to be able to distinguish spam data from genuine data in a real-time fashion. The conventional spam classifiers are centralized and examine an offline log of message histories \cite{xu2018oases}. However, such designs are not adaptable to P2P architecture where data gets generated in a distributed manner. For a P2P environment, a spam classifier must adhere to a decentralized architecture for the data collection and classification, and be real-time, efficient, and scalable. DHT-based group management techniques provide a key component toward timely data aggregation and classification. To address spam protection, Oases \cite{xu2018oases} and Sifter \cite{xu2018harnessing} utilize DHT-based group management.  Groups are formed around OSN users who are willing to participate in spam protection. The purpose of groups is to perform data collection, aggregation, and classification in a collaborative and coordinated manner. Ultimately, the result of classification, indicating a data item being spam or not, is released to the entire group members, and potentially the entire network. 

\begin{figure}
\centering{
   \includegraphics[scale = 0.40]{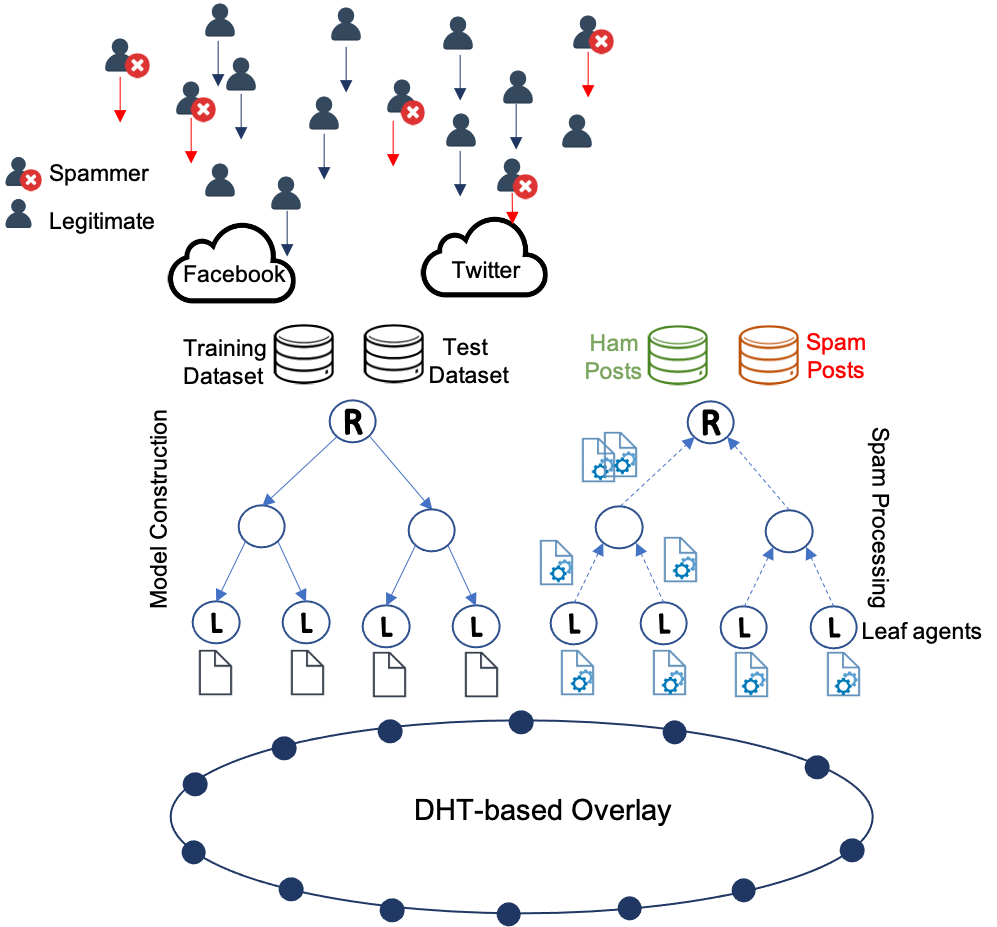}}
\caption{Overview of spam detection through DHT-based group management. }
\label{OSN:fig:spam}
\end{figure}

Oases \cite{xu2018oases} and Sifter \cite{xu2018harnessing} utilize Pastry DHT to manage OSN users connections. In specific, OSN users form the nodes of DHT and the published social contents constitute the data items stored in the DHT. To enable group management, a DHT-based aggregation tree is deployed called Scribe.  Scribe is a multi-cast tree and an application-level group communication system built upon Pastry. Users in a Scribe tree are divided into two types, the root, and the leaves. The root acts as the coordinator for the leaves whereas leaves are accountable for the data collection and processing.  Each group is associated with an identifier that is derived from the group topic. The membership to a group is done by sending a join message in the DHT overlay to the group identifier.  
An overview of spam protection through DHT-based group management, inspired by Oases design,  is shown in Figure \ref{OSN:fig:spam}.

The utilization of  Scribe, as a DHT-based hierarchical tree, allows the Oases root to disseminate datasets and instructions through $O(\log{n})$ hops, instead of $n$ point-to-point connections for $n$ leaf agents. Scribe enables a balanced workload where all the nodes in overlay have an equal possibility to take different roles (i.e., a root, parent, leaf agent, or any combination of the above) in different groups. Moreover, the Scribe structure allows multiple groups to be supported in one single DHT overlay. As such, the overhead of maintaining a complex overlay gets amortized over all the groups. Scribe can also enjoy a fast recovery process or low latency data dissemination by adjusting the value of the tree fan-out (a high fan-out value yields a lower latency whereas a lower fan-out enables a fast failure recovery).

\subsubsection{Efficient and Trustworthy Routing and File Discovery}
In a DHT-based routing overlay, nodes are assigned random identifiers which results in a random connection among the peers. This imposes communication costs on the nodes who get involved in the search process of queries that are not interested in. Moreover, having the search being traversed through unknown nodes introduces security issues and reliability concerns where misbehaving nodes can misroute or drop a query \cite{boshrooyeh2017guard, taheri2020proof}. The existing studies aim at enhancing the DHT routing efficiency and security by incorporating social factors such as users' mutual trust, and common interest in the routing process so that queries get deliberately routed through more trusted and reliable peers. A summary of these methods is provided in the following. 

\textbf{Socially-aware DHT \cite{nasir2015socially}:} 
In this solution the users' social proximity is integrated into the DHT routing overlay. The idea is that users with more mutual friends, as the social proximity measure, will be placed closer in the identifier space. Closeness is defined as the Euclidean distance of the identifiers as well as the number of hops in the overlay. Nodes initially form a DHT from their random identifiers, then they step into a refinery process during which each node swaps her identifier with one of her immediate peers, evaluates the cost of the identifier swapping, and accepts the new identifier in case it brings her closer to her social connections. Upon a successful swap, finger tables also get exchanged between the two nodes. 
% The underlying DHT utilized in \cite{nasir2015socially} is Symphony that features flexibility, fault-tolerance, and a stable lookup with the average complexity of $O(\frac{1}{k}\log{n})$ where $n$ signifies the total number of nodes. The chord is a special case of Symphony whose look-up operation is of $\log(n)$ complexity.

\textbf{Auxiliary lookup tables (ROUTIL) \cite{badis2016routil}:} 
ROUTIL \cite{badis2016routil} is a P2P routing algorithm that leverages users' common interests and social links to resolve search queries.
In a nutshell, social network users form the underlying DHT connections and each node is associated with a DHT identifier.  In addition to the DHT lookup tables, nodes maintain two auxiliary tables reflecting nodes' friendships and common interests. The friendship table is the outcome of the SPROUT algorithm \cite{marti2004sprout} in which the table entries are the identifier of the node's online friends. The interest table contains the identifier of nodes that share or look for similar content as the table owner. The interest table gets populated incrementally, i.e., when a data search query of a peer is resolved, the identifier of the data holder gets inserted into the querying peer's interest table.
Moreover, the nodes in the interest table are weighted based on the number of quires for which they hold the queried data item. The higher value of the weight indicates a higher degree of similarity. 
To resolve a query, a node consults its interest table, then invokes the SPROUT algorithm, and finally proceeds with the regular DHT lookup.
ROUTIL adopts Chord as the DHT routing overlay and enhances its performance through the utilization of interest tables where the probability of a query being resolved in a lower number of steps gets higher. Moreover, ROUTIL improves the security of Chord by incorporating the SPROUT routing algorithm to use social links to forward queries.

\begin{figure}
    \centering{
       \includegraphics[scale = 0.34]{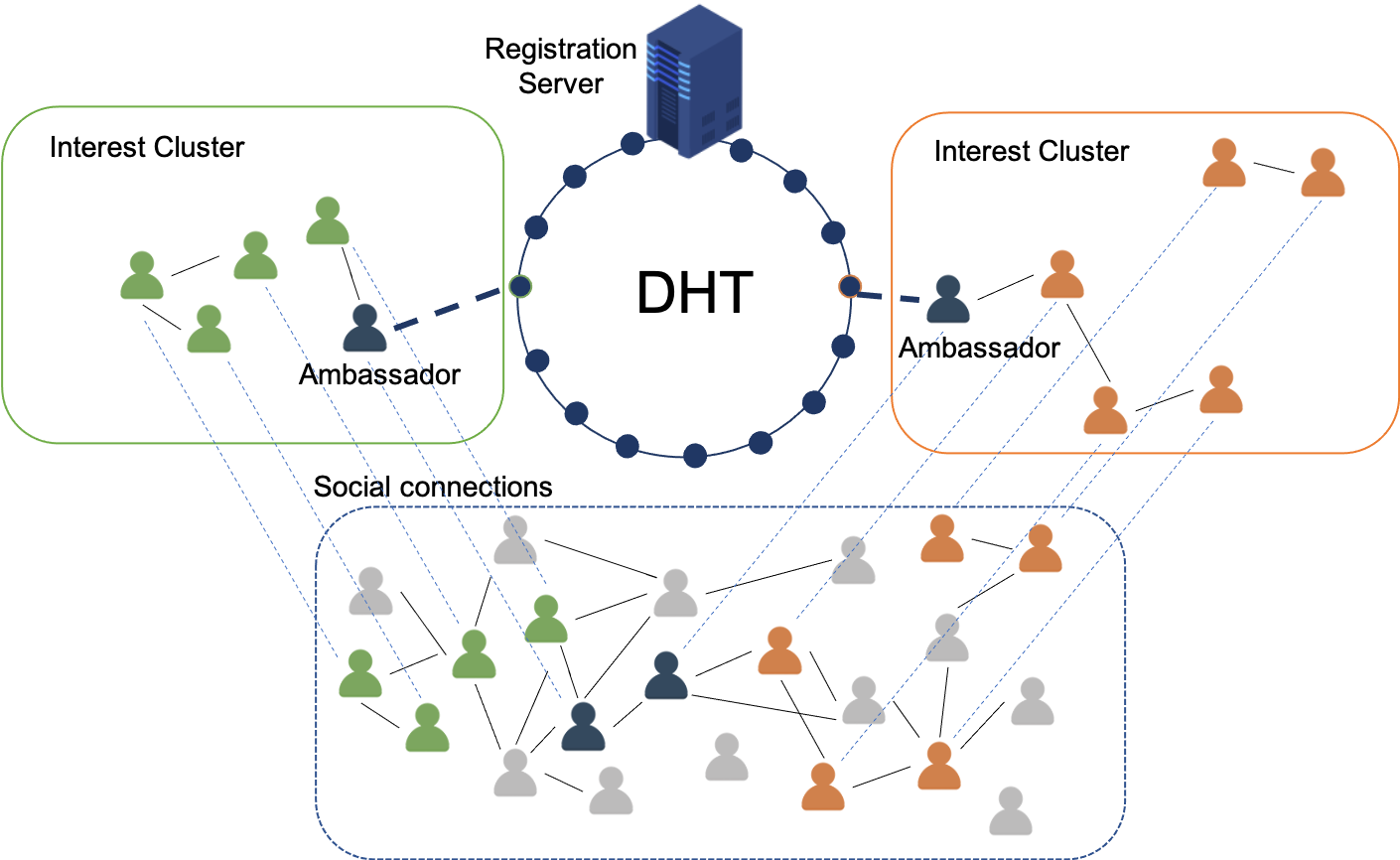}}
    \caption{Overview of DHT-based one-dimensional clustering \cite{liu2013efficient}. }
    \label{OSN:fig:onewayclustering}
\end{figure}

\textbf{Clustering in DHTs \cite{liu2013efficient, shen2014social}:}
Further studies towards a reliable and efficient DHT-based routing propose clustering methods to identify groups of peers with common features, e.g., interests and trust relationships. Once such groups are identified, the data sharing and message routing get restricted among the nodes within the same or similar clusters.  This way, the routing process enjoys better reliability and efficiency as well as becomes resilient against nodes' malicious actions such as providing faulty files or dropping query messages. Clusters can be further divided into smaller groups based on various factors to empower various levels of granularity on the security and efficiency of the routing paths. Clustering schemes usually utilize methods to sample users' identifiers from a $k$-dimensional space where each dimension represents an attribute and $k$ is the desired number of attributes. As such, identifier closeness indicates the similarity in the attribute space. Clusters are formed around the peers with similar identifiers based on some similarity metrics. Social P2P \cite{liu2013efficient} features one-dimensional clustering by using Hilbert curve. Each user picks an interest vector and derives a numerical identifier called Hilbert value using the Hilbert curve. The users with a close Hilbert value are then clustered together. The connections inside a cluster are usually based on social links, e.g., friendship. A DHT overlay is constructed from clusters ambassadors,  i.e., nodes featuring higher stability compared to the other members of the cluster. The search queries are then resolved through the intra- and inter-cluster routing algorithms. The intra-cluster routing is based on a random walk, and the inter-cluster routing is done through the DHT of clusters' ambassadors. An overview of this one-dimensional clustering is depicted in Figure \ref{OSN:fig:onewayclustering}. A similar clustering approach is proposed in \cite{shen2014social} using a two-dimensional identifier space of interest and physical proximity. 

\subsubsection{Managing Data Dependency}
The ability of the users to control, trace, and claim ownership of the content they share is vital given the recent wide development of digital assets. Users must be able to trace the number of copies/reshares made out of their data. Enforcing this becomes non-trivial when content is encrypted. 
The randomization used in encryption schemes breaks the tie between the original and the copied data. To enable tractability over encrypted content, Ushare \cite{chakravorty2017ushare} augments the DHT lookup tables with an additional column holding the hash of the original data. For a key-value pair $H(C'),C'$ where $c'$ is the re-encryption of $C$, an extra identifier $H(C)$ will be retained. This is to reflect the link between the original data and its replicas. The tuple $(H(C'),H(C))$ will constitute the DHT identifier of $C'$. Data items whose first component of their identifiers is identical are copies of the same data.

%In the augmented lookup table, three columns reside, namely, $(Col_1,Col_2,Col_3)$, $Col_1$ holds the hash of the original encrypted data (called old hash identifier), $Col_2$ contains the hash of the data item decrypted and re-encrypted for re-sharing (called the new hash identifier) and $Col_3$ stores the re-encrypted data item. In specific, when an authorized user wants to reshare an encrypted content $C=Enc_{pk}(x)$, she decrypts $C$ to $x$ using the decryption key, re-encrypts $x$ under the public key $pk'$ of her intended circle as $C'=Enc_{pk'}(x)$, and inserts $(H(C),H(C'),C')$ to the DHT overlay. Notice that in this new structure, data insertion and query is based on the conjunction of the first and second column. In other words, data items are addressed by the  tuple of keys that correspond to $(Col_1,Col_2)$. As such, the data traceability is preserved  since all the re-encrypted copies of a data share the same old hash identifier and hence are linkable to the original version. 

\subsection{Super-peers: A Common Architectural Pattern}
In DHT-based P2P social networks,  a large portion of OSN clients is mobile devices that are resource-constrained and cannot afford bandwidth, liveliness, or storage capacity for maintaining a DHT structure. As such, super-peer P2P architecture is widely adopted \cite{gondor2015sonic, gondor2016distributed, nasir2015socially,liu2013efficient,franchi2019blogracy,gondor2016distributed,shen2014social} to address peers heterogeneity. In this architecture, more stable peers with adequate storage, bandwidth, and processing power maintain the DHT structure and support other peers for complex operations. Thereby, a lower number of peers get involved in the DHT overlay which yields more efficient churn handling and lighter DHT maintenance. Due to this light maintenance overhead, this super-peer architecture also features scalability, reliability in content distribution and autonomy in administration \cite{tran2016decentralized}.

\subsection{Open Problems}
\textbf{Authenticated Dependency Management:} The data dependency management methods can address contents' copyright in a P2P environment. However, the existing solutions 
\cite{chakravorty2017ushare} rely on the trustworthiness of nodes in admitting the dependency of their files to the existing contents. However, nodes can deliberately misbehave and do not disclose the origin of the re-shared files. More robust and reliable solutions can utilize incentivization to encourage the resharing transparency e.g., financial rewards for re-sharing content. Alternatively, zero-knowledge proof of non-membership can be leveraged where a data publisher proves her content is not a copy of any existing OSN content.

\textbf{Reliable Storage and Retrieval: } The reliable storage and retrieval of data in a DHT-based OSN become a concern when data hosts are picked randomly. Such hosts have no mutual trust with the data owner and may refuse to serve the recent content, and pose an availability and accessibility threat to the system. One can adopt data replication over multiple non-colluding hosts to enable better reliability. 
% The consequences of such misbehavior in the distributed directory service of Blogracy \cite{franchi2019blogracy} is more severe where data is the entry point to the user's activity stream and manipulating the update log makes the user's profile inaccessible to the network. One can adopt data replication over multiple non-colluding hosts to enable better reliability. 

% In DHT-based P2P social networks, having the data items stored by random peers raises issues regarding the integrity of the updates log of the data. In particular, storing nodes may drop updates on their hosting key-value pairs and serve the system with the older versions of the data. This issue becomes more severe in the  distributed directory service of Blogracy \cite{franchi2019blogracy} where data is the entry point to the  user's activity stream and manipulating the update log  makes the user's profile unaccessible to the network. One way to alleviate this issue is through data replication in which multiple random peers get to house the same data.  This way, manipulating the integrity of the update log is much more costly since  the attacker requires the conspiracy of all the replicas  (otherwise her wrongdoing will get caught by honest replicas). The number of replicas can be configured depending on the fraction of malicious peers and the required efficiency and consistency-management overhead.

\textbf{Interest-based Identifiers:} The interest-based routing is adopted by  ROUTIL \cite{badis2016routil} to enable reliable and secure and efficient routing. Two nodes have a common interest if one of them retains the data item that is queried by the other node. However, this relation would not hold when storage is assigned to DHT nodes based on their random identifiers. To cope with this issue, interest-driven identifiers like Hilbert values \cite{bayer1992computation} can be utilized to enable interest-based identifier assignment.

\section{Mobile Ad Hoc Networks}
\label{dht_survey:sec_manet}
\begin{table*}
\centering
\setlength{\tabcolsep}{3pt}
\scalebox{1}
{
    \begin{tabular}{ |l|l|l|l|l|l| }
    \hline
    Ref.&
    Solution &
    Communication Type &
    Nodes of DHT &
    Purpose of DHT\\
    \hline
    \cite{abid2015merging, shah2016cross, abid20153d, shah2014efficient} &
    Network Partitioning \& Merging &
    IEEE 802.11 &
    Information of Nodes &
    Routing\\
    
    \cite{ramya2016deterring, manikandan2016stratified} &
    Selfish Node Detection &
    IEEE 802.11 &
    Reputation Score of Nodes &
    Data Transmission\\
    
    \cite{tahir2017logical} &
    Clustering &
    IEEE 802.11 &
    Information of Nodes &
    Dynamic Topology\\
    
    \cite{ibrar2016stability, arunachalam2017broadcast} &
    Multi-path Routing &
    IEEE 802.11 &
    Information of Nodes &
    Dynamic Topology\\
   
    \cite{zahid2018distributed} &
    Network Partitioning &
    IEEE 802.11 &
    Information of Neighbour &
    Traffic Overhead\\
    \hline

    \end{tabular}
    
}
\caption{A summary of DHT-based solutions in MANET}
\label{dht_survery:table_manet_comparison}
\end{table*}

\subsection{System Model}
\begin{figure}[ht]
\begin{center}
    \includegraphics[scale = 0.24]{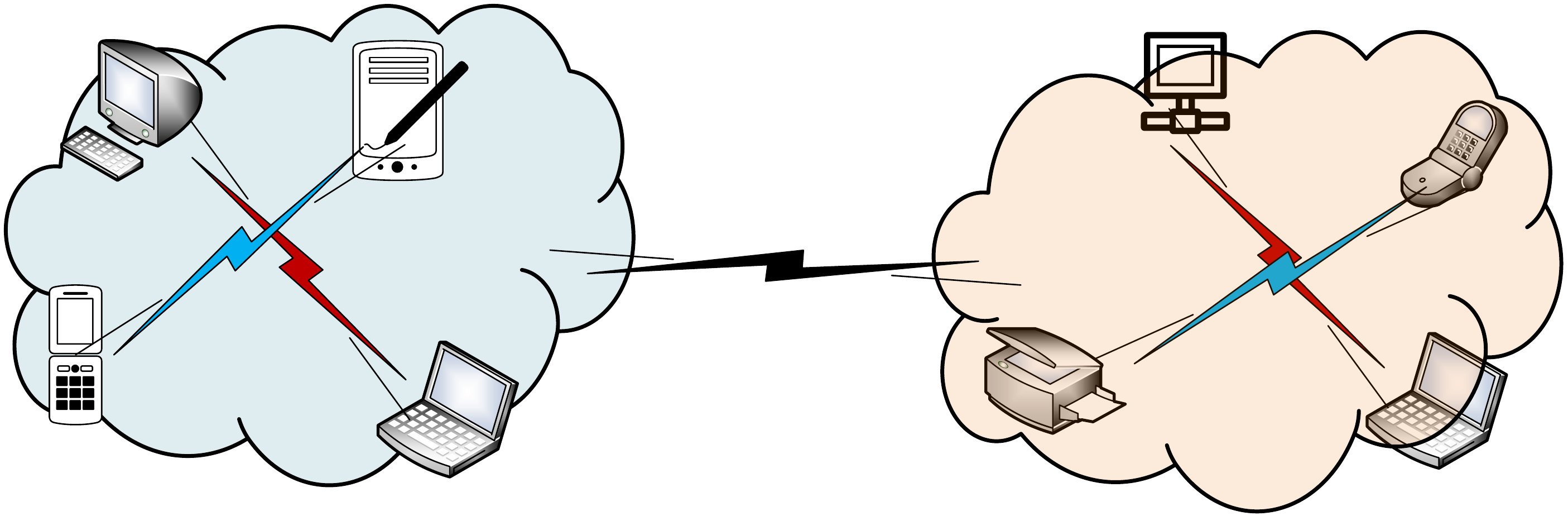}
\caption{Architecture of Mobile Ad Hoc Networks} \label{fig_general_manet}
\end{center}
\end{figure}

Advances in networking technologies expedite the deployment of self-organized wireless ad hoc networks. With the proliferation of the Internet and IoT-enabled devices, wireless ad hoc networks gain momentum and their first-of-their-kind applications come into reality. In typical wireless ad hoc networks, nodes communicate with each other through wireless links and collaborate to perform specific tasks. Mobile Ad Hoc Networks (MANETs) are kind of wireless ad hoc networks in which the participating nodes set up a temporary network that is capable of self-configuration and self-healing without any infrastructure. Nodes in MANET can be any device that is equipped with storage, sensing, communication, and computation. In MANETs, the nodes freely move and perform networking at any time and anywhere. The applications of MANETs are present in various domains such as Wireless Personal Area Networks (WPANs), Body Area Networks (BANs), Flying Mobile Ad Hoc Networks (FANETs). MANETs can be formed by any kind of wireless device as demonstrated in Fig.~\ref{fig_general_manet}. The wireless devices are equipped with wireless transmitters and receivers including radio interfaces (e.g., IEEE 802.11, IEEE 802.11p, IEEE 802.15, IEEE 802.15.4, IEEE 802.16, Wi-Fi, Bluetooth, ZigBee and WiMAX). 
%For example, vehicles can be the nodes in MANET, and the temporary network formed by the vehicles via IEEE 802.11p is called Vehicular Ad Hoc Networks (detailed in Section~\ref{section_vanet}).

In MANET, the topology changes frequently and all participating nodes need to participate in routing and equally contribute to maintenance. The MANET protocols should be adaptive to such dynamic topological change which is challenging. Nodes are communicating with each other through intermediate nodes in a multi-hop manner to deliver messages from source to destination. Although MANETs have been investigated for more than a decade, limited communication range, communication link breakage, efficient routing, and management protocols are still major challenges in many applications. To address these issues in different environments and settings, many solutions have been proposed in the literature. These solutions propose different approaches and/or methods to address unique MANET challenges in which they commonly share the usage of DHT. The DHT is mainly used for storing information and stored information is leveraged to come up with efficient routing and management protocols. However, there are still several challenges that need to be addressed. In this part of the paper, we want to shed light on this and want to explore more on the following. First, we want to understand how DHT is utilized to address the unique challenges of MANET in existing applications. Second, we aim to analyze the current limitation of DHT-based solutions and present several open problems and opportunities for future research. Table~\ref{dht_survery:table_manet_comparison} summarizes DHT applications in MANET and the following subsections will discuss them in detail next.

\subsection{DHT Applications in MANET}
The research attempting to use the DHT in MANET can be grouped according to four perspectives as; efficient routing \cite{abid2015merging, shah2016cross, abid20153d, shah2014efficient}, data transmission \cite{ramya2016deterring, manikandan2016stratified}, handling the dynamic topology \cite{tahir2017logical, ibrar2016stability, arunachalam2017broadcast} and mitigation of the traffic overhead \cite{zahid2018distributed}.

\subsubsection{Routing}
Among the challenges in MANET, routing is the most popular one. Due to varying topologies and infrastructure-less architecture, routing becomes the center of many works in MANET literature. The DHT is used for different purposes for the sake of achieving scalable and efficient routing. A survey that compares the features, strengths and weaknesses of existing DHT-based routing protocols is presented in \cite{abid2014survey}. In DHT-assisted routing, nodes have a logical identifier (LID) in addition to the universal identifier (UID) (e.g., IP/media access control (MAC) address). The nodes are interconnected with each other via their LIDs on a logical network. However, if the logical structure does not represent the physical proximity of nodes then the topology mismatch problem rises. The topology mismatch problem is further exacerbated by the network partitioning and merging. Due to the limited transmission range and self-organizing nature of MANET, network partitioning and merging can happen frequently. The DHT is leveraged to keep the relation between the logical and physical address of nodes in \cite{abid2015merging, shah2016cross, abid20153d, shah2014efficient} to solve the network partitioning and merging.  Through the simulation-based evaluation, it has been shown that DHT-assisted network partitioning and merging is beneficial in terms of achieving efficient and scalable routing in MANET.

\subsubsection{Data Transmission}
In MANET, each node acts autonomously and takes its own decision. Nodes have direct access to nearby other nodes, which are located in the communication range. When the data needs to be disseminated to a destination node, which is out of the transmission range, then nodes depend on each other in forwarding the data for reliable communication. However, some nodes do not want to share resources while delivering the data at larger distances. These nodes, on the other hand, are defined as selfish nodes and it is a major problem in cooperative communication-based MANET applications. To address the selfish node detection, reputation scores of the nodes are computed and kept in DHT to decide the next hope while relaying the data from source to destination in \cite{ramya2016deterring, manikandan2016stratified}. Locality-aware DHT is constructed collaboratively to gather the node's reputation which constitutes the global reputation. Based on the tracked reputation scores, selfish nodes are identified and blacklisted for the sake of reliable data transmission. 

\subsubsection{Dynamic Topology}
The high mobility and frequent network topology changes are the distinguishing characteristics of MANETs. Handling the high mobility and topology changes, on the other hand, are the most popular research problems that have been worked on for more than a decade.  Researchers leveraged the DHT and proposed many solutions and/or protocols to have a scalable and efficient MANET application. For example, a cluster-based DHT routing protocol is proposed in \cite{tahir2017logical} to handle the high mobility of nodes. Nearby nodes are clustered and information of nodes is kept in DHT. The gateway nodes (e.g., single or multiple nodes) which are in the radio range of cluster heads act as a relay node while forwarding data packet among clusters. \cite{ibrar2016stability} investigates multipath routing to solve the link failures which is caused by the dynamic network topology. Multi-path Dynamic Address Routing (M-DART) is proposed in which DHT is leveraged for two main reasons: (i) mapping the node's identifier and routing address in a dynamic environment and (ii) distributing the node's location information is distributed throughout the network.

\subsubsection{Traffic Overhead}
MANET is capable of self-configuration and self-healing without any infrastructure. However, both the limited transmission range and high mobility of nodes cause network dis-connectivity which results in loss of Logical Space (LS) information. The LS of nodes stores the mapping information of the LID and UID node pairs. The LS is important especially in network partitioning where the loss of it may cause recurrent network partitioning and traffic overhead. Detecting the partitioning beforehand and replication of important mapping information may be the solution. \cite{zahid2018distributed} addressed this issue and proposed distributed partitioning management to improve the performance of DHT-based routing in MANET. In their proposed solution, the critical nodes are identified without creating additional traffic overhead and mapping information of them is replicated to minimize the information loss and communication disruption.

\subsection{Open Problems}
Utilizing DHTs to address the unique challenges of MANET brings various research questions in which we believe they can be further improved in the following three perspectives.

% @seyhan: this part has been unitemized to save space
\textbf{How to enable energy-efficient DHTs?} With the proliferation of mobile devices, more research effort has been put into scalable and efficient MANET protocols. The larger MANET grows, the more overhead incurs which makes maintenance, self-configuration, and self-healing challenging. If we consider the battery constraint of MANET nodes, power consumption and energy efficiency became crucial. Despite many solutions rely on DHT usage to address the unique challenge of MANET, the energy efficiency of DHTs is orphaned. MANET has lots of applications and serves many users at the same time. Ignoring the power consumption of DHT may put MANET into services unavailability. An energy-efficient DHTs which dynamically adjust the DHT participating nodes is emerging technology for future MANET application.

\textbf{How to preserve DHTs in lossy networks?} MANET is a Low power and Lossy Networks (LLNs) that consists of many embedded devices with limited power, memory, and processing resources. The network formation may be triggered based on events and nodes start to sends their local state to the network. However, due to limited bandwidth and communication channels shared by different applications, a small set of nodes can send information through the network within some specified time interval. DHTs are utilized to come up with routing strategies including multi-path route generation with an assumption of DHTs are managed by nodes through communication. However, a lossy network is challenging and it is not realistic to assume that there exists a 100\% packet delivery ratio. A detailed study is needed to understand the DHTs in the lossy network and a brand new approach is a must to ensure the MANET service availability even in high lossy networks.

\textbf{Can DHT itself be standardized?} The general trend of DHT usage in MANET is nodes collaboratively keep and manage the DHT according to underlying network topology. Researchers come up with a different version of implementation according to needs in which there is no standard DHT in the literature. Many studies claim the usage of DHTs but the implementation details are hidden. A standardized DHT implementation with open-source implementation is still an open issue that should be addressed before any practical deployments of DHT-assisted MANET application.

\section{Vehicular Ad Hoc Networks} \label{section_vanet}
\subsection{System Model}

\begin{figure*}[t]
\centering
   \includegraphics[scale = 0.24]{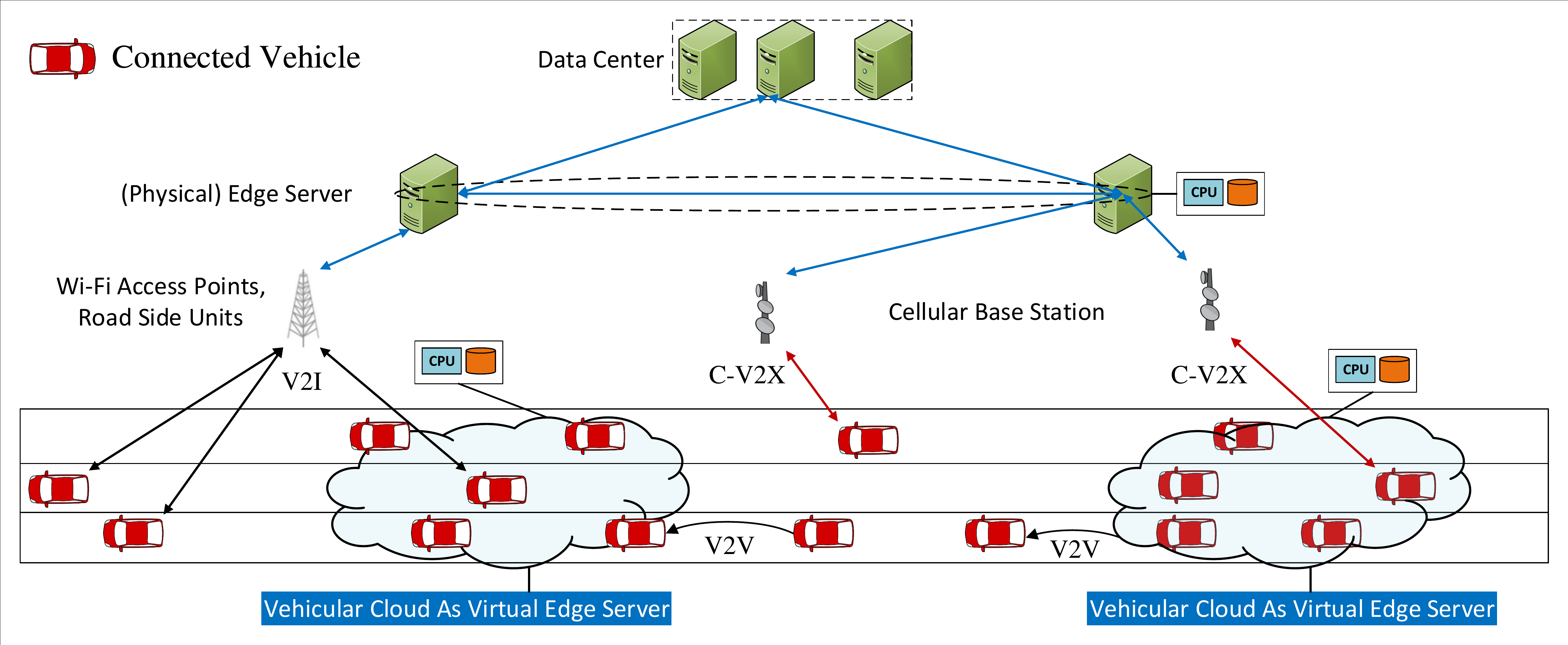}
\caption{Architecture of Vehicular Ad Hoc Networks}
\label{fig_general_vanet}
\end{figure*}

\begin{table*}
\centering
\setlength{\tabcolsep}{3pt}
\scalebox{0.9}
{
    \begin{tabular}{ |l|l|l|l|l|l|l| }
    \hline
    Ref.&
    Solution &
    Communication Type &
    Management of DHT&
    Nodes of DHT &
    Purpose of DHT\\
   \hline
    \cite{dressler2019virtual} &
    Virtual Edge Server &
    V2V, V2I and C-V2X &
    Vehicle Layer&
    Provided Services of Vehicles &
    Service Directory \& Discovery\\
    
    \cite{dressler2014towards} &
    Virtual Network \& Storage &
     V2V, V2I and C-V2X  &
     Edge/Cloud Layer&
    Storage Capabilities of Vehicles &
    Service Directory \& Discovery\\
   
    \cite{hagenauer2016poster, hagenauer2018vehicular, ucar2019platoon} &
    Virtual Network Infrastructure &
    V2V and V2I &
    Edge/Cloud Layer&
    Provided Services of Vehicles &
    Service Directory \& Discovery\\
    
    \cite{higuchi2019cooperative} &
    Cooperative Download &
    V2V and V2I &
    Vehicle Layer&
    Data Interest of Vehicles &
    Lightweight Coordination\\
    
    \cite{malik2019trust} &
    Trust\&Reputation Computation &
    V2V &
    Vehicle Layer&
    Trustworthy of Vehicles &
    Storage \& Look up\\
    
    \cite{rowan2017securing} &
    Secure Communication &
    V2V and V2I &
    Edge/Cloud Layer&
    Hash Value of Data &
    Storage \& Look up\\
    
    \cite{raj2016descriptive} &
    Intrusion Detection &
    - &
    -&
    IP Addresses of Vehicles &
    Routing\\
   
    \cite{jain2016rsu} &
    Name Data Networking &
    V2V and V2I &
    Edge/Cloud Layer&
    Name of Vehicles &
    Routing\\
   
    \cite{lu2020federated} &
    Federated Learning &
    V2V and V2I &
    Edge/Cloud Layer&
    Metadata of Vehicles &
    Secure \& Private Vehicle Election\\
    \hline
    
    \end{tabular}
    
}
\caption{A summary of DHT-based solutions in VANET}
\label{dht_survery:table_vanet_comparison}
\end{table*}

Cars today are equipped with a set of rich resources consisting of storage, sensing, communication, and computation  \cite{altintas2015making}. Through Vehicle-to-Vehicle (V2V), Vehicle-to-Infrastructure (V2I), and Vehicle-to-Cloud (V2C) communications, vehicles form Vehicular Ad Hoc Networks (VANETs) and collaborate over vehicle-to-X (V2X) networks as demonstrated in Fig.~\ref{fig_general_vanet}. Despite such heterogeneous V2X networking brings many challenges, it also paves the way for great opportunities. For example, heterogeneous communication capabilities of vehicles are leveraged to come up with a hybrid architecture for delivering the safety messages with shorter delay and high packet delivery ratio in \cite{ucar2015multihop}. Similarly, radio frequency and visible light hybrid communication-based architecture are proposed in \cite{ucar2018ieee} to ensure the safety of traveling and maneuvering under different security attacks.

Although vehicular networking technologies have been investigated for more than a decade, increasing data demand is still a problem. Today, a major part of the provided services for connected vehicles relies on interaction with remote servers. The communication with the data center, on the other hand, is achieved by cellular communication over the underlying backbone network which may suffer from large end-to-end delay and capacity limits. To overcome these issues, Mobile Edge Computing (MEC) is recently proposed and standardized over the years. The basic idea in MEC is to deploy small-scale computing facilities near the edge of the network to partially take over the computational tasks of cloud servers. MEC is beneficial in terms of reducing the end-to-end communication latency as well as improving the capacity of backbone networks. However, the high demand and huge data traffic in the radio access network are still problems. Recently, leveraging vehicle resources in such a heterogeneous networking environment has given birth to the Vehicular Cloud (VC) concept \cite{higuchi2017feasibility}. VCs are a promising solution in which connected vehicles collaborate with each other through V2X networks and offer their processing, storage and communication services as virtual edge servers (see Fig.~\ref{fig_general_vanet}). Through such virtual edge servers, the data can be aggregated and/or cached and the latency and communication overhead in cellular networks could be significantly reduced. Furthermore, VCs can be complemented with conventional cloud and edge computing infrastructure and vehicle resources can be rented out by other users. Such architecture can be an enabler of a wide range of novel applications such as collaborative data storage \cite{ucar2019collaborative}.

Fig.~\ref{fig_general_vanet} demonstrates the general architecture of VANET which consists of VCs as virtual edge servers. In such an architecture, the collaboration among vehicles is coordinated to generate advanced vehicular cloud services, which an individual vehicle cannot make alone. It has been demonstrated that collaboration enabled vehicular architecture is an emerging paradigm to utilize the ever-growing computational resources of intelligent vehicles to form small-scale virtual edge servers \cite{higuchi2019offloading}. Although conventional cloud technologies are mature enough to provide services to vehicles, the high mobility of vehicles is the limiting factor in achieving scalable and efficient VANET applications. To provide robust service provisioning under different road topology and vehicle densities, on the other hand, many solutions have been proposed in the literature. One shared point in most of the proposed solutions is the utilization of DHT. DHT is mainly used for storing/retrieving location and time-sensitive information and it provides a lookup service with name and value pairs. Vehicles can search for a service with the name associated with a provided service. However, there are still several challenges that need to be addressed. In this part of the paper, we want to go one step further and shed light on this by exploring the following. First, we want to understand how DHT is utilized to address the unique challenges of VANET in existing applications. The DHT has been studied for some key VANET applications as summarized in Table~\ref{dht_survery:table_vanet_comparison}. Second, we aim to analyze the current limitation of DHT-based solutions and present several open problems and opportunities for future research.

\subsection{DHT Applications in VANET}
 DHT is leveraged for service directory and discovery \cite{dressler2019virtual, dressler2014towards, hagenauer2016poster, hagenauer2018vehicular, ucar2019platoon}, routing and scalability \cite{higuchi2019cooperative, malik2019trust, rowan2017securing, raj2016descriptive, jain2016rsu}, security and privacy \cite{lu2020federated}. Following subsections will discuss them in detail.

\subsubsection{Service Directory and Discovery}
 One key functionality of DHT usage in VANET is the so-called service directory and discovery. In the service directory and discovery, vehicles in and around the vehicular network can register their available services as well as look up the services offered by other vehicles. Recent research works have studied some key features in the realization of DHT-assisted service directory and discovery. Authors in \cite{dressler2019virtual} proposed to form a VC and vehicle's resources are provided as virtual edge servers to others. Vehicles are organized in interconnected groups and the DHT keeps the vehicles' provided services. The members of VCs leveraged the DHT-assisted lookup service to reduce the delay of service discovery and data transfer. \cite{dressler2014towards} established connected clouds of parked vehicles as temporary network and storage infrastructure to maintain network connectivity and provide storage capabilities as a service to other road users. A Virtual Chord Protocol (VCP) is proposed to track the connected vehicles as well as their available storage capabilities. The vehicle resources are utilized as virtual network infrastructures and moving service directories are proposed in \cite{hagenauer2016poster, hagenauer2018vehicular, ucar2019platoon} to bridge service providers and service users. Vehicles collaboratively keep a DHT-based service directory to provide media for other vehicles for registering their services.  Cooperative downloading in vehicular heterogeneous networks is proposed in \cite{higuchi2019cooperative}. Vehicles collaboratively maintain a DHT. The DHT is leveraged as a lightweight coordination mechanism for the vehicles to agree on non-overlapping subsets of data segments that they are responsible for downloading on behalf of other members.

\subsubsection{Routing and Scalability}
Routing and scalability are heavily investigated topics in VANET \cite{karagiannis2011vehicular}. Simple broadcasting algorithms are proposed for local information and geo-routing is leveraged for reaching more distant destinations. Proposed routing methods work quite well for shorter and highly local transmission. However, vehicular data on roads are highly dynamic. Moreover, the system load is much higher when routing is performed under tens of thousand vehicles at the same time. The DHT is leveraged and utilized to come up with scalable and reliable routing strategies. DHT-assisted storage and lookup methods are proposed to evaluate the trustworthiness and credibility of vehicles in \cite{malik2019trust}. The trust and reputation values are computed and updated as the vehicles live inside the vehicular network. Anyone in the network can query the DHT to retrieve the trust and reputation values in a scalable manner. A scalable secure communication method is proposed in \cite{rowan2017securing}. Block-chain-assisted DHT-based secure communication architecture is proposed to keep the hash values of transmitted data. A survey that explores the intrusion detection system in different domains is presented in \cite{raj2016descriptive}. An intrusion detection system requires extensive search and computation. The survey demonstrates that tracking the identified intrusions by DHT is beneficial in terms of detection accuracy and predicting future intrusions. A name data networking architecture is combined with DHT to improve the routing and communication efficiency in \cite{jain2016rsu}. The roadside units keep track of the names of the vehicles (a.k.a. usernames) and the data is forwarded according to the usernames of vehicles dynamically.
\subsubsection{Security and Privacy}
Advances in artificial intelligence and MEC accelerates the Vehicular Cyber-Physical Systems (VCPS). Services such as content caching, dynamic resource assignment, efficient data sharing, and distributed computing become the enabler methodologies in reducing the cost and enhancing the utility of VCPS. Since the data is cached and computed in a distributed way, security and privacy become a major vulnerabilities in VCPS. To address this issue privacy preserved learning method is proposed in \cite{lu2020federated}. DHT is kept by roadside units and all related vehicles are determined by searching the DHT before the learning process starts.

\subsection{Open Problems}
The rapid growth of VANET opens a specific line of research. The proposed concept of combining DHT with the vehicle and cloud/edge layers brings various research questions in which we believe are central in the following four questions.

% @seyhan: this part has been unitemized for sake of space. 
\textbf{Which vehicles should keep the DHT?} There exist many applications that utilize the DHT in the vehicle layer. However, it is not clear how they elect the vehicles to keep DHT. Cars today have various sets of computing, storage, communication resources. Leveraging too many cars might cause overhead and the consistency/synchronization of DHT becomes problematic. Electing a fewer number of vehicles, on the other hand, results in a bottleneck in terms of reaching DHT nodes. An intelligent vehicle election mechanism that handles the dynamic join and leave of vehicles is an emerging enabler technology for future VANET applications.
    
\textbf{Which layer should manage the DHT?} As illustrated in Fig.~\ref{fig_general_vanet}, the VANET applications leverage not only the vehicle but also the cloud/edge layer. Despite it is application-oriented, there is no clear distinction under what conditions which layer the DHT should rely on. Collaborative DHT in the vehicle layer is beneficial in terms of latency via short-range communications but it becomes problematic while handling the dynamic join and leave of vehicles. The cloud/edge layer DHTs is more stable compared to the collaborative approach in the vehicle layer. However, exchanging large amounts of data may require longer-lasting sessions which is currently one of the limiting factors in cloud/edge layer DHT applications. A more detailed study is needed to understand the advantages and drawbacks of DHT in different layers.  Moreover, a hybrid DHT protocol that functions in both vehicle and cloud/edge layers is another direction of DHT applications that can potentially improve the benefits.
    
\textbf{How to connect to the DHT?} In existing solutions, elected vehicles (e.g. the Cloud Leader) become a gateway and function as a bridge between the DHT and the other vehicles. Gateway vehicles periodically send messages to announce their role and other vehicles request services through these gateways. If gateway vehicles are distributed uniformly within the VANET then the connectivity to DHT is good. However, this could cause congested channel conditions. A study that explores the election of gateway vehicles is needed to further understand the trade-offs between performance and cost.
    
\textbf{How to connect DHTs with each other?} Due to the distributed nature of VANET, many DHTs may be kept not only in the vehicle but also in cloud/edge layers. Sometimes, messages need to be sent from one DHT to another for service discovery. The DHTs are not necessarily disconnected from each other therefore the service discovery messages can be delivered and responded immediately. An efficient communication protocol to maintain a stable connection among DHTs in both vehicle and cloud/edge layers is still an open problem that should be addressed before any practical deployments of DHT-assisted VANET applications. 

\section{Conclusion}
\label{dht_survey:conclusion}
As distributed key-value stores, DHTs are among the most significant P2P overlay mechanisms enabling efficient services such as data storage, replication, query resolution, and load balancing. With the advances in various distributed system technologies, the design of novel and efficient DHT-enabled solutions becomes more and more important. In this paper, we presented the first survey on the state-of-the-art DHT-based solutions in the emerging technological domains of edge, fog, and cloud computing, blockchains, IoT, MANETs, VANETs, and OSNs from system architecture, communication, routing, and technological perspectives, and identified their open problems along with future research guidelines. 

In the context of edge and fog computing, we studied DHT utilization as decentralized object storage platforms for a variety of data objects ranging from simple plain data objects to intermediary computation results and access control lists. In the domain of cloud computing, we studied the applications of DHTs in sharing data objects among the federated clouds, providing low-latency and highly-available storage and streaming services, and task managements on the cloud of containers. When coming to the blockchains, we showed that DHTs are utilized to support authentication, scalable storage management, efficient information dissemination, reliable communication, and augmenting the decentralized trust in the system. In the area of IoT, our survey presented DHT-based solutions addressing scalable service and information discovery. In the OSN domain, we surveyed DHT-based solutions enabling friend discovery, data sharing, spam protection, OSNs federation, reliability, and trustworthiness. Finally, in the context of MANETs and VANETs, we studied DHT-based solutions to improve the quality of sensing the environment and sharing the observations, provide distributed service directory, and adapt to the high mobility and frequent network topology changes.

\bibliographystyle{IEEEtran}
\bibliography{references}
\end{document}